\documentclass[12pt, twocolumn]{aastex701}
\usepackage{graphicx} 
\usepackage{float}
\usepackage{booktabs}
\usepackage{multirow}
\usepackage[T1]{fontenc}

\newcommand{\Mearth}{$M_\oplus$}

\begin{document}

\title{A JWST, ALMA and VLA survey of the Ophiuchus-A star-forming region: \\ Unveiling hidden dust mass and connecting infrared outflows to their radio origins}

\author[0009-0007-2837-8207]{Isaac C. Radley}
\affiliation{School of Physics and Astronomy University of Leeds, LS2 9JT, Leeds, UK}
\email[show]{py17icr@leeds.ac.uk}

\author[0000-0003-1008-1142]{John D. Ilee}
\affiliation{School of Physics and Astronomy University of Leeds, LS2 9JT, Leeds, UK}
\email[show]{J.D.Ilee@leeds.ac.uk}

\author[0000-0002-2189-6278]{Gemma Busquet}
\altaffiliation{Serra H\'unter Fellow}
\affiliation{Departament de Física Quàntica i Astrofísica (FQA), Universitat de Barcelona, Martí i Franquès 1, E-08028  Barcelona, Catalonia, Spain}
\affiliation{Institut de Ciències del Cosmos (ICCUB), Universitat de Barcelona, Martí i Franquès 1, E-08028 Barcelona,  Catalonia, Spain}
\affiliation{Institut d'Estudis Espacials de Catalunya (IEEC), Esteve Terradas 1, edifici RDIT, Parc Mediterrani de la Tecnologia (PMT) Campus del Baix Llobregat - UPC 08860 Castelldefels (Barcelona), Catalonia, Spain}
\email[]{}

\author[0000-0003-2300-2626]{Hauyu Baobab Liu}
\affiliation{Department of Physics, National Sun Yat-Sen University, No. 70, Lien-Hai Road, Kaohsiung City 80424, Taiwan, R.O.C.}
\affiliation{Center of Astronomy and Gravitation, National Taiwan Normal University, Taipei 116, Taiwan}
\email[]{}

\author[0000-0001-7552-1562]{Klaus M. Pontoppidan}
\affiliation{Jet Propulsion Laboratory, California Institute of Technology, 4800 Oak Grove Drive, Pasadena, CA 91109, USA}
\email[]{}

\author[0000-0003-3133-3580]{Alvaro Ribas}
\affiliation{Astronomy Unit, School of Physics and Astronomy, Queen Mary University of London, London E1 4NS, UK}
\email[]{}

\author[0000-0003-4721-034X]{Marc Audard}
\affiliation{Department of Astronomy, University of Geneva, Chemin Pegasi 51, 1290 Versoix, Switzerland}
\email[]{}

\author[0000-0001-9249-7082]{Eleonora Bianchi}
\affiliation{INAF, Osservatorio Astrofisico di Arcetri, Largo E. Fermi 5, 50125 Firenze, Italy}
\email[]{}

\author[0000-0001-7491-0048]{Tyler L. Bourke}
\affiliation{SKA Observatory, Jodrell Bank, Lower Withington, Macclesfield, SK11 9FT, UK}
\email[]{}

\author[0000-0003-1514-3074]{Claudio Codella}
\affiliation{INAF, Osservatorio Astrofisico di Arcetri, Largo E. Fermi 5, 50125 Firenze, Italy}
\email[]{}

\author[0000-0003-1805-3920]{Audrey Coutens}
\affiliation{Univ Toulouse, CNES, CNRS, IRAP, Toulouse, France}
 \email[]{}

\author[0000-0002-3829-5591]{Josep M. Girart}
\affiliation{Institut d'Estudis Espacials de Catalunya (IEEC), Esteve Terradas 1, edifici RDIT, Parc Mediterrani de la Tecnologia (PMT) Campus del Baix Llobregat - UPC 08860 Castelldefels (Barcelona), Catalonia, Spain}
\affiliation{Institut de Ciències de l'Espai (ICE-CSIC), Campus UAB, Carrer de Can Magrans S/N, E-08193 Cerdanyola del Vallès, Catalonia}
\email[]{}

\author[0000-0003-2684-399X]{Melvin G. Hoare}
\affiliation{School of Physics and Astronomy University of Leeds, LS2 9JT, Leeds, UK}
\email[]{}

\author[0000-0003-4493-8714]{Izaskun Jim\'enez-Serra}
\affiliation{Center of Astrobiology (CAB), CSIC-INTA, Ctra. de Ajalvir km 4, E-28850, Torrej\'on de Ardoz, Madrid, Spain}
\email[]{}

\author[0000-0002-6773-459X]{Doug Johnstone}
\affiliation{NRC Herzberg Astronomy and Astrophysics: Victoria, BC, CA}
\email[]{}

\author[0000-0002-5635-3345]{Laurent Loinard}
\affiliation{Instituto de Radioastronomía y Astrofísica, Universidad Nacional Autónoma de México, Morelia 58341, México}
\affiliation{Black Hole Initiative at Harvard University, 20 Garden Street, Cambridge, MA 02138, USA}
\email[]{}

\author[0000-0002-6648-2968]{Olja Pani\'c}
\affiliation{School of Physics and Astronomy University of Leeds, LS2 9JT, Leeds, UK}
\email[]{}

\author[0000-0002-3972-1978]{Jaime E. Pineda}
\affiliation{Max-Planck-Institut f\"ur extraterrestrische Physik, Giessenbachstrasse 1, D-85748 Garching, Germany}
\email[]{}

\author[0000-0003-2733-5372]{Linda Podio}
\affiliation{INAF, Osservatorio Astrofisico di Arcetri, Largo E. Fermi 5, 50125 Firenze, Italy}
\email[]{}

\author[0000-0002-6195-0152]{John J. Tobin}
\affiliation{National Radio Astronomy Observatory, Charlottesville, Virginia, USA}
\email[]{}

\author[0000-0003-1526-7587]{David J. Wilner}
\affiliation{Center for Astrophysics, Harvard \& Smithsonian, 60 Garden Street, Cambridge, MA 02138-1516, USA}
\email[]{}


\begin{abstract}
We present an infrared, millimetre, and radio survey of 20 Class~0--III young stellar objects in the Ophiuchus A L1688 star-forming cluster, combining high-resolution (7--25\,au) VLA and JWST observations with archival ALMA data.  We implement physically motivated models to derive dust and ionised gas properties, spectral behaviour and their relative contributions through the millimetre-centimetre radio spectral energy distribution. Our models reveal circumstellar dust disks that are, on average, tens to hundreds of times more massive than millimetre-only estimates (subject to uncertainties arising from the choice of dust opacity) and contain millimetre-sized grains even at the Class~0 stage. Owing to the VLA's high resolution we are able to connect outflows to their origins, detecting protostellar jet emission on scales of 10s--1000s\,au. Our results represent a homogeneous characterisation of the dust and ionised gas properties in Ophiuchus and present a potential solution to the long-standing `missing disk mass' problem. However, our understanding is still limited by resolution and sensitivity at frequencies $<40$\,GHz. Future facilities like the SKA and ngVLA are needed to provide the necessary capabilities to fully spatially resolve this emission ($<$\,0\farcs18) even in one of the closest star-forming regions. 
\end{abstract}


\section{Introduction}\label{sec:Intro}
The traditional paradigm of planet formation requires the oligarchal growth of micron-sized dust grains into km-sized planetary bodies, known as the core accretion model \citep[e.g.][]{Lissauer1993,Pollack1996}. This growth occurs in circumstellar disks formed during the embedded Class~0 and I protostellar phases which, as the surrounding envelope disperses through accretion onto the protostar, and through outflows, reveals the star-disk system commonly known as the Class~II phase \citep[][]{Lada87,Andre1993}. As the system evolves, several barriers hinder dust growth, including bouncing, fragmentation and radial drift, with the latter facilitating the inward transport of large grains through the disk and onto the protostar on timescales of $<\,$1\,Myr \citep[][]{Weidenschilling1977MNRAS.180...57W,Blum2008,Zsom2010A&A...513A..57Z,Brauer2008}. Planet formation, therefore, necessitates sufficient dust growth before or during the Class II phase in order to preserve the available solid material and overcome these barriers \citep[e.g. $<10^{5.5} - 10^7$ years,][]{Pascucci22}. 

Theoretical dust growth models such as the streaming instability offer a potential solution to these barriers, enabling the rapid formation of planetesimals up to 100s\,km in size \citep[see e.g.][]{Johansen2014,Simon2016}. Observationally, the growing catalogue of ring and gap structures in disks may be due to the presence of pressure bumps which would locally confine millimetre-sized dust grains in the outer disk, increasing the available timescale for dust growth and thus, planet formation \citep[][]{Pinilla2012,Benisty2015,Andrews_2018,Lau2022}. Such structures may have originated from dynamical interactions with embedded planets, a scenario supported by recent observations of planet-disk systems like PDS 70 and WISPIT 2 \citep[][]{Keppler2018,vanCapelleveen2025,Facchini2026}. The finding of similar substructures in a growing number of Class~0 and Class I objects may point to planet formation taking place in the first million years of the disk's lifetime \citep[e.g.][]{Nakatani2020,Segura-Cox2020,Michel23,Maureira2024,Hsieh2025}.

In addition, Class II young stellar objects (YSOs) appear to have reduced circumstellar disk masses compared to what would be expected from the broader exoplanet demographic \citep[see e.g.][]{Manara18}. Several explanations for this apparent deficit in solid mass have been proposed including high optical depths \citep[e.g.][]{Beckwith1990,Ballering2019,Ribas2020,Chung2024}, unresolved or hidden substructures \citep[e.g.][]{Tripathi_2017,Liu2022} and even early planet formation \citep[][]{Najita2014}. To better understand the existence of potential planet-induced substructures and mass evolution within disks we need to determine the level of dust growth and available solid mass contained in protoplanetary disks around YSOs spanning multiple evolutionary stages.

Characterising dust evolution observationally requires measurements spanning infrared, (sub-)mm and cm wavelengths, as the emissivity at the observing wavelength is closely tied to the maximum grain size being probed (e.g. $a_{\rm max} \sim \frac{\lambda}{2\pi}$). Infrared continuum observations trace $\sim$\micron-sized dust grains in the upper disk layers, often revealing disk geometry through features such as dark lanes and outflow cavities originating from disk winds and jets \citep[e.g.][]{Seale2008, Avenhaus2018, Duchene2024, Mullin2024}. In addition, infrared observations are sensitive to shock heated molecular emission, allowing us to probe the interface between outflowing material and the ambient ISM \citep[e.g. H$_2$ pure rotational lines,][]{Ray2023,Nakamura2025}. Observations in the (sub-)mm probe millimetre-sized dust grains which appear to gradually settle to the disk midplane as the system evolves \citep[e.g.][]{Villenave2023,Ohashi2023,Lin2023}. However, recent studies suggest that this emission is optically thick, leading to underestimated flux densities and potentially obscured substructures \citep[][]{Ballering2019,Liu2019ApJ...877L..22L,Zhu2019,Ribas2020,Macias2021,Rilinger2023,Chung2024,Radley2025,Hsieh2025}. Centimetre observations probe cm-sized dust grains and are typically more optically thin, offering a means of overcoming these effects \citep[][]{CarrascoGonzalez2019,Viscardi2025}. Unfortunately, dust emission at centimetre wavelengths is inherently faint due to its low emissivity \citep[e.g. $\kappa_\nu\propto\nu^{\beta}$;][]{Beckwith1990} with the interpretation of continuum observations further complicated by contributions from ionised gas emitting in the inner disk through disk winds \citep[][]{Pascucci12}, protostellar jets \citep[][]{Anglada18} and magnetospheric accretion funnels \citep[][]{Dzib13,Liu2014ApJ...780..155L}.

At low frequencies ($\lesssim$\,30\,GHz), the spectral energy distribution (SED) is often dominated by emission from such ionised gas mechanisms. These mechanisms can be broadly separated into thermal free-free emission and non-thermal gyro-emission, distinguished through their spectral index, $\alpha$, where F$_\nu \propto \nu^{\alpha}$. Thermal free-free emission can arise from protostellar jets, described by a spectral index between 0.1--1 and, often exhibiting highly collimated morphologies aligned with large-scale outflows \citep[][]{Anglada1995,Tychoniec2018_extendedEmission,Anglada18}. Alternatively, thermal free-free can originate in magnetohydrodynamic (MHD) or photoevaporative disk winds \citep[see][for reviews]{Pudritz2007,Alexander2014}. These winds are believed to have a turnover frequency at $\sim1$\,GHz resulting in optically thin free-free emission at higher frequencies \citep[e.g. $\alpha\sim-0.1$,][]{Reynolds1986,Pascucci12}. 
In contrast, non-thermal emission, typically arising from magnetospheric gyrosynchrotron radiation has been observed to have $\alpha < -0.1$ with a theoretical minimum of --2 \citep[][]{Dulk1985,Dzib13,Dzib2015}. However, each of the above mechanisms exhibits intrinsic variability on timescales of minutes to years \citep[e.g.][]{Liu2014ApJ...780..155L,Ubach2017,Forbrich2017}. Protostellar jets may activate intermittently through episodic accretion events, while non-thermal gyrosynchrotron radiation varies much more rapidly due to active magnetospheres which produce stellar flares \citep[e.g.][]{Machida2014,Lovell2024ApJ...962L..12L}. Therefore, whilst we may infer certain physical mechanisms from the spectral index, its observed value may depend on the local stellar environment and active mechanisms at the time of observation. In addition, ionised gas processes dominate emission in the inner disk ($\lesssim$10\,au), necessitating high angular resolution observations in order to accurately characterise their emission \citep{Rota2024}.

Consequently, a high angular resolution, multiwavelength approach is necessary to disentangle contributions from dust and ionised gas emission, enabling a more complete recovery of the dust growth properties and available solid mass reservoir. To this end, we focus on the L1688 core of the Ophiuchus molecular cloud, one of the closest star forming regions at a distance of 138.4 pc \citep[][]{OrtizLeon2018}. The proximity of Ophiuchus has previously enabled high-resolution studies of both individual YSOs and systems (e.g. VLA~1623~AaAb, B and W, \citealt{Harris18}; GSS30-IRS3, \citealt{SantamariaMiranda_2024}) as well as broader surveys such as the Ophiuchus DIsc Survey Employing ALMA \citep[ODISEA,][]{Cieza_2019}. The present work extends the frequency coverage of \citet{Coutens19} and sample size of \citet{Radley2025} to enable a multiwavelength survey of Class~0--III YSOs in Ophiuchus.

In Section~\ref{sec:Observations} we present the data reduction and calibration of observations from the Karl G.\ Jansky Very Large Array (VLA), James Webb Space Telescope (JWST) and, Atacama Large Millimetre Array (ALMA). Morphologies and flux densities for our VLA observations are shown in Section~\ref{sec:VLA_obs} and connected to near-infrared (NIR) observations in Section~\ref{sec:JWST_obs}. In Section~\ref{sec:SpIx} we present a robust characterisation of the dust and ionised gas spectral indices for each object, whilst in Section~\ref{sec:DustPopSED} we use a physical model to interpret the radiative contributions of dust and ionised gas in each protostellar system. We discuss our results in Section~\ref{sec:discussion} and summarise our conclusions in Section~\ref{sec:Conclusions}.

\begin{deluxetable*}{llcccccl}
\tablecaption{Observed sample, positions and protostellar characteristics.\label{tab:introTable}}
\tabletypesize{\footnotesize}
\tablehead{
\colhead{Object} &
\colhead{Position} &
\colhead{Class} &
\colhead{A$_{\rm V}$} &
\colhead{T$_{\rm bol}$} &
\colhead{L$_{\rm bol}$} &
\colhead{M$_{*}$} &
\colhead{Other Names}  \\
\colhead{} &
\colhead{J2000} &
\colhead{} &
\colhead{mag} &
\colhead{K} &
\colhead{L$_{\odot}$} &
\colhead{M$_{\odot}$} &
\colhead{} 
}
\startdata
VLA~1623~B & 16:26:26.30 $-$24:24:30.84 & 0 & 0.0 & $\cdots$ & $\cdots$ & 1.9 & $\cdots$ \\
VLA~1623~Ab & 16:26:26.38 $-$24:24:31.00 & 0 & $\cdots$ & $\cdots$ & $\cdots$ & $\cdots$ & $\cdots$ \\
VLA~1623~Aa & 16:26:26.40 $-$24:24:30.93 & 0 & $\cdots$ & $\cdots$ & $\cdots$ & $\cdots$ & $\cdots$ \\
SM1 & 16:26:27.85 $-$24:23:59.73\tablenotemark{a} & 0 & $\cdots$ & $\cdots$ & $\cdots$ & $\cdots$ & $\cdots$ \\
CRBR12 & 16:26:17.23 $-$24:23:45.91 & I & 9.8 & 570 & 0.28 & $\cdots$ & ISO-Oph 21 \\
GSS30-IRS1 & 16:26:21.36 $-$24:23:05.03 & I & 9.8 & 300 & 14.0 & $\cdots$ & ISO-Oph 29; Oph-emb 8 \\
GSS30-IRS3 & 16:26:21.72 $-$24:22:51.10 & I & $\cdots$ & $\cdots$ & $\cdots$ & 0.35 & ISO-Oph 31 \\
LFAM3 & 16:26:23.57 $-$24:24:40.14 & I & 9.8 & 810 & 0.97 & 0.77 & ISO-Oph 37; GY 21 \\
CRBR36 & 16:26:25.47 $-$24:23:01.98 & I & 9.8 & $\leq$280 & $\geq$0.2 & $\cdots$ & Oph-emb9 \\
VLA~1623~W & 16:26:25.63 $-$24:24:29.63 & I & 9.8 & $\leq$120 & $\geq$0.066 & $\cdots$ & $\cdots$ \\
VSSG27 & 16:26:30.45 $-$24:22:57.90 & I & $\cdots$ & $\cdots$ & $\cdots$ & 0.66 & ISO-Oph 46 \\
GSS26 & 16:26:10.33 $-$24:20:55.40 & II & 27.0 & 3300 & 6.7 & 0.69 & ISO-Oph 17 \\
GSS29 & 16:26:16.85 $-$24:22:23.75 & II & 10.8 & 3100 & 2.1 & $\cdots$ & ISO-Oph 19 \\
DoAr24 & 16:26:17.06 $-$24:20:22.18 & II & 4.2 & 3300 & 1.2 & 0.83 & ISO-Oph 20 \\
CRBR15 & 16:26:18.98 $-$24:24:14.83 & II & 9.8 & 1000 & 0.11 & 0.14 & ISO-Oph 26 \\
GSS30-IRS2 & 16:26:22.39 $-$24:22:53.58 & II & $\cdots$ & $\cdots$ & $\cdots$ & $\cdots$ & ISO-Oph 34 \\
DoAr24Ea & 16:26:23.36 $-$24:21:00.16 & II & 8.5 & 3000 & 7.8 & $\cdots$ & ISO-Oph 36 \\
DoAr24Eb & 16:26:23.43 $-$24:21:01.90 & II & 8.5 & 3000 & 7.8 & $\cdots$ & ISO-Oph 36 \\
S2 & 16:26:24.04 $-$24:24:48.66 & II & 15.9 & 3000 & 7.8 & 1.12 & ISO-Oph 39 \\
GSS35 & 16:26:34.17 $-$24:23:28.86\tablenotemark{a} & III & $\cdots$ & $\cdots$ & $\cdots$ & $\cdots$ & ISO-Oph 48; S1 \\
\enddata
\tablenotetext{a}{Not detected at 44\,GHz so measured from 22\,GHz object position.}
\tablecomments{Object names and protostellar classes are consistent with the \citet{Coutens19} sample, while positions are derived from Gaussian fits to the Q-band continuum presented in this work, unless stated otherwise. We take A$_{\rm V}$, T$_{\rm bol}$ and, L$_{\rm bol}$ from \citet{Evans2009}, where available. Stellar masses are taken from \citet{Sadavoy2024} (VLA~1623~B, W), \citet{SantamariaMiranda_2024} (GSS30-IRS3) with the remaining taken from \citet{Shoshi2025}.}
\end{deluxetable*}

\section{Observations}\label{sec:Observations}
\begin{figure*}[h]
    \centering
    \includegraphics[height=0.86\textheight,width=\textwidth,keepaspectratio]{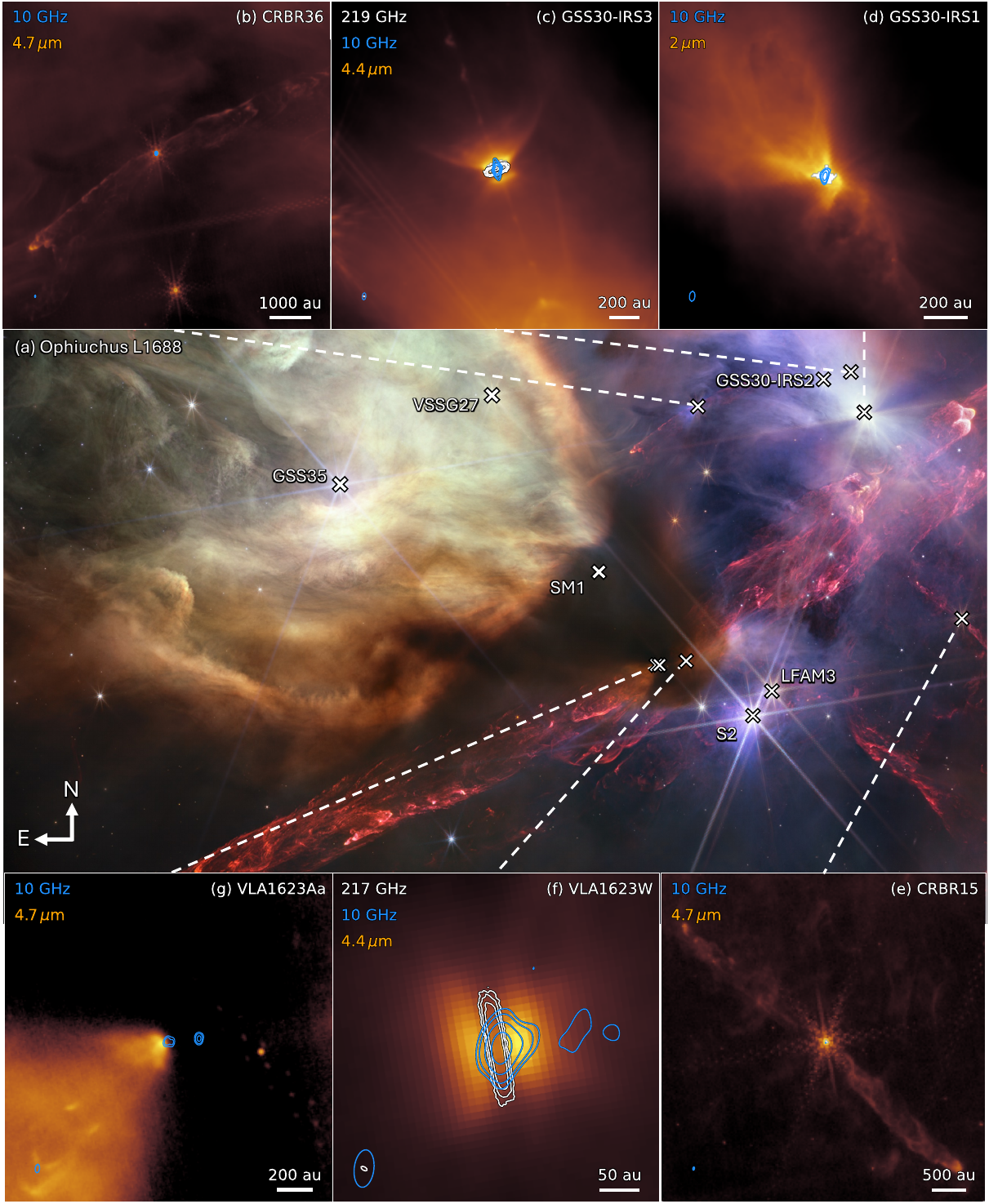}
    \caption{(a) NIRCam first anniversary image of the Ophiuchus L1688 core (NASA/ESA/CSA/STScI) overlaid with white crosses representing the positions of a subset of objects in our sample. Image colours correspond to emission from the following filters: F187N (blue), F200W (light blue) F335W (cyan), F444W (yellow) and F470N (Red). Inset panels (b)--(g) show zoomed in 2\,\micron\,, 4.4\,\micron\, and 4.7\,\micron\, images corresponding to the F200W, F444W and F470N filters, respectively. In each panel we overlay 10\,GHz continuum contours (blue) corresponding to 3-, 5-, 10-, and 20-$\sigma$, while panels (c) and (f) additionally show ALMA 217-219\,GHz contours (white) corresponding to 5-, 10-, 20-, and 100-$\sigma$ with $\sigma$ as defined in Table~\ref{tab:AllUsedFluxes}. Note the central white regions of panels (c) and (d) are saturated in NIRCam containing no data.}
    \label{fig:ImpactImage}
\end{figure*}

We have targeted 20 objects in the Ophiuchus L1688 core spanning Class~0--III drawn from the initial sample of YSOs studied in \citealt{Coutens19}.  These objects were selected on the basis of having both follow-up VLA observations (described in Section 2.1) and ALMA observations in the literature or archive, allowing us to assemble a mm-cm spectral energy distribution.  Table~\ref{tab:introTable} gives they key information for the sample including positions, alternative nomenclature and other physical parameters found in the literature for each object. In Figure~\ref{fig:ImpactImage} we present the JWST first anniversary image of L1688 overlaid with several JWST, ALMA and VLA observations. These images demonstrate the synergies between each facility as well as the range of scales (7--1000s of\,au) and star forming environments covered in this survey. In the following subsections we describe the observational set up for each telescope and instrument along with the steps taken to calibrate and image each observation.

\subsection{VLA}\label{sec:VLA_DataReduction}

Observations of the Ophiuchus L1688 core were carried out across a total of 18 epochs (3 March -- 18 April 2022) using the VLA of the National Radio Astronomy Observatory. These observations used the Q- (44\,GHz), K- (22\,GHz) and X- (10\,GHz) bands for each epoch (project code 22A-164, P.I. Busquet). Nine individual pointings were observed across two epochs for the Q- and K-band observations and two pointings across 18 epochs for the X-band (see Appendix~\ref{apx:VLA_ObsParams} Table~\ref{tab:obsparams}). All observations were taken with the array in A-configuration leading to a maximum baseline range of $\sim$34--37~km resulting in a theoretical synthesised beam of 0\farcs04, 0\farcs08, and 0\farcs17 at Q, K and X band. Our minimum baselines range between $\sim$500--800m correlating to a theoretical maximum recoverable scale of 1\farcs1--1\farcs7 (150--240\,au), 2\farcs2--3\farcs5 (300--500\,au) and 4\farcs7--7\farcs5 (650--1000\,au) for the Q-, K- and X-band observations respectively. Data was acquired using 3-bit samplers, with bandwidths of 2$\times$2~GHz in the X band and 4$\times$2~GHz for Q and K bands, in full polarization mode. 

We processed the data using the VLA Calibration Pipeline\footnote{https://science.nrao.edu/facilities/vla/data-processing/pipeline} within CASA \citep[version 6.5.2,][]{CASA}. Additionally, we applied one round of phase self-calibration for five of the Q-band observations with model images created by masking regions which contained strong emission from YSOs, incorporating this emission into the model. We combined spectral windows and scans, and used a solution time interval of infinity for all Q-band self calibration. Signal-to-noise typically increased by $\sim10\%$ in each observation. In addition, we carried out self-calibration for 15 out of 18 K-band observations following a similar approach to the Q-band. Once again, we combined spectral windows and scans but used a range of solution intervals from 10s to infinity depending on the amount of data flagged. We find typical signal-to-noise ratio (SNR) increases of $\sim$ few percent. Despite the modest increases in the SNR, the act of self-calibration mainly serves to  improve image fidelity by suppressing the impact of sidelobes. Finally, due to the large number of X-band observations and increased computational cost of self-calibration, we do not self-calibrate each epoch instead opting to improve signal-to-noise through combining measurement sets as described below.

\subsubsection{Mitigating variability}\label{sec:variability}

YSOs have been observed to be inherently variable at low frequencies ($\lesssim30$\,GHz) with significant flux density variations reported over timescales of minutes to weeks \citep[e.g.][]{Dzib13,Liu2014ApJ...780..155L,Ubach2017,Vargas-Gonzales2023}. To ensure such variability does not impact our derived flux densities significantly we must assess the level of variability for each object across each band. Using the self-calibrated datasets for Q- and K-band and the calibrated X-band datasets, we create initial images of each field and band using a Briggs robust of 2 to maximise the sensitivity. We then apply the same approach as \citet{Radley2025} to determine variability across each field and band. Briefly, we extract flux densities from each object using single 2D Gaussian fits to the continuum calculated by the Python Blob Detector and Source Finder package \citep[PyBDSF,][]{PyBDSF}. Next, we implement the variability criteria from \citet{DiazMarquez24} such that the maximum flux density deviation does not exceed 3 times the combined uncertainty of each measurement to determine whether each object is variable across each band.

To produce the fiducial images for each Q- and K-band pointing we combine epochs where objects are found to be non-variable, minimising the effect of variability on the derived flux densities. At Q and K band we find GSS29 exceeds the above variability criteria whilst GSS30-IRS2 exceeds it only at K band. GSS30-IRS2 is detected in two separate pointings which are each observed for two epochs. The object remains non-variable in one pointing over both epochs while in the second pointing shows signs of variability between the two epochs. Therefore, we opt to combine the non-variable pointing for the fiducial image. GSS29 exhibits strong variability across both Q- and K-band which are both observed with 2 epochs each. We therefore use the epoch in which the object has a lower extracted flux density as our fiducial images. This approach minimises the impact of YSO flaring events which can lead to order of magnitude flux density increases compared to more quiescent states \citep[][]{Forbrich2017}. In addition, in the field of DoAr24 there is a strongly flaring source, with the flare present in only one epoch. We choose to use the non-flaring epoch for our fiducial image to reduce the risk of potential flux contamination and improve image fidelity. Finally for objects that are not detected in both individual epoch images nor the combined epoch images such as SM1, we report 3$\sigma$ upper limits based on the local RMS of the combined image.

For the X band, we have 18 epochs at two individual pointings. Following a similar procedure as for Q and K band, we combine all epochs for a pointing in which an individual object is non-variable, leading to a maximum stacking of 18 epochs (e.g. for VLA~1623~AaAb). For example, if an object satisfies the variability criteria for 10/18 observations, then we combine the 10 in which the object is non-variable. Where an object is covered in both X-band pointings, we choose the detection with the highest SNR. We note that a more detailed analysis of the X-band variability will be presented in Busquet et al. (in prep.).

After implementing the above variability checks, we combined non-variable measurement sets, imaging the combined observations with a Briggs robust of 0.5 and the single epoch images with a robust of 2. Finally, each image was primary beam corrected using CASA \textit{impbcor} to ensure accurate fluxes could be extracted. We present our fiducial image parameters for all Q-, K-, and X-Band observations in Appendix~\ref{apx:VLA_ObsParams} (Table~\ref{tab:obsparams}). On average, our final images have minor beam axes 0\farcs05 (7\,au), 0\farcs09 (12\,au) and 0\farcs18 (25\,au) for Q (44\,GHz), K (22\,GHz) and X (10\,GHz) bands respectively.

\subsection{JWST/NIRCam}\label{sec:JWST_DataReduction}

The L1688 core of the Ophiuchus star forming region was observed using the near infrared camera (NIRCam) instrument \citep{Rieke23} on the 7$^{\text{th}}$ March, 5$^{\text{th}}$ April and 6$^{\text{th}}$ April 2023 with six filters: F115W, F187N, F200W, F335M, F444W, and F470N (PID: 2739, P.I. Pontoppidan). These observations resulted in the production of the JWST first anniversary image, which can be seen in panel (a) of Figure~\ref{fig:ImpactImage} and covers $\sim$49 square arcminutes containing the majority of our sample. The observational program and calibration followed the strategy of \citet[][]{Pontoppidan2022} using a 3$\times$2 mosaic and 71.5\% row overlap ensuring uniform depth across the image. The FULLBOX+6TIGHT dither pattern was employed for each tile yielding maximum exposure times of 2834\,s for the narrow-band filters (N) and 1416-1674\,s for wide- and medium-band filters (W and M). All images were processed using the JWST calibration pipeline \citep{Bushouse23} version 11.16.21 with CRDS context \texttt{jwst$\_$1077.pmap}.

In this work we focus solely on the F200W, F444W and F470N filters from the above dataset. The F200W and F444W wideband filters have a central wavelength of 2\micron\ and 4.4\micron, respectively, and are primarily composed of continuum emission from scattered light and dust thermal emission. The F470N filter is centred on 4.7\micron\ and is dominated by emission from the H$_2$S(9) rotational line. The final image resolution is wavelength dependent and ranges from approximately $0\farcs06$--$0\farcs12$ between 2\micron\ and 4.7\micron\ (or 8--17\,au at the distance of Ophiuchus).

\begin{figure*}[htbp]
    \centering
    \includegraphics[width=0.9\textwidth,keepaspectratio]{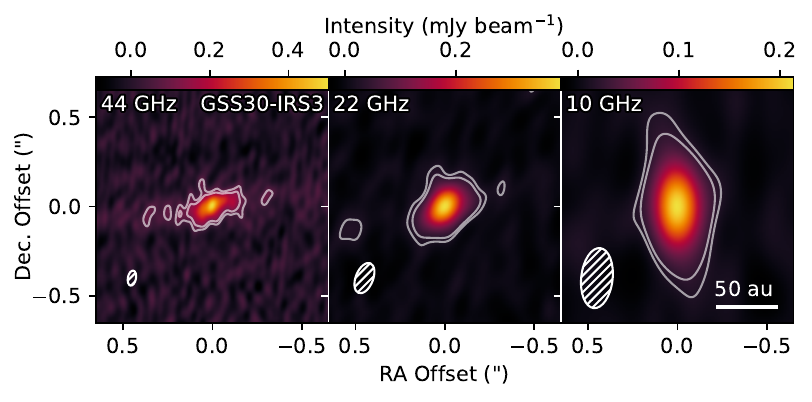}
    \caption{VLA 44\,GHz (left), 22\,GHz (middle) and 10\,GHz (right) continuum images for GSS30-IRS3 centred on the position shown in Table~\ref{tab:introTable}. The transition in morphology from disk-like (44\,GHz) to extended emission perpendicular to the disk (10\,GHz) reveals a changing dominant emission mechanism with frequency. We overlay 3- and 5-$\sigma$ contours in white with $\sigma=$ RMS as defined in Table~\ref{tab:AllUsedFluxes}. Beam sizes are shown in the bottom left of each panel with a shared 50\,au scale bar shown in the bottom right of the final panel.}
    \label{fig:VLACont_GSS30_VLA1623W}
\end{figure*}

\subsection{ALMA}\label{sec:ALMA_DataReduction}
To help characterise the spectral behaviour of the objects we searched for complementary \mbox{(sub-)mm} flux density measurements. In the first instance, these flux densities are obtained from literature results (e.g. \citealt{Kirk2017}, \citealt{Cieza_2019}). We also searched the ALMA Science Archive for observations of our targets at millimetre wavelengths. We used the ALMA User-Defined Imaging (AUDI\footnote{\url{https://science.nrao.edu/srdp}}) service to generate continuum images for our targets with available data in the ALMA Science Archive.  AUDI reprocesses pipeline-calibrated measurement sets using the current ALMA imaging pipeline, thereby incorporating the latest improvements in continuum determination, automasking, and self-calibration. The service delivers both non-self-calibrated and self-calibrated images, with self-calibration applied when feasible. Verification steps include ensuring that the images achieve the expected RMS sensitivity, the requested synthesized beam size, and are free of obvious artifacts. We found suitable observations for 9 objects across 3 ALMA Bands under project codes 2017.1.00107.S (P.I. Hirano) and 2019.A.00034.S (P.I. Tobin). All imaging was performed with multi-term multi-frequency synthesis with two terms \citep{Rau2011A&A...532A..71R}.

From ALMA project code 2017.1.00107.S we extracted flux densities from primary beam corrected images for several objects at ALMA Bands 3 (85\,GHz), 4 (148\,GHz) and, 6 (224\,GHz) using PyBDSF as described in Section~\ref{sec:variability}. In addition, for GSS30-IRS1 and IRS3, we similarly extract Band 6 (219\,GHz) flux densities from project code 2019.A.00034.S. We summarise the image parameters in Appendix~\ref{apx:ALMA_ObsParams} (Table \ref{tab:ALMAProjects}) and the extracted fluxes and morphologies in Table~\ref{tab:AllUsedFluxes}. Observations from project 2017.1.00107.S have a typical beam major axis between 1\farcs16 and 1\farcs23 compared to 0\farcs38 for ALMA project 2019.A.00034.S and 0\farcs09--0\farcs34 for our VLA observations. At this coarser resolution, the disk morphologies are less clear, making morphological comparisons with the higher resolution data unfeasible. We therefore restrict our use of this data to focus solely on their extracted flux densities.

\section{Results \& Analysis}

\subsection{VLA morphology \& flux densities}\label{sec:VLA_obs}
\raggedbottom
We present all fiducial VLA images in Appendix~\ref{apx:VLA_Obs} Figures~\ref{fig:VLA_Cont_0}-\ref{fig:VLA_Cont_2} with image parameters as described in Table~\ref{tab:obsparams}. Integrated flux densities and deconvolved sizes were extracted using 2D Gaussian fits with PyBDSF which we present in Appendix~\ref{apx:MCMC_Fluxes} Table~\ref{tab:AllUsedFluxes}. Uncertainties in the measured integrated flux density are calculated using the uncertainty from PyBDSF added in quadrature with the calibration uncertainty for each Band (10\% for Q and K band; 5\% for X band). The local noise level, RMS, is estimated with CASA \textit{imstat} using an inner-outer annuli of 0\farcs6-0\farcs9, 0\farcs6-1\farcs1 and, 0\farcs9-1\farcs9 centred on the peak of the emission for Q, K and X band respectively. The high angular resolution of our observations allow us to resolve or marginally resolve the majority (16/18) of our sample at 44\,GHz. We define a source as resolved if both its deconvolved major and minor axes are non-zero, and marginally resolved if at least one deconvolved axis is non-zero. However, the fraction of marginally or fully resolved objects quickly decreases with decreasing frequency going from 15/20 at 22\,GHz, down to 10/20 at 10\,GHz. Interestingly, two objects, GSS30-IRS2 and GSS29, remain unresolved across all frequencies indicating a very compact emission region. 

We find a range of deconvolved source diameters at 44\,GHz with the largest being the edge-on disk VLA~1623~W with a diameter of 77\,au. If we exclude this particularly large object we find that the average source diameter appears to tentatively decrease from 13\,$\pm\,$3\,au for Class~0 objects, down to 9\,$\pm\,$1\,au for Class~II objects, although they remain consistent within errors. Interestingly, Class~I sources have an average diameter of 15\,$\pm\,$3\,au sitting slightly above Class~0 but consistent within errors. This same trend can be seen for the 22\,GHz observations with an average diameter of 14\,$\pm\,$2\,au for Class~0, 16\,$\pm\,$1\,au for Class~I and 10\,$\pm\,$2\,au for Class~II. However, it should be noted that low frequency observations are increasingly dominated by ionised gas mechanism resulting in a combination of morphologies from large dust grains in the disk and ionised gas processes such as winds and jets \citep[e.g.][]{Macias16}. The specific emission contributions for each object at each frequency are discussed further in Section \ref{sec:SpIx}. Furthermore, given the larger spatial extents of dust disks at ALMA frequencies \citep[e.g.][]{CarrascoGonzalez2019}, there may exist faint continuum emission from larger grains in the outer disk which remain undetected below our sensitivity limits.

Whilst most of our observations at 10\,GHz are unresolved, we can gain some morphological insight for marginally resolved sources through extended emission at the 3- to 5-$\sigma$ level.
In Figure~\ref{fig:VLACont_GSS30_VLA1623W} we see GSS30-IRS3 which is resolved at 10\,GHz along both major and minor axes and has a notable vertical extension, perpendicular to the source at 44\,GHz. Interestingly, the extension to the north also appears at 22\,GHz and very marginally at 44\,GHz, hinting that the emission mechanism responsible emits with decreasing strength for increasing frequency.  The collimated morphology of the 10\,GHz extension is consistent with previous observations of ionised protostellar jets at lower frequencies \citep[e.g.][]{Tychoniec2018_extendedEmission}. The emission at 10\,GHz is expected to be more dominated by processes such as protostellar jets and winds whereas the emission at 44\,GHz is more likely to probe dust emission from the disk. Protostellar jet emission is observed to align perpendicular to the dust disk \citep[][]{Anglada18} whereas disk winds have been observed to launch angled at 10s of degrees above the circumstellar disk (e.g. 36\degr, \citealt{Duchene2024}; 55\degr, \citealt{Arulanantham2024}). Finally, magnetospheric emission is expected to be extremely compact due to its stellar origin. For objects with resolved emission at both 44\,GHz and 10\,GHz we can compare the difference in position angles, $\Delta$PA = PA$_{\rm 44\,GHz}-$PA$_{\rm 10\,GHz}$, where individual position angles are taken from Table~\ref{tab:AllUsedFluxes}, in order to morphologically infer which emission mechanism may be contributing at 10\,GHz. We find six objects (GSS30-IRS3, LFAM3, GSS26, VLA~1623~AaAb and VLA~1623~W) which have $\Delta$PA$\sim$\,90\degr\, indicating potential jet-like emission at 10\,GHz.

\begin{figure*}[t]
    \centering
    \includegraphics[width=\textwidth,trim=0cm 0cm 0cm 0.5cm,keepaspectratio]{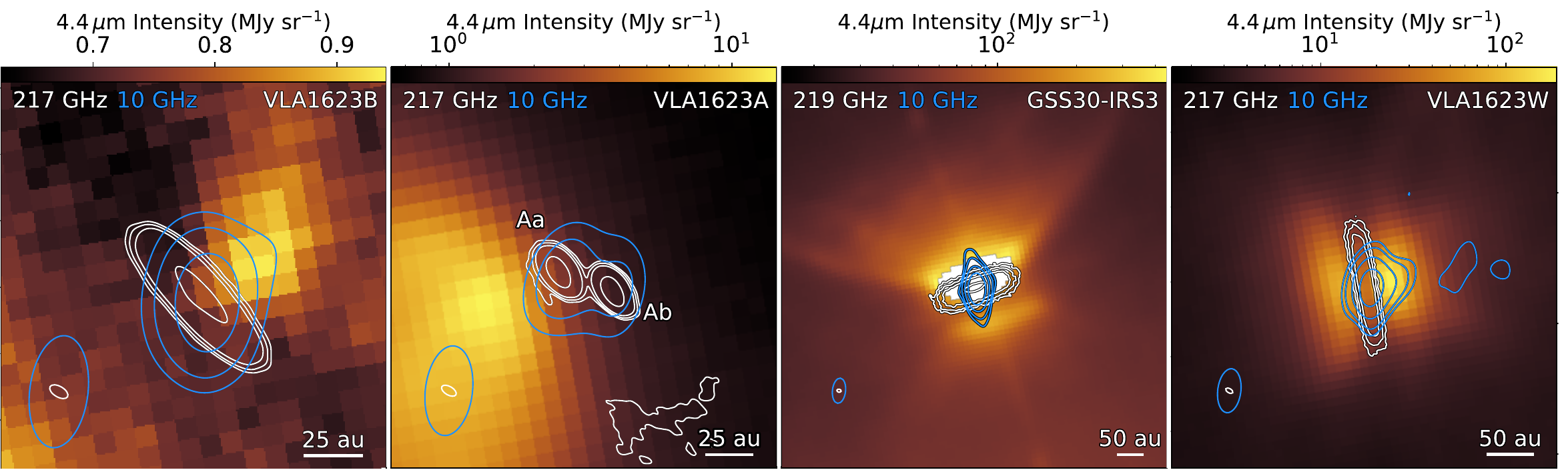}
    \caption{Zoomed 4.4\,$\mu$m NIRCam image with ALMA 217-219\,GHz continuum (white contours) tracing the disk morphology and VLA 10\,GHz continuum (blue contours) revealing extended emission along outflow cavities. Contours correspond to 5-, 10-, 20-, and 100-$\sigma$ (ALMA) and 3-, 5-, 10-, and 20-$\sigma$ (VLA), with $\sigma$ as defined in Table~\ref{tab:AllUsedFluxes}. Note: the central region of GSS30-IRS3 is highly saturated meaning there is no data in the white region.}
    \label{fig:10GHzExtensions}
\end{figure*}

\subsection{Connecting radio and NIR morphologies} \label{sec:JWST_obs}

The high extinction of the Ophiuchus L1688 core (see Table~\ref{tab:introTable}) and embedded nature of Class~0 and I objects can significantly impact the detection of infrared sources and their potential outflow cavities. However, the high resolution and sensitivity of JWST NIRCam enables an unprecedented view of YSO disks, outflow cavities and collimated jet emission in the NIR. While the combination of our VLA survey and archival ALMA observations allows us to probe emission mechanisms on $<$10s\,au scales, the use of NIR observations present a way to understand their impact on the broader astrophysical environment (e.g. 100--1000s\,au). The synergy between these three facilities allows us to connect protostellar disks and outflows across multiple spatial scales, dust populations and frequency regimes.

\subsubsection{Emission on 10--100s au scales}

We start by considering the smallest spatial scales ($<100$\,au), which primarily probe the disk and dense inner regions of outflows. In Figure~\ref{fig:10GHzExtensions} we present a subset of NIRCam wideband 4.4\,$\mu$m observations which are sensitive to both thermal continuum emission and, scattered light from micron-sized dust grains. The outflow cavities associated with VLA~1623~A and B, as well as the edge-on disk of VLA~1623~W have previously been reported in \citet[][]{Radley2025} while GSS30-IRS3 has been presented in the appendix of \citet[][]{SantamariaMiranda_2024}. The observations shown in Figure~\ref{fig:10GHzExtensions} indicate the presence of micron-sized grains within outflow cavity walls as well as the upper layers of the protoplanetary disks.

As discussed previously, we detect several 10\,GHz continuum extensions originating from several YSOs. We find that some of these extensions are aligned with outflowing material from the central star on 10s to 100s of au scales as seen in the JWST NIRCam 4.4\,$\mu$m observations. The alignment of the emission at 10\,GHz with the outflow cavities detected at 4.4\,$\mu$m indicates that we are likely tracing the mechanism responsible for creating these outflows e.g. a jet or a wind. Assuming the emission at 10\,GHz is dominated by outflow processes, we can estimate the potential size of the launching region for each object by considering their deconvolved sizes (see Table~\ref{tab:AllUsedFluxes}). We find the smallest launching region for VLA~1623~Aa which has a radius of $\sim$2.5\,au whereas VLA~1623~B, VLA~1623~W and GSS30-IRS3 have radii between 7 and 10\,au. These radii should be regarded as upper limits due to potential contamination from dust emission and their marginally resolved nature. Nevertheless, the generally small radii indicate an origin within the innermost regions of the disk, consistent with previously inferred jet launching regions \citep[e.g.][]{CarrascoGonzalez2019,Ohashi22}.

In the case of VLA~1623~W, our combination of multiple epochs leading to improved sensitivities have allowed us to confirm previously detected extended 10\,GHz emission at the $\geq5\sigma$ level \citep[see e.g.][]{Radley2025}. We present these upgraded observations alongside ALMA 217\,GHz and JWST 4.4\micron\, observations in Figure~\ref{fig:10GHzExtensions}. Similarly to GSS30-IRS3, the PA at 10\,GHz is perpendicular to the dust continuum probed by ALMA, indicative of a protostellar jet. However, as yet, VLA~1623~W has no confirmed detections of a jet-like component in either (sub-)mm or NIR wavelengths. In contrast, we detect extended emission to the West and several 3-$\sigma$ blobs aligned with the 10\,GHz PA and extended emission. This asymmetric extended emission has been previously inferred to explain westward astrometric shifts in the low-frequency morphology (e.g. Hern\'andez Garnica et al. submitted.) Interestingly, we find that the 10\,GHz extension is aligned with the brighter lobe of the edge-on disk seen with JWST, potentially indicating this is the nearside of the disk.

\subsubsection{Emission on 100--1000s au scales}
Moving to larger spatial scales ($>$100\,au) we find a diversity of large-scale NIR structures such as outflow knots and collimated emission at a range of wavelengths from 2\micron--4.7\micron. The wide band emission at 2\micron\ and 4.4\micron\ mostly originates from thermal and scattered light emission of dust grains. These tracers, therefore, can be used to infer the distribution of small dust grains in the surrounding envelope and cloud. Conversely, the narrow-band emission at 4.7\micron\ originates from shock-excited molecular hydrogen making it a useful probe for protostellar jet working surfaces \citep[e.g.][]{NoriegaCrespo2004}.

\begin{figure*}[htbp]
    \centering
    \includegraphics[height=0.84\textheight,width=\textwidth,keepaspectratio,trim=1cm 0 1cm 0cm]{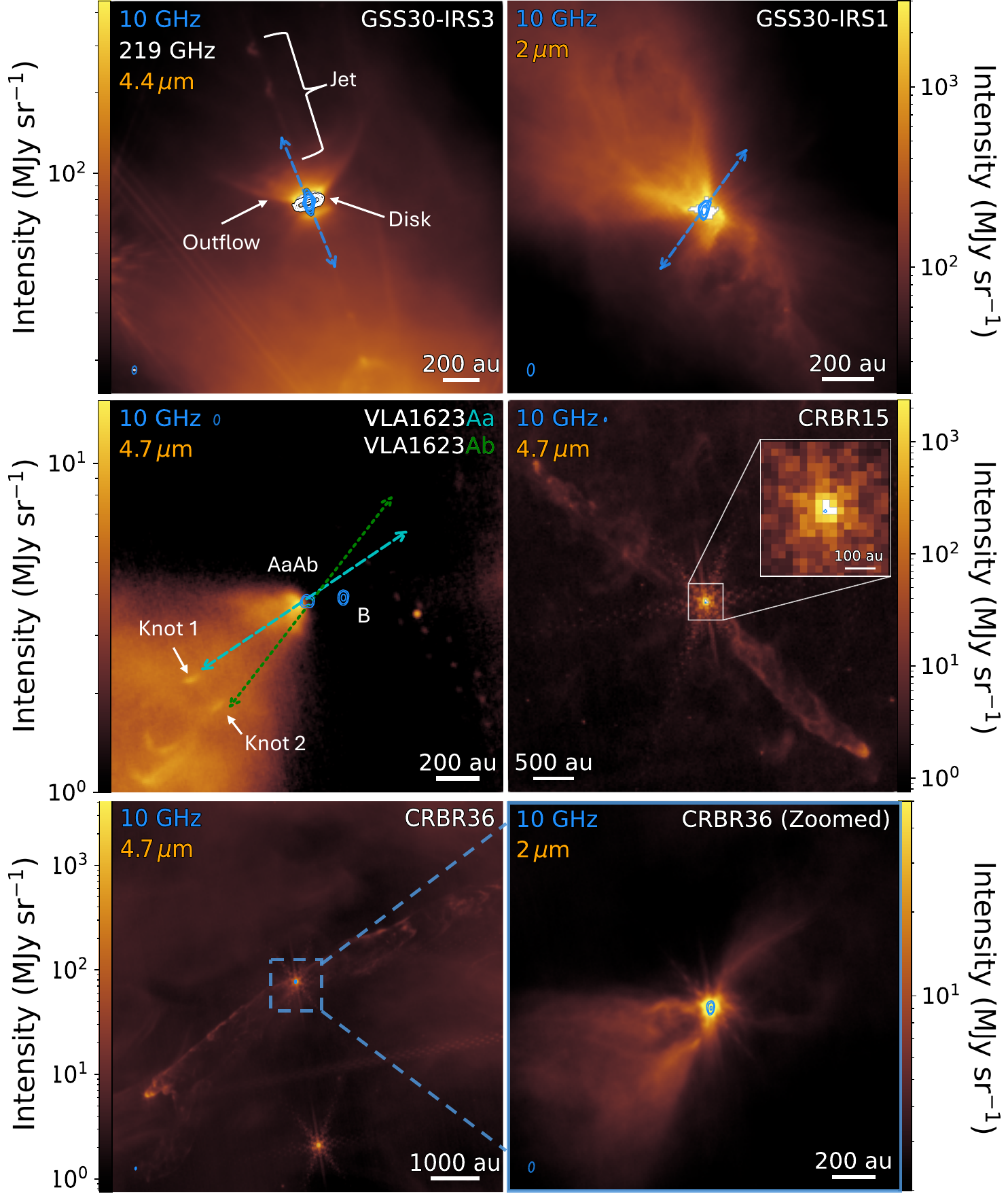}
    \caption{Large-scale NIRCam images reveal extended outflows, protostellar jets and knot-like features on 100s--1000s au scales with each panel showing one of three filters: 2\,\micron\ (continuum, scattered light), 4\,\micron\ (continuum, scattered light) and, 4.7\,\micron\ (H$_2$S(9) rotational line). Overlaid ALMA continuum emission (white contours) traces the disk morphology while VLA 10\,GHz continuum (blue contours) traces ionised gas emission, illustrating the connection of these three facilities across multiple spatial scales. Position angles of the 10\,GHz morphology are overlaid as blue (GSS30-IRS1, IRS3), cyan (VLA~1623~Aa) and green (VLA~1623~Ab) dashed arrows. Excluding GSS30-IRS1, NIRCam images use lower data-range limits of 0.75 (CRBR15, CRBR36), 1 (VLA~1623~AaAb), 2 (CRBR36 Zoomed) and, 15\,MJy\,sr$^{-1}$ (GSS30-IRS3). Continuum contours correspond to 5-, 10-, 20-, and 100-$\sigma$ (ALMA) and 3-, 5-, 10-, and 20-$\sigma$ (VLA), with $\sigma=$\,RMS as defined in Table~\ref{tab:AllUsedFluxes}.}
    \label{fig:JWSTLargeScale}
\end{figure*}

Considering the correlation between extended 10\,GHz emission and the inner regions of outflow cavities, we can attempt to trace this emission from its origin and connect it to larger scale structures. While on smaller spatial scales the 10\,GHz emission tends to extend towards the outflow direction, on larger scales the connection between radio and NIR appears more diverse. In Figure~\ref{fig:JWSTLargeScale} we present the 10\,GHz continuum of several objects, alongside NIRCam observations of the wider protostellar environment. Each object presents a unique connection between the local protostellar environment probed by the VLA and the wider star forming region probed by JWST. Below, we present additional details on the operational mechanisms and unique features for each object.

\textbf{\textit{GSS30-IRS3}} --- In Figure~\ref{fig:JWSTLargeScale} we present JWST 4.4\,\micron\ observations, overlaid with contours of the disk continuum at 217\,GHz in white \citep[taken from][]{SantamariaMiranda_2024}. The disk continuum aligns well with the dark central lane seen at 4.4\,\micron, commonly seen in infrared observations of inclined disks \citep[e.g.][]{GalvanMadrid2018,Tazaki2025}. As discussed previously, this object has a 10\,GHz position angle that is perpendicular to the dust disk at both 217\,GHz and 44\,GHz, indicating a potential jet-like origin. From Figure~\ref{fig:JWSTLargeScale} we see both the 10\,GHz continuum and position angle (indicated by blue contours and arrows) are aligned with the collimated outflow seen at 4.4\,\micron. Due to the highly collimated morphology and knot-like features of this outflow, we believe it originates from a protostellar jet. Previous JWST observations at 4.4\micron\ and 4.7\micron\ have also revealed evidence of a bipolar outflow originating from the central protostar \citep[e.g.][]{SantamariaMiranda_2024,Nakamura2025}. Similarly, observations in CO have detected multi-component outflows comprising of a slow moving extended component and a much narrower high velocity component \citep[e.g.][]{Friesen2018}. The low velocity, wide-angled component has been inferred to indicate outwardly propagating shells driven by disk winds \citep[e.g.][]{Lee2000} whilst the high velocity, collimated component has been attributed to protostellar jet emission \citep[][]{Raga1993}. Therefore, our results appear to further confirm the jet-like origin of the fast moving, collimated molecular line component. The alignment and similar collimation of emission at 10\,GHz indicates that we are tracing the origin of this outflow and thus the base of the jet with our VLA observations. 

In addition, we also see evidence of a potential disk wind at NIR wavelengths, indicated by the characteristic X-shape morphology \citep[see e.g.][]{Duchene2024}. The wide-angle of the northern 4.4\,\micron\ emission ($\sim106\degr$, measured by eye) is consistent with the previously detected $^{12}$CO(2$-$1) opening angle of 120$\degr$ \citep[][]{SantamariaMiranda_2024}. The similarities between these two tracers indicates we are likely tracing dust grains at 4.4\,\micron\ that are well coupled to the gas, as one might expect in a disk wind \citep[e.g.][]{Giacalone2019,Rodenkirch2022}. While synthetic NIRCam images of dust entraining winds at 1.8\,\micron\ present qualitatively similar morphologies, a direct comparison at 4.4\,\micron\ is currently unavailable \citep[e.g.][]{Franz2022}. Therefore, we would require dedicated radiative transfer modelling in order to confirm this disk wind interpretation. Alternatively, this structure may arise from scattered light in the outflow cavity walls, having been excavated by the bipolar jet \citep[e.g.][]{Federman2024}.

\textbf{\textit{GSS30-IRS1}} --- Unlike GSS30-IRS3, we find that the emission at 10\,GHz broadly follows the disk, with a position angle perpendicular to the outflow direction. The misalignment of the emission with the outflow may indicate we are tracing the ionised disk surface as opposed to a jet component \citep[e.g.][]{Tychoniec2018_extendedEmission}. Curiously, previous H$_2$ observations indicate collimated emission along the outflow direction, consistent with a jet-like origin \citep[][]{Skretas2025}. In Figure~\ref{fig:JWSTLargeScale} we find very faint evidence of potential bow shock structures in the northern regions which are much more apparent at 4.4\micron\, and 4.7\micron\, (see Appendix~\ref{apx:IRS1_Bowshock} Figure~\ref{fig:GSS30-IRS1_BowShocks}). These bow shocks cover similar spatial scales to the collimated H$_2$ emission, hinting at a previous ejection from a jet-like component. Therefore, GSS30-IRS1 may previously have had an active protostellar jet which was inactive at the time of our VLA observations. Alternatively, if the jet was still in operation, it may have been below our resolution and/or sensitivity limits.  

\textbf{\textit{VLA~1623~AB}} --- In Figure~\ref{fig:JWSTLargeScale} we present JWST 4.7\,\micron\, narrowband observations which reveal two knot-like structures along the VLA~1623~A outflow cavity. The position angles of VLA~1623~Aa and Ab at 10\,GHz, indicated by cyan and green arrows respectively, appear to approximately align with each of these knots. However, it should be noted that only Aa is resolved at 10\,GHz while Ab remains only marginally resolved. If we assume that our observations at 10\,GHz trace jet-like emission, as suggested by the extended emission and position angle of Aa, then these two sources may be driving independent jets responsible for the creation of Knot~1 and Knot~2. Further along the outflow, however, there appears to be two consecutive bow-shock features, which are consistent with episodic ejection from a single driving source \citep[e.g.][]{Lora2024}. In this scenario, the two knots would represent shock-excited emission along the outflow cavity walls. We discuss the potential origin of these features further in Section~\ref{sec:VLA1623}.

\textbf{\textit{CRBR15 and CRBR36}} --- In Figure~\ref{fig:JWSTLargeScale} we present the large scale, collimated outflows of CRBR15 and CRBR36 at 4.7\,\micron\, as shown previously by \citet[][]{Nakamura2025}. The highly collimated nature of both outflows over large scales (1000s au) indicates a jet-like origin. In addition, we also present the outflow cavity for CRBR36 seen at 2\,\micron, yet no such structure is detected for CRBR15. Both of these objects remain unresolved at 10\,GHz indicating a compact emission region ($<$\,24\,au). However, we note that the outflow direction of the jet aligns perpendicularly to the dust disk detected at 44\,GHz for each object.

\subsection{Potential contamination from envelope emission}\label{sec:mmOrigin}

The integrated millimetre and centimetre flux densities of the youngest sources may include emission from both compact disk material and remnant protostellar envelopes. This effect is most relevant for the Class~0/I objects, where extended envelope material is expected to be present \citep[e.g.][]{Jorgensen2009}. To assess the possible contribution of such emission, we can compare measurements obtained at similar frequencies but with different angular resolutions and maximum recoverable scales.

GSS30--IRS3 provides a useful example of this test, with multiple measurements close to 220~GHz (see Table \ref{tab:AllUsedFluxes}). The measured flux densities are $124\pm12$~mJy at 225~GHz with a $0\farcs05$ beam and maximum recoverable scale of $2\arcsec$; $161\pm16$~mJy at 219~GHz with a $0\farcs3$ beam and maximum recoverable scale of $4\arcsec$; and $166\pm17$~mJy at 224~GHz with a $0\farcs9$ beam and maximum recoverable scale of $11\arcsec$. The latter two measurements agree within their uncertainties, despite probing different spatial scales, suggesting that the observations sensitive to larger-scale structure do not recover a significant additional envelope component at these frequencies.  While there is a lower flux density measured for the highest angular resolution data, this is likely to be caused by spatial filtering of compact emission, since the source extent is comparable to the maximum recoverable scale of those observations.

We perform the same comparison for the Class~I objects in our sample and find consistent ALMA integrated fluxes across observations with different angular resolutions and maximum recoverable scales. For the Class~0 objects, the ALMA observations were obtained at very high angular resolution, with maximum recoverable scales of order $0\farcs8$ \citep{Radley2025}. These data are therefore expected to filter out smooth envelope emission on larger angular scales. Similarly, the VLA observations were taken in the extended A configuration, with maximum recoverable scales of approximately $1$--$5\arcsec$ over 44--10~GHz, and should also be insensitive to extended, smoothly distributed emission.  We therefore conclude that large-scale envelope contamination is unlikely to dominate the integrated millimetre/centimetre fluxes reported here.

\subsection{Radio spectral energy distributions}\label{sec:SEDAnalysis}

To better constrain the dominant emission mechanisms and their underlying physical properties in each YSO, we can decompose the source-integrated radio spectral energy distribution (SED; e.g.\ flux density vs.\ frequency) into its individual dust and ionised gas emission components. 

\subsubsection{Measuring dust and ionised gas spectral indices}\label{sec:SpIx}
In order to accurately characterise the radio SED and thus, spectral index, we require a model which minimises residual scatter without adding unnecessary complexity. To this end, we consider three radio SED models of increasing complexity, applied to each object in our sample. The simplest model uses a single power law component, $\alpha$, characterising the spectral index across the entire data range:

\begin{equation}\label{eq:SinglePowerLaw}
    F_\nu = A_1\left(\frac{\nu}{10\text{\,GHz}}\right)^{\alpha},
\end{equation}

\noindent where $F_\nu$ is the flux density at frequency $\nu$ and $A_1$ is the scaling amplitude. The second model follows the same methodology as \citet{Radley2025}, utilising two power laws to characterise the dust ($\alpha_{\text{dust}}$) and ionised gas ($\alpha_{\text{ionised}}$) spectral indices, individually:

\begin{equation}\label{eq:DoublePowerLaw}
    F_\nu = A_{\text{ionised}}\left(\frac{\nu}{10\text{\,GHz}}\right)^{\alpha_{\text{ionised}}}+A_{\text{dust}}\left(\frac{\nu}{10\text{\,GHz}}\right)^{\alpha_{\text{dust}}},
\end{equation}

\noindent where $A_{\text{dust}}$, $A_{\text{ionised}}$ are the scaling amplitudes for the ionised gas and dust components, respectively. The final model utilises a piecewise approach fitting a single power law component (Eq~\ref{eq:SinglePowerLaw}.) to the millimetre and centimetre portion of the radio SED, delineated at 45\,GHz, which we refer to as the broken power law model: 

\begin{equation}\label{eq:cm-mmPowerlaw}
    F_\nu  =\left\{
    \begin{array}{ll}
    A_{\text{ionised}}\left(\frac{\nu}{10\text{\,GHz}}\right)^{\alpha_{\text{ionised}}} & \text{if } \nu < 45\,\text{GHz}\\
    &\\
    A_{\text{dust}}\left(\frac{\nu}{10\text{\,GHz}}\right)^{\alpha_{\text{dust}}} & \text{if } \nu > 45\,\text{GHz } \\
    \end{array}
    \right.,
\end{equation}

\noindent The broken power law model allows for the estimation of spectral indices which may change significantly across the (sub-)mm to centimetre regime. Previous studies have observed a steepening of the spectral index at frequencies above 200\,GHz compared to those derived below which may be explained by a variation in the optical depth properties \citep[e.g.][]{Painter2025,Garufi2025A&A...694A.290G}. Through fine spectral sampling of a sample of Class~II disks, \citet{Chung2025} found evidence of distinct emission regimes at frequencies $>200$\,GHz and $<50$\,GHz which do not smoothly transition between each other. Therefore, this model allows us to measure the spectral index in both regimes without adding additional degeneracy that would arise from fitting further distinct model components such as a purely optically thin dust component. We examine the implications and potential origin of a such a spectral model in Section~\ref{sec:SEDDiscont}.

For each model we follow a Markov Chain Monte Carlo (MCMC) approach with emcee \citep{Emcee} fitting all available data for each object from 10--400\,GHz (see Table~\ref{tab:AllUsedFluxes}). The dust spectral index may vary between $\sim$ 3.8, consistent with observations of ISM-like dust distributions i.e. no dust growth, and 1.5 if both optical depth and dust self-scattering are significant \citep[see e.g.][]{Draine06,Sierra20}. The ionised gas component may have a spectral index between $-0.1$ and 1 if it originates from thermal free-free emission within protostellar jets or disk winds \citep[see e.g.][]{Pascucci12,Anglada18}. However, contributions from non-thermal magnetospheric gyrosynchrotron emission may result in spectral indices between 2.5 in the optically thick limit, and $-2$ in the optically thin limit, depending on the energetic distribution of electrons \citep[e.g.][]{Dulk1985,Condon16}. Previous modelling of stellar and solar flares indicate a turnover frequency between 1-10\,GHz with optically thin emission expected at higher frequencies \citep[][]{Gudel2002}. Given that our observations focus on frequencies $\geq$10\,GHz we can assume we are in the optically thin regime, further supported by previously inferred spectral indices $<-0.1$ between 4.5\,GHz and 7.5\,GHz, consistent with a non-thermal origin \citep[][]{Dzib13}. Therefore, in order to separate dust contributions from ionised gas emission, we allow the ionised gas spectral index to vary between 1.5, the lower limit expected from dust self scattering, and --2 representing optically thick non-thermal emission. 

Considering the above, for the single power law model we use a Gaussian likelihood with flat priors and allow the spectral index to vary within the total parameter space  i.e. $-2 < \alpha < 3.8$. For the double and broken power law models we also use a Gaussian likelihood and flat priors, $\vec{\theta}$, with the following limits:

\begin{equation}
    P(\, \vec{\theta} \, ) =\left\{
    \begin{array}{ll}
    1, & \text{if } 0 < A_{\rm ionised} < 1.4. \\
    1, & \text{if } -2 < \alpha_{\text{ionised}} <  1.5. \\
    1, & \text{if } 0 < A_{\rm dust} < 1.4.  \\
    1, & \text{if } 1.5 < \alpha_{\text{dust}} < 3.8.  \\
    -\infty, & \text{Otherwise.}
    \end{array}
    \right.
\end{equation}

\renewcommand{\arraystretch}{1}
\begin{deluxetable*}{lcccccc}
\tablecaption{Measured spectral indices and ionised gas contamination fractions. \label{tab:SEDFits}}
\tablehead{\colhead{Object} & \colhead{$\alpha_{\text{dust}}$} & \colhead{$\alpha_{\text{ionised}}$} & \colhead{f$_{\text{ion}}$ (\%)} & \colhead{f$_{\text{ion}}$ (\%)} & \colhead{f$_{\text{ion}}$ (\%)} & \colhead{f$_{\text{ion}}$ (\%)}    \\
\colhead{} & \colhead{} & \colhead{} & \colhead{218\,GHz} & \colhead{100\,GHz} & \colhead{44\,GHz} & \colhead{10\,GHz}     
}
\startdata
\sidehead{\textbf{Single Power Law}}
GSS30-IRS1 & 1.7$_{-0.1}^{+0.1}$ & $\cdots$ & $\cdots$ & $\cdots$ & $\cdots$ & $\cdots$  \\
DoAr24Eb & 1.8$_{-0.1}^{+0.1}$ & $\cdots$ & $\cdots$ & $\cdots$ & $\cdots$ & $\cdots$  \\
\sidehead{\textbf{Double Power Law}}
VLA1623Aa & 2.3$_{-0.1}^{+0.1}$ & -0.5$_{-0.8}^{+0.6}$ & 0.1 & 0.6 & 5.2 & 75.6   \\
VLA1623Ab & 2.1$_{-0.1}^{+0.1}$ & -0.2$_{-1.0}^{+1.2}$ & 0.1 & 0.4 & 2.3 & 42.3  \\
VLA1623W & 2.2$_{-0.1}^{+0.1}$ & -1.3$_{-0.5}^{+0.7}$ & 0.0 & 0.1 & 1.3 & 71.4  \\
DoAr24 & 2.0$_{-0.1}^{+0.2}$ & -1.0$_{-0.7}^{+1.0}$ & 0.0 & 0.3 & 3.3 & 76.4  \\
DoAr24Ea & 2.1$_{-0.2}^{+0.4}$ & 0.6$_{-0.9}^{+0.4}$ & 3.3 & 9.7 & 26.8 & 76.9  \\
GSS29 & 1.9$_{-0.2}^{+0.4}$ & -0.3$_{-0.7}^{+0.4}$ & 0.9 & 5.1 & 25.0 & 89.9  \\
GSS30-IRS2 & 3.1$_{-0.5}^{+0.4}$ & -0.5$_{-0.2}^{+0.2}$ & 0.7 & 10.0 & 68.0 & 99.8  \\
S2 & 1.6$_{-0.1}^{+0.1}$ & -0.7$_{-0.7}^{+0.6}$ & 0.3 & 1.6 & 9.9 & 77.6  \\
\sidehead{\textbf{Broken Power Law}}
SM1 & 2.2$_{-0.1}^{+0.1}$ & 0.8$_{-0.2}^{+0.2}$ & $<$1.2 & $<$3.3 & $<$9.4 & $<$42.9   \\
VLA1623B & 2.0$_{-0.1}^{+0.1}$ & 1.1$_{-0.1}^{+0.1}$ & $<$8.0 & $<$14.9 & $<$27.0 & $<$58.8  \\
CRBR12 & 1.9$_{-0.1}^{+0.1}$ & 1.2$_{-0.1}^{+0.1}$ & $<$5.7 & $<$9.5 & $<$15.8 & $<$34.4  \\
CRBR36 & 2.0$_{-0.1}^{+0.1}$ & 1.7$_{-0.1}^{+0.1}$ & $<$20.2 & $<$24.1 & $<$28.8 & $<$38.6  \\
GSS30-IRS3 & 2.0$_{-0.1}^{+0.1}$ & 1.3$_{-0.1}^{+0.1}$ & $<$10.0 & $<$16.0 & $<$24.9 & $<$47.7   \\
LFAM3 & 2.2$_{-0.1}^{+0.1}$ & 0.9$_{-0.1}^{+0.1}$ & $<$1.6 & $<$4.4 & $<$12.0 & $<$48.8  \\
VSSG27 & 1.6$_{-0.2}^{+0.2}$ & 2.1$_{-0.3}^{+0.3}$ & $<$37.4 & $<$29.8 & $<$22.8 & $<$13.3  \\
CRBR15 & 1.8$_{-0.1}^{+0.1}$ & 0.8$_{-0.3}^{+0.3}$ & $<$2.3 & $<$5.0 & $<$11.2 & $<$37.4  \\
GSS26 & 2.3$_{-0.2}^{+0.1}$ & 1.2$_{-0.1}^{+0.1}$ & $<$3.6 & $<$8.3 & $<$19.0 & $<$56.6  \\
\enddata
\end{deluxetable*}

For each model we use 32 walkers, 1$\times10^5$ steps and remove a burn in of 1500 to derive our model parameters and sample the parameter space. From these fits we estimate the individual spectral indices ($\alpha,\alpha_{\text{dust}},\alpha_{\text{ionised}}$) as well as their contributions to the overall emission profile.

\begin{figure*}[ht]
    \centering
\includegraphics[height=0.899\textheight,width=0.899\textwidth,keepaspectratio]{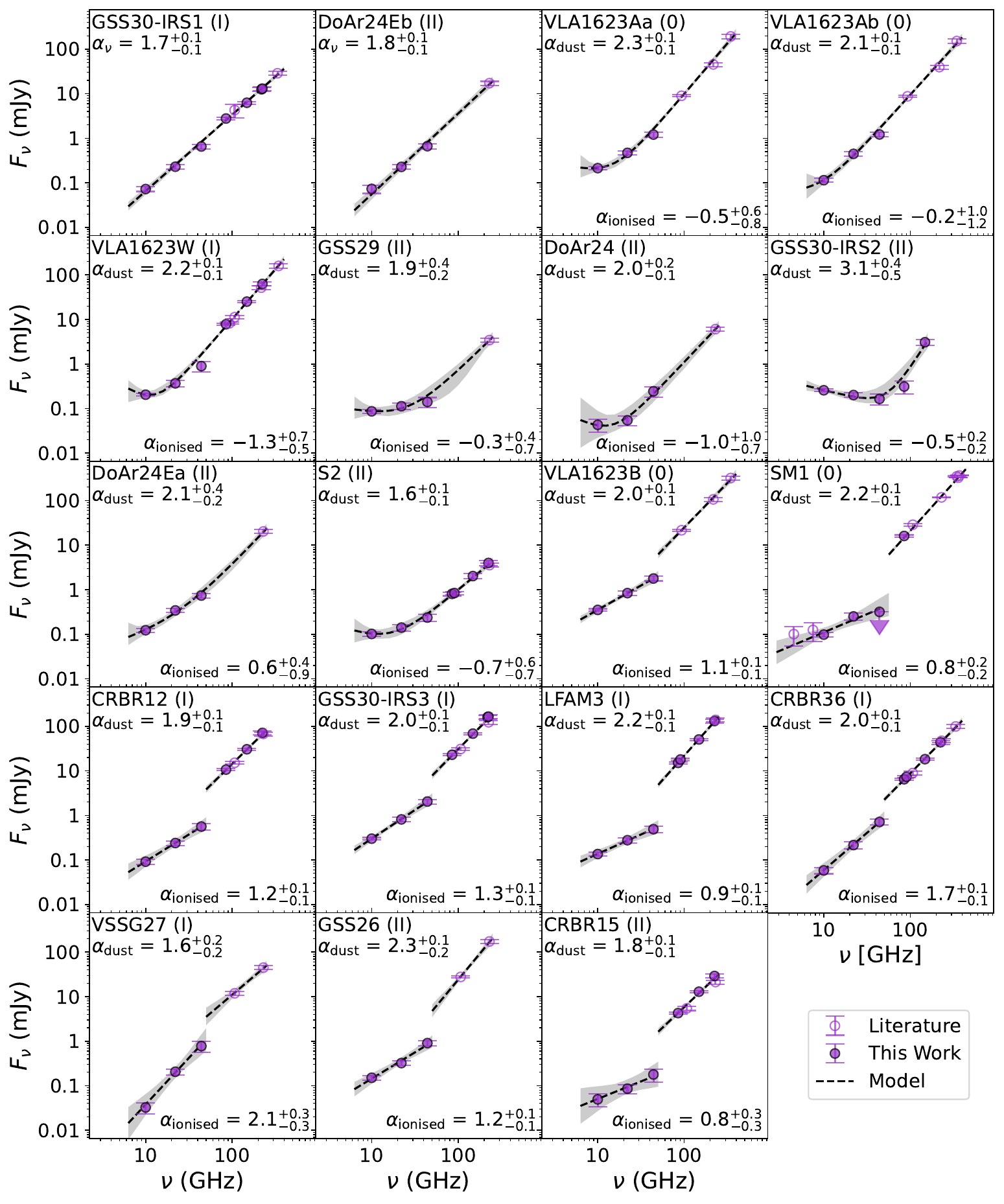}
    \caption{Radio spectral energy distributions and their derived spectral indices for all objects in our sample, annotated in the top left alongside their Class. 
    Flux densities taken from this work are presented with filled circles and flux densities from the literature are shown as open circles. Posterior realisations are shown as the grey shaded region with the posterior median indicated as the black dashed line. For GSS30-IRS1 and DoAr24Eb, we use a single power law fit to the radio SED and so only present $\alpha$. Spectral indices for $\alpha_{\text{dust}}$ are shown in the top left and $\alpha_{\text{ionised}}$ is shown in the bottom right.}
    \label{fig:SpIx}
\end{figure*}

To decide which model most accurately represents the data whilst avoiding overfitting we employ the Akaike information criterion (AIC, \citealt{Akaike1973}). The AIC balances a model's goodness of fit to its complexity, implementing a penalty for more complex models. We calculate the pairwise differences in AIC, defined as $\Delta\text{AIC}=\text{AIC}_{\text{complex}}-\text{AIC}_{\text{simple}}$, such that negative values indicate support for the more complex model. Following the criteria described in \citet{Burnham2004}, we choose the simplest, single power law model unless substantial support ($\Delta$AIC\,$<-4$) was found for the double power law model. As the broken power law model does not account for the full data range, it was only selected when there was strong support ($\Delta$AIC\,$<-10$) relative to both single and double power law models. We find only two objects, GSS30-IRS1 and DoAr24Eb, to be consistent with the single power law model, each with a spectral index less than 2. We present the results of our radio SED fitting in Figure~\ref{fig:SpIx}.

Considering both the double and broken power law models we find that all objects except GSS30-IRS2 have $\alpha_{\text{dust}}\leq2.3$ indicating either dust growth or potentially optically thick emission at (sub-)mm wavelengths. GSS30-IRS2 has $\alpha_{\text{dust}} = 3.1$, potentially indicating a comparatively lower level of dust growth and lower optical depth. However, it is important to note the high level of uncertainty associated with GSS30-IRS2 ($\sim \pm 0.5$). The second highest $\alpha_{\text{dust}}$ comes from GSS26 which has been found to have ring and gap features \citep[][]{Cieza2021}. Excluding GSS30-IRS2, we find that $\alpha_{\text{dust}}$ is on average consistent across evolutionary stages going from 2.2\,$\pm$\,0.1 for Class~0 to 2.1\,$\pm$\,0.1 and 2.1\,$\pm$\,0.2 for Class~I and II respectively. Spectral indices approaching 2 indicate that in general, our sample contains optically thick emission, primarily in the (sub-)mm.

For the ionised gas component we find that, on average, Class I objects have the highest $\alpha_{\text{ionised}}\sim0.4\,\pm\,0.5$. Whereas, Class~0 and II objects have typically more negative spectral indices with $\alpha_{\text{ionised}}=$ $-$0.2\,$\pm$\,0.4 and $-$0.2\,$\pm$\,0.2, respectively. The difference in $\alpha_{\text{ionised}}$ may show evidence of varying dominant emission mechanisms with evolutionary stage, with Class~II objects being more dominated by non-thermal gyro-emission and Class~I objects being more consistent with thermal free-free emission from jets or winds \citep[][]{Dzib13,Anglada1998AJ....116.2953A}. However, given such high uncertainties this trend is purely speculative and cannot be confidently confirmed.

Using the spectral index we can estimate the contribution of each emission component to the total flux density in order to understand which frequencies are most dominated by dust emission, and which are dominated by ionised gas emission. In Table~\ref{tab:SEDFits} we present the fraction of emission contributed by ionised gas, f$_{\text{ion}}$, at a range of commonly observed frequencies from ALMA (218\,GHz) to VLA (10\,GHz). We derive f$_{\text{ion}}$ by comparing the total flux density predicted by the model to the flux density contributed solely from the ionised gas component. We note that values derived from the broken power law model will underestimate the true fraction of contaminating emission at high and low frequencies, as we estimate $\alpha_{\text{ionised}}$ using all VLA flux densities. As can be seen from the double power law sample, there is a large range of contamination values at 44\,GHz ranging from $\sim1\%$ up to $68\%$ in the most extreme case. If we consider the fraction of ionised emission at 10\,GHz, we find that all but one object (VLA~1623~Ab) have f$_{\text{ion}}>70\%$.

As expected, emission at ALMA frequencies (218 and 100\,GHz) is typically dominated by emission from dust over ionised gas processes. Conversely, emission at 10\,GHz primarily originates from ionised gas mechanisms as found previously \citep[e.g.][]{Coutens19,Guidi2022,Radley2025}. Interestingly, if we consider the average properties of the double power law and broken power law sample, we find that more evolved objects tend to have higher fractions of non-dust emission at low frequencies. Class~0 objects have f$_{\text{ion}}=11\pm6\%$ and $55\pm8\%$ at 44\,GHz and 10\,GHz, respectively, whereas Class~II objects have f$_{\text{ion}}=23\pm8$\% and $74\pm8$\% for the same frequencies. 

Finally, while our dust spectral indices are relatively well constrained, the ionised gas spectral index and, therefore f$_{\text{ion}}$, remains largely uncertain across the majority of our sample. This uncertainty arises primarily from increased variability and lower angular resolutions at frequencies $<10$\,GHz, resulting in potentially unreliable flux density measurements. To overcome this, future surveys should incorporate contemporaneous observations spanning the 1-40\,GHz range, enabling an instantaneous characterisation of the low frequency spectral index. This could be achieved through a coordinated effort combining A-configuration VLA observations, as presented here, with SKA-MID for frequencies $<10$\,GHz, or alternatively through the ngVLA alone, which would provide comparable spectral coverage and increased resolution across its full frequency range \citep[][]{Braun2019,Selina2018}.

\subsubsection{A physical model for dust and ionised gas emission}\label{sec:DustPopSED}

Whilst measurements of the dust and ionised gas spectral indices provide us with insight into which physical mechanisms may be operating within YSOs, they do not directly reveal anything about the properties of the emitting material. We therefore adopt the strategy of \citet{Liu2019ApJ...884...97L,Liu2021ApJ...923..270L} fitting the observed radio SED of each source considering the individual contributions of dust and ionised gas emission. This method enables us to infer maximum grain sizes, dust temperatures, surface densities and, dust masses for each source based on its radio SED. Our model assumes that the dust temperature ($T_{\rm dust}$) and maximum grain size ($a_{\rm max}$) are uniform in the dust emission source and the electron temperature ($T_{e}$) is uniform in the ionised gas emission source. It is important to note that we do not consider mutual obscuration between the dust and ionised gas emission sources.

Recent surveys towards Class~II disks have confirmed that dust emission is likely to be optically thick at frequencies $>$\,200 GHz (\citealt{Chung2024,Painter2025}), such that dust self-scattering may have significant effects on the radio SED (\citealt{Liu2019ApJ...877L..22L,Zhu2019}). For this reason, we model the dust emission (incorporating the effects of dust self-scattering) using the formulation presented in \citet{Birnstiel2018ApJ...869L..45B}. Our dust model assumes an MRN grain size distribution \citep[e.g. q = $-$3.5,][]{Mathis1977ApJ...217..425M} and DSHARP dust compositions \citep{Birnstiel2018ApJ...869L..45B}. However, it is important to note that the dust opacity remains a significant source of uncertainty in modelling approaches, with a strong dependence on dust composition, porosity and size distribution \citep[][]{Woitke2016A&A...586A.103W,Andrews2020ARA&A..58..483A}. Presently, there are no strong claims to a physically preferred opacity model and so our adoption of the often used DSHARP opacities allows for direct comparison to the literature, acknowledging the caveat that different opacity laws impact the scaling of absolute values such as dust mass \citep[e.g.][]{Macias2021,Liu2022,GuerraAlvarado2024_HLtau,Guerra-Alvarado2024,Painter2025}. However, we note that recent studies of Class~II disks have shown that dust emission is likely to be optically thick at frequencies $>$\,200 GHz (\citealt{Chung2024,Painter2025}), such that dust self-scattering may have significant effects on the radio SED (\citealt{Liu2019ApJ...877L..22L,Zhu2019}). For this reason, we model the dust emission (incorporating the effects of dust self-scattering) using the formulation presented in \citet{Birnstiel2018ApJ...869L..45B}. Our dust model assumes an MRN grain size distribution \citep[e.g. $q = -3.5$,][]{Mathis1977ApJ...217..425M} and DSHARP dust compositions \citep{Birnstiel2018ApJ...869L..45B}. However, it is important to note that the dust opacity remains a significant source of uncertainty in modelling approaches, with a strong dependence on dust composition, porosity and size distribution \citep[][]{Woitke2016A&A...586A.103W,Andrews2020ARA&A..58..483A}. For example, the radial SED analysis of CI Tau presented in \citet[][]{Zagaria2025} found that DSHARP dust compositions typically resulted in larger inferred surface densities and maximum grain sizes when compared to dust compositions taken from \citet{Zubko1996} and those adapted from \citet{Ricci2010A&A...512A..15R}. These differences were attributed to variations in the dielectric constants of each composition, altering the dust opacity, its spectral dependence, and thus inferred grain properties. 
However, it should be noted that the observed spectral behaviour results from a combination of line of sight optical depth and dust opacity.  While the former can be measured observationally \citep[see, e.g.,][]{Isella2018} the latter requires an assumption on the dust composition to be made.  Ultimately, while the compositions presented in \citet{Zagaria2025} appear to better match CI Tau, there are no strong claims to a physically preferred opacity model for our sample. Therefore, our adoption of the often used DSHARP opacities allows for direct comparison to the literature with the acknowledgment that different opacity laws will impact the scaling of absolute values such as dust mass \citep[e.g.][]{Macias2021,Liu2022,GuerraAlvarado2024_HLtau,Guerra-Alvarado2024,Painter2025}.

In addition, ionised gas emission may have significant contributions at frequencies between 4--50 GHz \citep[e.g.][]{Chung2024,Garufi2025A&A...694A.290G,Painter2025}, while synchrotron emission may be prominent only in a few exceptional cases \citep[e.g.][]{Liu2014ApJ...780..155L,Chung2025}. In order to account for this emission component we adopt a modified black body model with a frequency-dependent opacity based on free-free emission within an ionised gas source, similar to a compact \ion{H}{2} region. Specifically, we follow the approach of \citet{Mezger1967ApJ...147..471M}, adopting their Equation~A.1 to estimate the opacity. This formulation is independent of the ionisation mechanism and therefore applies equally to protostellar jets and disk winds as well as \ion{H}{2} regions, accurately recovering the expected spectral behaviour in both optically thick ($\alpha_\nu=2$) and optically thin regimes ($\alpha_\nu=-0.1$) \citep[e.g.][]{Reynolds1986,Pascucci12}.
Using \textit{emcee}, our fitting procedure optimized the following free parameters, $T_{\rm dust}$, $a_{\rm max}$, dust column density ($\Sigma_{\rm dust}$), solid angle of dust emission source ($\Omega_{\rm dust}$), emission measure of ionised gas emission source (EM), and solid angle of ionised gas emission source $\Omega_{\rm ionised}$. We note that SED fitting is inherently degenerate and alternative solutions may reproduce the observations equally well. To reduce the impact of these degeneracies, we constrain the parameter space through physically motivated priors, as detailed below.

In order to minimise degeneracy in the ionised gas source we choose to fix the electron temperature. Previous studies have derived electron temperatures in the range of 5000--11000\,K using radio recombination lines of galactic \ion{H}{2} regions, with lower values associated with less luminous ionising sources \citep[e.g.][]{Quireza2006,Balser2024}. Given that our sample consists of low mass YSOs, we choose to fix the electron temperature, $T_e$, to 8000\,K consistent with the lower to middle end of this range. It should be noted that the ionised gas component is only a weak function of temperature, particularly in the optically thin limit, with the spectral morphology largely driven by the emission measure.

For the majority of objects we allow EM to vary between 10$^6$\,cm$^{-6}$\,pc and 10$^{12}$\,cm$^{-6}$\,pc. However, for DoAr24 and DoAr24Eb we find that the lack of millimetre data in the radio SED can lead to solutions consistent with arbitrarily large EM. For these two objects we implement an upper limit of 10$^{10}$\,cm$^{-6}$\,pc for EM which provides solutions in line with the range of values derived for the rest of the sample (see Table~\ref{tab:DustGasPop}). Finally, we implement an upper limit to $\Omega_{\rm ionised}$ such that $\Omega_{\rm ionised}\leq\Omega_{\rm dust}$.

Similarly, we find that due to the multivariate nature of our dust component, this too is subject to degeneracy. We can mitigate these degeneracies by fixing components which have been accurately measured previously, such as the solid angle of the dust emission. Most of our target sources were spatially resolved in previous, high angular resolution ALMA observations, allowing us to fix $\Omega_{\rm dust}$ to these observed spatial extent \citep[e.g.][]{Cieza_2019,SantamariaMiranda_2024,Radley2025}. However, for objects without such observations in the literature, we estimate their spatially resolved extents from Gaussian fits to the highest frequency ALMA continuum image as discussed in Section~\ref{sec:ALMA_DataReduction}. For such sources, we fix the $\Omega_{\rm dust}$ constrained by the resolved ALMA image.
For spatially compact sources that have not yet been resolved by ALMA observations (DoAr~24, DoAr~24~Eb, S2, GSS29), we implement an upper limit to $\Omega_{\rm dust}$ based on their unresolved sizes taken from ODISEA \citep[][]{Cieza_2019} and set $\Omega_{\rm dust}$ as a free parameter. 

Furthermore, we can constrain $a_{\rm max}$ by considering the wavelength at which dust emission cannot be distinguished from ionised gas emission given the flux calibration uncertainty. Based on the spectral indices derived in Section~\ref{sec:SpIx}, most objects become dominated by ionised gas emission between 44\,GHz and 10\,GHz. We therefore choose a conservative uncertainty of 10\%, consistent with the calibration uncertainty of our 44\,GHz and 22\,GHz observations. Given this uncertainty, we can estimate the wavelength at which ionised gas contributes 90\% of the total flux density and thus the dust component falls within the calibration uncertainty. This wavelength is then converted to a grain size following $a_{\rm max} \sim \frac{\lambda}{2\pi}$, providing an upper limit to the maximum grain size that can be determined from our observations. We employ this upper limit to improve computational efficiency, noting that the prior is non-informative in practice, with all inferred maximum grain sizes lying well below the $a_{\rm max}$ upper limit. Unfortunately, we cannot implement the same methodology for objects where the radio SED is best described by a single-component (GSS30-IRS1, DoAr24Eb). Previous multiwavelength analyses have implemented upper limits of 10\,cm for $a_{\rm max}$ justified by low emission contributions from large grains at millimetre wavelengths \citep[][]{Viscardi2025}. However, through including observations in the centimetre, \citet{Zhang2021} found potential evidence of dust growth to 10\,cm sizes in a Class~I disk. Therefore, given that our inclusion of cm-wave observations extends our sensitivity to larger dust sizes, and the possibility of grains sizes $\gtrsim10$\,cm we adopt a conservative upper limit of 50\,cm for GSS30-IRS1 and DoAr24Eb.

In addition, we allow $T_{\rm dust}$ to vary between 20--300\,K. We choose a lower limit of 20\,K to be consistent with previously observed dust temperatures and, to avoid solutions where cold dust can lead to radio SEDs which deviate significantly from the Rayleigh-Jeans regime at millimetre wavelengths \citep[][]{Andrews_Williams_2005,Woitke2016A&A...586A.103W}. Young, Class 0/I disks are expected to have higher dust temperatures compared to their more evolved counterparts with measured disk temperatures around 200\,K \citep[e.g][]{VantHoff2020_WarmDisks,VantHoff2020_tempProfile}. The upper limit of 300\,K is therefore a conservative value well in excess of previously observed disk temperatures.

\startlongtable
\setlength{\extrarowheight}{2pt}
\begin{deluxetable*}{ l c c c c c c c  }
\tablecaption{Dust and ionised gas properties inferred from the SED modelling. \label{tab:DustGasPop}}
\tabletypesize{\footnotesize}
\decimals
\tablehead{
&
\multicolumn{5}{c}{\textbf{Dust}}&
 \multicolumn{2}{c}{\textbf{Ionised Gas}}
\\
\cmidrule(l{10pt}r{10pt}){2-6}
\cmidrule(l{10pt}r{10pt}){7-8}
Object &
$a_{\rm max}$ &
$T_{\rm dust}$ &
$\Sigma_{\rm dust}$ &
$\Omega_{\rm dust}$  &
$M_{\rm dust}$ &
EM  &
$\Omega_{\rm ionised}$  \\
&
(mm) &
(K) &
(g\,cm$^{-2}$) &
(au$^2$) &
(\Mearth) &
(10$^8$ cm$^{-6}$\,pc) &
(au$^2$)
}
\startdata
\sidehead{\textbf{Class 0}}
VLA1623B & $0.5_{-0.1}^{+0.1}$ & $91_{-6}^{+6}$ & $4.6_{-1.0}^{+1.1}$ & 1330 & $231_{-48}^{+56}$ & $7.4_{-1.8}^{+2.0}$ & $13_{-1}^{+1}$ \\
VLA1623Aa & $3.4_{-2.7}^{+2.3}$ & $134_{-7}^{+9}$ & $22.1_{-17.2}^{+20.4}$ & 380 & $315_{-245}^{+291}$ & $1.0_{-0.8}^{+5.9}$ & $20.0_{-10}^{+120}$ \\
VLA1623Ab & $0.6_{-0.1}^{+3.4}$ & $151_{-13}^{+10}$ & $6.5_{-2.2}^{+6.3}$ & 300 & $73_{-25}^{+71}$ & $21.7_{-8.3}^{+13.8}$ & $3.6_{-0.5}^{+0.4}$ \\
SM1 & $0.6_{-0.1}^{+0.1}$ & $24_{-1}^{+1}$ & $2.2_{-0.2}^{+0.2}$ & 6400 & $523_{-57}^{+50}$ & $0.3_{-0.3}^{+2.1}$ & $30_{-20}^{+240}$ \\
\sidehead{\textbf{Class I}}
CRBR12 & $0.8_{-0.1}^{+0.1}$ & $20_{-1}^{+1}$ & $2.6_{-0.3}^{+0.4}$ & 4300 & $417_{-56}^{+57}$ & $7.2_{-2.6}^{+3.0}$ & $3.4_{-0.6}^{+0.9}$ \\
GSS30-IRS1 & $0.5_{-0.2}^{+356.5}$ & $29_{-2}^{+2}$ & $22.9_{-10.1}^{+16.3}$ & 430 & $371_{-164}^{+263}$ & $15.4_{-5.6}^{+15.8}$ & $2.1_{-0.5}^{+0.4}$ \\
GSS30-IRS3 & $0.8_{-0.1}^{+0.1}$ & $21_{-1}^{+1}$ & $3.3_{-0.5}^{+0.5}$ & 8800 & $1090_{-169}^{+171}$ & $9.2_{-2.3}^{+2.5}$ & $10_{-1}^{+1}$ \\
LFAM3 & $2.3_{-0.3}^{+0.1}$ & $20_{-1}^{+1}$ & $0.5_{-0.1}^{+0.1}$ & 11000 & $216_{-10}^{+10}$ & $2.1_{-1.8}^{+2.2}$ & $9_{-3}^{+46}$ \\
CRBR36 & $0.7_{-0.1}^{+0.1}$ & $24_{-1}^{+1}$ & $3.8_{-0.9}^{+1.1}$ & 2000 & $287_{-66}^{+86}$ & $21.2_{-10.3}^{+24.2}$ & $1.8_{-0.3}^{+0.4}$ \\
VLA1623W & $1.6_{-0.6}^{+2.2}$ & $21_{-1}^{+1}$ & $2.2_{-0.6}^{+1.4}$ & 3100 & $258_{-74}^{+167}$ & $3.5_{-2.6}^{+2.6}$ & $10_{-2}^{+15}$ \\
\sidehead{\textbf{Class II}}
GSS26 & $2.6_{-1.7}^{+12.4}$ & $21_{-1}^{+1}$ & $0.5_{-0.1}^{+0.2}$ & 12000 & $243_{-31}^{+104}$ & $5.6_{-3.4}^{+3.2}$ & $6_{-1}^{+4}$ \\
GSS29 & $2.7_{-1.4}^{+1.4}$ & $56_{-29}^{+86}$ & $24.3_{-17.2}^{+17.6}$ & $50_{-32}^{+65}$ & $45_{-32}^{+33}$ & $1.1_{-0.8}^{+1.3}$ & $9_{-4}^{+19}$ \\
DoAr24 & $3.3_{-2.5}^{+2.3}$ & $56_{-29}^{+106}$ & $24.1_{-15.2}^{+17.0}$ & $90_{-70}^{+120}$ & $81_{-51}^{+57}$ & $0.9_{-0.8}^{+7.6}$ & $4_{-4}^{+25}$ \\
CRBR15 & $0.7_{-0.1}^{+0.1}$ & $20_{-1}^{+1}$ & $2.8_{-0.4}^{+0.4}$ & 1700 & $183_{-25}^{+27}$ & $0.5_{-0.5}^{+3.4}$ & $10_{-10}^{+120}$ \\
GSS30-IRS2 & $0.6_{-0.3}^{+0.5}$ & $67_{-38}^{+88}$ & $7.7_{-0.1}^{+0.1}\times10^{-3}$ & 11000 & $3_{-2}^{+5}$ & $0.5_{-0.1}^{+0.2}\times10^{-1}$ & $440_{-340}^{+940}$ \\
DoAr24Ea & $5.1_{-2.9}^{+2.8}$ & $42_{-4}^{+4}$ & $26.2_{-17.4}^{+16.5}$ & 400 & $393_{-262}^{+247}$ & $9.0_{-2.7}^{+3.5}$ & $4.0_{-0.6}^{+0.7}$ \\
DoAr24Eb & $239.1_{-160.3}^{+175.0}$ & $65_{-38}^{+119}$ & $27.4_{-16.2}^{+15.4}$ & $200_{-140}^{+340}$ & $211_{-124}^{+118}$ & $12.5_{-7.2}^{+18.7}$ & $2_{-1}^{+1}$ \\
S2 & $0.3_{-0.1}^{+0.1}$ & $41_{-17}^{+76}$ & $27.7_{-13.0}^{+14.1}$ & $74_{-49}^{+65}$ & $79_{-37}^{+40}$ & $2.0_{-1.5}^{+2.0}$ & $7_{-3}^{+16}$ \\
\enddata
\tablecomments{$\Omega_{\rm dust}$ and $\Omega_{\rm ionised}$ are converted to physical areas at a distance of 138.4\,pc such that $\Omega\,[\rm au^2]=\Omega \,D^2$, where D is the distance.}
\end{deluxetable*}

Dust surface densities recovered from radially resolved multiwavelength radio SED analyses imply decreasing surface densities with increasing evolutionary stage. Class~II YSOs have previously been observed to have $\Sigma_{\rm dust}<10$\,g\,cm$^{-2}$, whilst the majority of the disk in Class~0 objects have been found to be below $\Sigma_{\rm dust}\lesssim50$\,g\,cm$^{-2}$ \citep[e.g.][]{CarrascoGonzalez2019,Guidi2022,Guerra-Alvarado2024}. Furthermore, studies modelling ring structures in disks have adopted upper limits of 20\,g\,cm$^{-2}$ reflecting the locally enhanced surface densities in these regions \citep[][]{Han2023}. As we take a disk integrated approach, considering Class~0-II YSOs, we implement a conservative upper limit of 50\,g\,cm$^{-2}$, consistent with the higher end of recovered dust surface densities.

\begin{figure*}[ht]
    \centering
    \includegraphics[height=0.91\textheight,width=\textwidth,keepaspectratio]{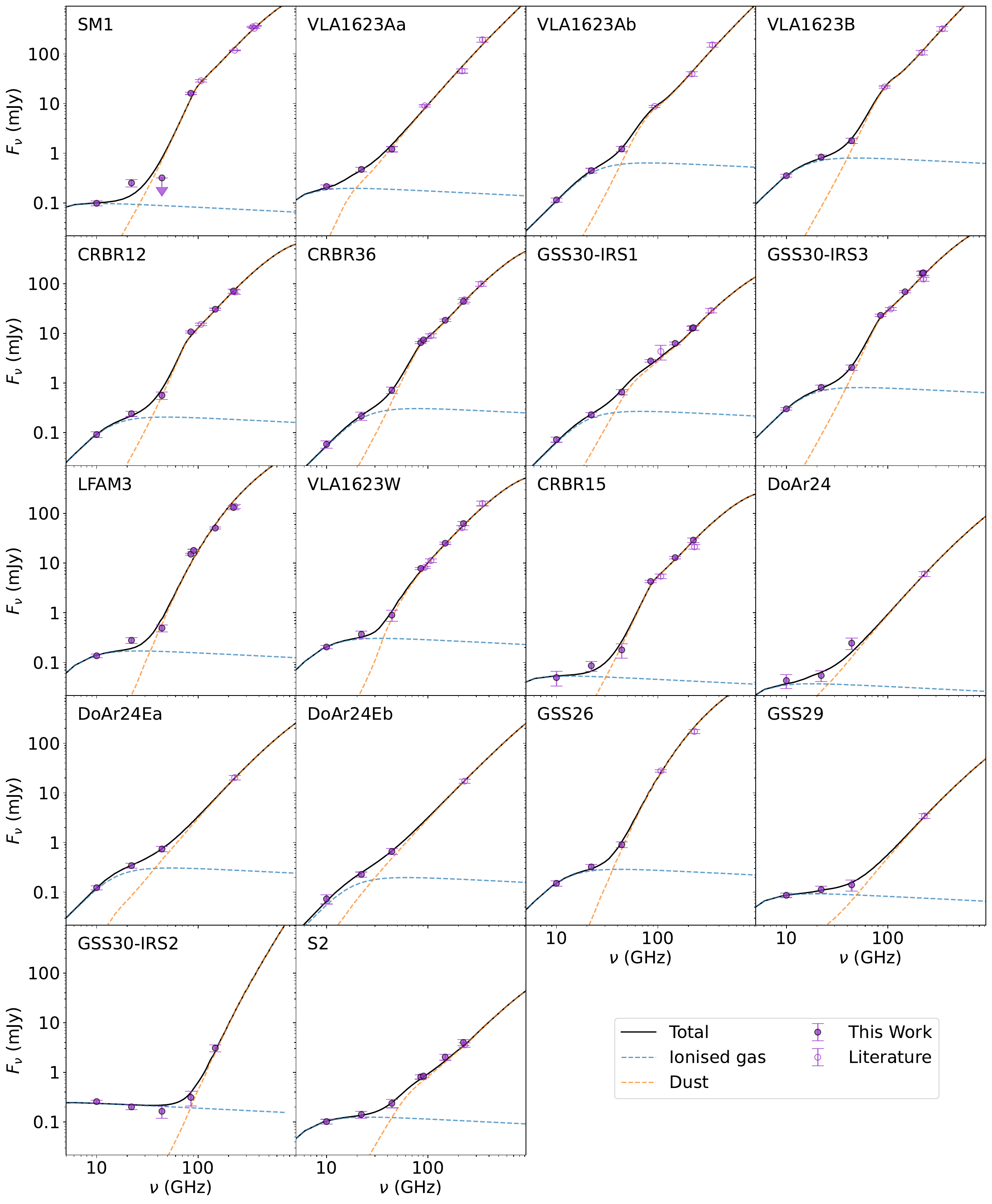}
    \caption{SED models for each object as presented in Table~\ref{tab:DustGasPop}. Orange and blue dashed lines show the contributions of dust and ionised gas emission components, respectively. The total integrated flux density is shown as a black solid line. Flux densities from this work are shown as filled, purple circles and those from literature are shown as open circles. Upper limits are denoted with downward arrows.}
    \label{fig:DustGasPopMCMC}
\end{figure*}

Each model is run for 1$\times10^5$ iterations using 50 walkers. We remove a burn in of 2000 iterations and use the 50th percentile as our posterior median model, with uncertainties derived from the 16th and 84th percentiles. Unfortunately, no valid solution was found for VSSG27 likely due to the strict $a_{\rm max}$ prior from our spectral index estimation. We note that GSS30-IRS2 has an ionised gas spectral index consistent with non-thermal gyro-emission ($\alpha_{\rm ionised}\sim-0.5\pm0.2$), which is not captured in our thermal free-free model. Whilst the uncertainty means a thermal origin cannot be completely excluded, the ionised gas parameters should be interpreted with this caveat in mind. We present our model radio SED fits in Figure~\ref{fig:DustGasPopMCMC} and posterior parameters in Table~\ref{tab:DustGasPop}.

All objects show signs of grain growth beyond ISM-like sizes ($\sim$0.1\,\micron) ranging between 0.27\,mm for S2 and 24\,cm for DoAr24Eb. However, it is important to note that due to the less constrained $a_{\rm max}$ prior and minimal (sub-)mm data of DoAr24Eb, this result is largely uncertain, thus we do not consider it in the following averages. We find that dust growth appears to be, on average, greater for Class~II objects ($a_{\rm max}=2.2\pm0.7$\,mm) than for Class~I ($a_{\rm max}=1.1\pm0.3$\,mm) as one might expect from successive dust growth over time. Excluding VLA~1623~Aa, Class~0 objects are comparable to Class~I objects in terms of maximum grain size. 

We find that the majority of objects are consistent with dust masses greater than 100\Mearth. Class~II objects appear to have the lowest dust disk masses with an average of 147\,$\pm$\,51\Mearth, just less than half the average for Class~I (440\,$\pm$\,133\Mearth), whilst Class~0 objects generally occupy the lower end of the masses derived for Class~I (200--500\Mearth), with the exception of VLA~1623~Ab. The overall trend of increasing dust mass between Class~0 and I may be explained by the accretion of the surrounding envelope in which these YSOs are embedded (e.g. Figure~\ref{fig:10GHzExtensions}, VLA~1623~A and B). We discuss potential explanations for the reduced masses in Class~II objects in Section~\ref{sec:MissingMass}.

Finally, we find that Class~0 disks appear to have the highest dust temperatures in the sample, with an average of 100$\pm$30\,K, in line with previous estimates of their brightness temperature \citep[e.g.][]{Maureira2026}. Class~I disks appear to be typically cooler, with average dust temperatures of 23$\pm$1\,K, slightly below the 30\,K often associated with Class~I objects \citep[e.g.][]{Tychoniec2020}. Unfortunately, the majority of our Class~II objects lack high angular resolution ALMA data and therefore do not have fixed $\Omega_{\rm dust}$ values, resulting in a degeneracy between the dust solid angle and dust temperature, yielding more uncertain $T_{\rm dust}$ estimates. For the two objects with well constrained dust temperatures, GSS26 and CRBR15, we find that they occupy the lower end of values observed in Class~I disks. We note that our analysis assumes a uniform disk temperature, whereas radial temperature gradients are commonly observed in molecular line emission and multiwavelength continuum studies \citep[e.g.][]{VantHoff2020_tempProfile, CarrascoGonzalez2019}. Our derived $T_{\rm dust}$ values should therefore be interpreted as a disk-averaged dust temperature.

\section{Discussion}\label{sec:discussion}

\subsection{The nature and emergence of outflows in the VLA~1623 system}\label{sec:VLA1623}

The proximity of VLA~1623~A and B, combined with typically low resolution gas observations has led to ambiguity on the precise origins of these outflows. Using CO(3--2) observations, \citet{Hsieh20} inferred that VLA~1623~A and B are driving one outflow each, with CS observations from \citet{Ohashi22} indicating that the VLA~1623~A outflow originates close to VLA~1623~Aa. Conversely, $^{12}$CO(2--1) observations from \citet{Hara21} point to both outflows originating from VLA~1623~A, inferring a misalignment of the protostellar disks in order to explain the jet morphology. However, the observed misalignment could equally be explained by outflows launching from A and B which are themselves, misaligned \citep[e.g.][]{Hsieh2024}. With our detection of multiple 10\,GHz continuum extensions aligned with NIR outflow cavities, our increased sensitivity provides new constraints on their potential origins. Using these results, we reassess which components of the VLA~1623 system are responsible for driving each outflow, building on the inferences of \citet[][]{Radley2025}.

Figure~\ref{fig:JWSTLargeScale} reveals two knot-like structures in the large scale outflow associated with VLA~1623~A. The 10\,GHz position angle of VLA~1623~Aa aligns directly with Knot~1, with continuum emission extended along the same direction, pointing to Aa as the driving source of this outflow feature. The position angle of VLA~1623~Ab shows a partial but offset alignment with Knot~2, however the absence of any continuum extension makes this association uncertain. An alternative interpretation arises from the presence of two consecutive bow-shock features further along the outflow cavity, which could indicate episodic ejection from a single driving source with both knots tracing shocked gas along the cavity walls \citep[e.g.][]{Lora2024,Dutta2025}. Nevertheless, given that VLA~1623~Aa exhibits a higher 10\,GHz flux density with extended emission along the outflow and a jet-like PA, it remains the most likely driver of this outflow. The misalignment between the PA and the bow-shock may indicate jet precession, which is broadly consistent with the rotating outflow reported by \citep[][]{Ohashi22}

On smaller scales, Figure~\ref{fig:10GHzExtensions} reveals a previously undetected 10\,GHz extension from VLA~1623~B, aligned along the direction of its outflow cavity. This provides direct evidence for a collimated jet as inferred by \citet[][]{Santangelo2015} and confirms the origin of the SiO jet reported by \citet[][]{Codella2024}. Together, our results support the outflow scenarios proposed by \citet[][]{Hsieh20} and \citet{Ohashi22} in which VLA~1623~Aa and B are the two main drivers of the outflow. Whilst we cannot rule out a jet origin from VLA~1623~Ab, the relatively low ionised gas fraction at 10\,GHz in Table~\ref{tab:SEDFits} indicates that our observations are not dominated by ionised gas emission. In addition, the higher emission measure ($21\times10^8\,\rm cm^{-6}pc$) and lower ionised emitting region (4$\times10^{-15}$\,Sr) compared to Aa and B may indicate either a still forming or weakly operating jet, which is below either our resolution ($\sim25$\,au) or sensitivity limits ($\sim$20$\mu$Jy). A complete characterisation of the outflow origins in this system will therefore require the next-generation angular resolution of facilities like the SKA and ngVLA \citep[][]{Selina2018,Braun2019}.

Finally, for VLA~1623~W, we find extended 10\,GHz emission aligned perpendicular to the dust continuum. \citet[][]{Radley2025} previously inferred that a jet may be present based on similar extensions, high flux density contributions of ionised gas and lobes of low spectral indices in high resolution images. Due to the fortunate beam orientation at 10\,GHz we have the highest spatial resolution along the direction of the outflowing jet. This has allowed us to confirm the previously mentioned extension at the $\geq5\sigma$ level, as well as detect two new compact regions of 3$\sigma$ emission. The morphology at 10\,GHz strongly resembles what one would naively expected from a protostellar jet with each compact region potentially tracing a knot caused by episodic ejection \citep[e.g.][]{Gomez2013}. However, high-sensitivity follow-up observations are required to confirm the nature of this emission.

Considering the perpendicular extension, potential ejecta, and lack of large scale outflow structure (e.g. knots, bow shocks), it is possible that VLA~1623~W may have recently launched this protostellar jet. However, MHD simulations predict the formation of a wide-angled, low velocity outflow prior to the emergence of a jet, which is not evident in the NIR morphology \citep[][]{Machida2013,Machida2014}. One explanation for this absence is the interaction of VLA~1623~W with the nearby VLA~1623~AB system which is connected to VLA~1623~W via streamers extending over 1000\,au \citep[][]{Mercimek23}. Such streamers are seen in several objects, acting as an asymmetric channel of accretion beyond the parental core \citep[e.g.][]{Pineda2020,Pineda2023}. This interaction may have stripped the surrounding envelope \citep[e.g.][]{Sadavoy2024}, producing a less dense environment and thus outflow emission below our sensitivity levels. While largely uncertain, the morphology of VLA~1623~W is consistent with the earliest stages of jet formation \citep[e.g.][]{Machida2014}, making it potentially one of the earliest observed examples of jet and outflow evolution.

\subsection{Alleviating the missing disk mass problem with multiwavelength observations}\label{sec:MissingMass}

While planet formation in protoplanetary disks is now well established \citep[e.g.][]{Keppler2018,vanCapelleveen2025}, there appears to be an observational disconnect between inferred dust masses and the masses of exoplanets \citep[e.g.][]{Najita2014}. Comparisons of the available dust mass in Class~II disks and solid content of
exoplanetary systems reveal a significant deficit, necessitating an almost 100\% planet formation efficiency \citep[e.g.][]{Mulders2015,Manara18}. This discrepancy is often referred to as the `missing mass' or `mass budget' problem. However, these dust masses were derived using ALMA observations which in some cases have been found to be optically thick \citep[e.g.][]{Ballering2019,Ribas2020,Rilinger2023,Xin2023,Chung2024,Radley2025}, therefore leading to reduced disk masses. Additionally, neglecting dust scattering can lead to a reduction in mass by a factor of 10 for compact ($<30$\,au) disks \citep[e.g.][]{Zhu2019}. Given that our radio SED analysis indicates that a portion of the millimetre emission may be optically thick (e.g. $\alpha_{\text{dust}}\sim2$) or potentially impacted by dust self-scattering \citep[e.g. $\alpha_{\text{dust}}\sim1.5$,][]{Sierra20} in half of our sample, our multiwavelength analysis will be able to investigate the impact of such effects.

\begin{figure*}[t]
    \centering
    \includegraphics[width=\textwidth,trim=0cm 0.5cm 0cm 0,keepaspectratio]{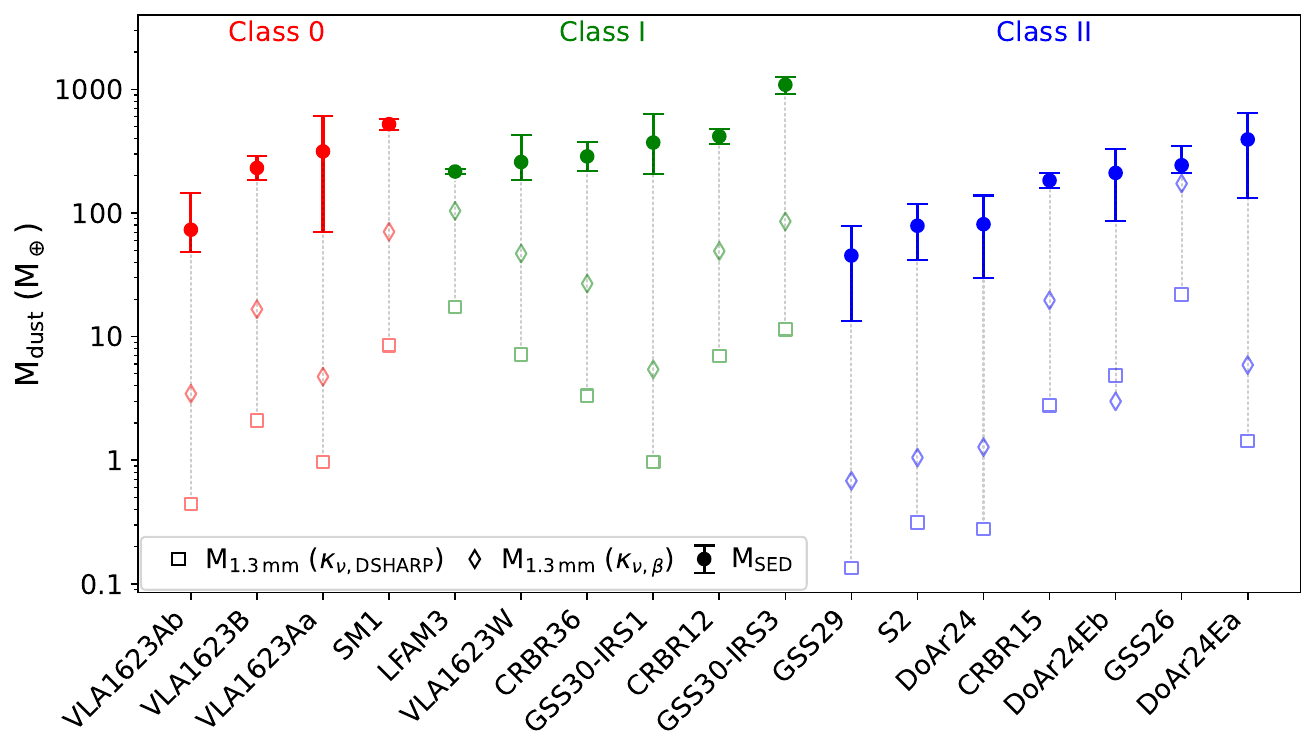}
    \caption{Dust masses derived for each object employing three methods, ordered by Class and increasing SED dust mass. Filled circles show masses derived using the multiwavelength radio SED analysis described in Section~\ref{sec:DustPopSED} with errors as shown in Table~\ref{tab:DustGasPop}. The remaining two methods use a singular ALMA 1.3\,mm (Band 6) flux density and Equation~\ref{eq:DustMass} but with differing opacities. Open squares signify masses calculated using opacities derived from our dust model ($\kappa_{\nu, \rm DSHARP}$), while open diamonds adopt the \citet{Beckwith1990} opacity power law prescription ($\kappa_{\nu, \beta}$).}
    \label{fig:SingFlux_Mass_Comparison}
\end{figure*}

Our SED-derived dust masses enable us to directly quantify the bias introduced from single wavelength mass estimates, which may be significantly impacted by optical depth effects.  In order to compare our results to previous approaches we use the standard dust mass equation from \citet[][]{Hildebrand1983QJRAS..24..267H}:

\begin{equation}
    M_{\rm dust} = \frac{F_\nu D^2}{\kappa_\nu B_\nu(T_{\rm dust})}
    \label{eq:DustMass}
\end{equation}

\noindent where $F_\nu$ is the flux density at frequency, $\nu$, $B_\nu(T_{\rm dust})$ is the Planck function calculated using the dust temperature, $T_{\rm dust}$, D is the distance to the disk and $\kappa_\nu$ is the opacity at frequency, $\nu$. For each object we use flux densities measured at frequencies corresponding to ALMA Band 6 observations (211--275\,GHz), as commonly chosen in the literature \citep[e.g.][]{Ansdell2017,Tychoniec2020,RuizRodriguez2025}. We consider two different opacities, the first includes dust scattering and absorption calculated from a$_{\rm max}$ as shown in Table~\ref{tab:DustGasPop}, assuming DSHARP dust compositions \citep[][]{Birnstiel2018ApJ...869L..45B} and an MRN distribution \citep[e.g. $q=-3.5$,][]{Mathis1977ApJ...217..425M} which we refer to as $\kappa_{\nu, \rm DSHARP}$. The second opacity assumes the \citet{Beckwith1990} prescription such that $\kappa_\nu\propto\nu^\beta$ where $\beta=\alpha_{\rm dust}-2$ using the spectral index, $\alpha_{\rm dust}$, from Table~\ref{tab:SEDFits}. Taking the reference opacity of 1.84 cm$^2$g$^{-1}$ at 345\,GHz from \citet{Ossenkopf1994}, we extrapolate to the Band 6 frequency using the calculated $\beta$ value, yielding $\kappa_{\nu, \beta}$. The strong dependence of $\kappa_\nu$ on a$_{\rm max}$ at millimetre and centimetre wavelengths, particularly when including scattering, necessitates this source by source approach \citep[][]{Woitke2016A&A...586A.103W,Sierra20}. Additionally, it is important to note that while Equation~\ref{eq:DustMass} is only valid in the optically thin limit, it provides a useful comparison to previous approaches. In Figure~\ref{fig:SingFlux_Mass_Comparison} we compare dust masses derived from 1.3\,mm flux densities assuming $T_{\rm dust}$ presented in Table~\ref{tab:DustGasPop} and opacities as discussed above, to those we derive from our multiwavelength radio SED analysis.

We find a significant increase in the available dust mass for each object when using our multiwavelength radio SED approach corresponding to a minimum increase of 11 times the mass derived from a 1.3\,mm flux density measurement for the $\kappa_{\nu, \rm DSHARP}$ opacity model, and a much more modest minimum increase of 1.4 for the $\kappa_{\nu, \beta}$ model. On average we find 150 and 34 times more mass for the $\kappa_{\nu, \rm DSHARP}$ and $\kappa_{\nu, \beta}$ models, with a maximum increase of 380 and 75 times the mass, respectively. The difference in inferred mass between each opacity model likely originates from the more complex function of opacity assumed in the DSHARP model. We infer typical $\kappa_{\nu, \beta}$ of $\sim1.8$\,cm$^2$\,g$^{-1}$ at 1.3\,mm whereas $\kappa_{\nu, \rm DSHARP}$ has an average of $\sim11$\,cm$^2$\,g$^{-1}$. As $M_{\rm dust}\propto\kappa_\nu^{-1}$ this order of magnitude difference results in much closer agreement between the masses inferred from the radio SED and those inferred from 1.3\,mm flux densities when using the \citet{Beckwith1990} opacity law. Nevertheless, both opacity laws produce reduced dust masses compared to our multiwavelength approach.

The significant increase in SED-derived dust masses relative to those derived from flux densities at 1.3\,mm may have several explanations. The most significant of which is high optical depths at millimetre wavelengths, as reported across a number of Class~0--II disks \citep[][]{Zhu2019,Chung2024}.  For the majority of our sample, we find dust spectral indices $\leq2.3$, consistent with the presence of optically thick millimetre emission, supporting high optical depths as a contributing factor \citep[][]{Garufi2025A&A...694A.290G}. 

Additionally, since the emissivity of dust grains peak at $\lambda\sim2\pi a_{\rm max}$, millimetre observations are less sensitive to cm-sized grains, potentially excluding a portion of the mass reservoir \citep[][]{Draine06,Viscardi2025}. Observations with ALMA Band 6 are therefore most sensitive to the emission of dust grains of the order 0.2\,mm, yet we infer grain growth beyond this for all objects in our sample. This suggests that a fraction of the dust mass may be unaccounted for in the mass derived from the 1.3\,mm flux density. Previous comparisons of dust masses derived from SEDs and masses derived from millimetre flux densities have reported underestimations up to a factor of $\sim$6, with the discrepancy increasing for more massive disks \citep[e.g.][]{Macias2021,Xin2023,Rilinger2023}. However, these studies did not include cm-wavelength observations and thus, may be insensitive to emission from larger grains representing a lower limit to the discrepancy. Consistent with this, \citet[][]{Painter2025} found an order of magnitude increase in the available dust mass when comparing optically thin $40$\,GHz emission to optically thick emission at 340\,GHz. Our inclusion of cm-wave observations, therefore, allows us to utilise optically thin tracers and, increase our sensitivity to larger dust grains, providing a more complete picture of the available dust mass.

Despite the large discrepancies of inferred masses in each method, our radio SED-derived dust mass estimates remain physically plausible. Previous multiwavelength approaches have inferred dust disk masses between 250--2000\,\Mearth\ for Class~II disks \citep[e.g.][]{Macias2021,Guidi2022,GuerraAlvarado2024_HLtau}. In our sample, the derived masses for Class~II disks are typically below 200\,\Mearth\ with DoAr24Ea having the highest inferred dust mass of 390\,\Mearth, occupying the lower end of the currently observed range. In addition, several of our Class~0 and I objects (VLA~1623AaAb,B,W, and SM1) have masses comparable to those found by \citet[][]{Maureira2026}. Class~0 and I objects are generally expected to have higher disk masses than Class~II objects as they have had less time to deplete their circumstellar disk through accretion \citep[][]{Xu2022}. However, with the exception of GSS30-IRS3, the dust masses we infer for Class~0-I objects are typically $\lesssim500$\,\Mearth, once again in agreement with previous multiwavelength dust mass estimates.

In summary, we find that our derived disk parameters and masses represent a physically plausible model of the observed radio SED emission. The substantially increased dust masses relative to those derived from millimetre flux densities alone can be explained by the combined effects of high optical depths, dust growth and the inclusion of cm-wave observations in our analysis. While these masses are indeed larger than estimates from millimetre flux densities, they primarily occupy the lower range of previous SED-derived dust masses, providing confidence in their reliability \citep[e.g.][]{Macias2021,Guidi2022,Painter2025}.

Consequently, these results have significant implications for planet formation and the missing mass problem. Masses inferred from millimetre flux densities using DSHARP opacities typically lack the sufficient solid material to build Earth-like cores, with the majority unable to achieve the 10\,\Mearth\ threshold required for runaway gas accretion to form Jupiter-like planets \citep[][]{Mizuno1980,Pollack1996}. This is consistent with the previously reported `missing mass' problem \citep[e.g.][]{Mulders2015,Manara18}. Alternatively, the \citet[][]{Beckwith1990} opacity prescription yields a higher average dust mass of $\sim$40\,\Mearth, which while beneficial for planet formation, would require very high planet formation efficiencies to create multi-planet systems \citep[e.g. 100\%][]{Mulders2021}. Therefore, our radio SED-derived dust masses present a partial alleviation to this tension, with all objects in our sample containing sufficient solid material to form multiple gas giant cores, with residual mass available to form smaller rocky bodies even through inefficient processes. 

\subsection{Is an evolutionary decline in dust mass evidence for early planetesimal formation?}

Through our multiwavelength radio SED analyses, we have found evidence of an increased solid mass reservoir for all objects when compared to typically inferred masses. However, there appears to be a stark separation in dust mass between Class~II objects and their less evolved counterparts in Class~0 and I. Previous studies utilising millimetre flux densities have reported similarly reduced masses for Class II objects in Ophiuchus compared to other star forming regions \citep[e.g.][]{Williams19}. Whilst high optical depths, dust growth and the inclusion of cm-wave observations can account for the discrepancy between radio SED-derived masses and those from millimetre flux densities, the separation in mass persists regardless of the method employed, pointing to an underlying physical cause. 

One explanation for the difference in observed masses may be higher optical depths at millimetre frequencies for Class~II disks. However, a comparison of the average dust spectral indices argues against this, with all evolutionary stages having $\alpha_{\rm dust}\sim2$, consistent with optically thick emission. The similarity between these spectral indices suggests that more evolved objects appear to be equally impacted by optical depths and thus cannot account for the large difference in mass. A more natural explanation could be the accretion of the disk onto the protostar, systematically reducing the disk mass over time \citep[][]{LyndenBell1974,Hartmann1998}. Class~0--I objects have been observed to have higher accretion rates than Class~II objects which would explain the reduced mass content in more evolved disks \citep[][]{Fiorellino2023}. Conversely, observations of infall through streamers onto Class 0 and I disks have been found to constitute a significant fraction of their total mass accretion rates, potentially providing an complementary mass reservoir from their natal cloud \citep[e.g.][]{Codella2024,Tanious2025}. However, given that these are estimates from molecular lines it is unclear how much of the dust content is replenished through streamers onto the disk. Considering we detect dust growth to mm-sizes for Class~II objects, the high levels of accretion place strong constraints on the formation timescales of planets. To counteract the depleting mass reservoir, dust growth must proceed rapidly in the Class~II phase in order to form km-sized rocky bodies.

Alternatively, dust in Class~II disks may have already grown to sizes which contribute negligibly at our observed wavelengths and, therefore, would be absent from our mass estimates \citep[][]{Ricci2010}. This is supported by the maximum grain sizes derived in Section~\ref{sec:DustPopSED} where Class II objects typically have larger dust grain sizes ($\sim2$\,mm) compared to Class 0 and Class I ($\sim1$\,mm), indicating that dust growth is well underway at these early evolutionary stages. The detection of ring/gap structures in millimetre continuum of Class~0 and I objects lends further support to this idea, potentially indicating early planet formation before the Class~II stage \citep[e.g.][]{Segura-Cox2020,Maureira2024}. The rapid formation of such large bodies would have to proceed through either the streaming instability or GI \citep[][]{Boss1997,Youdin2005}. Our radio SED-derived dust disk masses provide added support to the viability of the streaming instability, which requires high local dust-to-gas ratios in order to form 100\,km-size planetesimals, however, the true gas mass remains uncertain \citep[e.g.][]{Johansen2007}. While GI may form similarly large objects, it relies on a complex interplay between stellar mass, disk mass and local temperatures \citep[][]{Toomre1964,Gammie2001}. However, given that several objects in our sample may be gravitationally unstable \citep[$M_{\rm disk}\slash M_{\rm star} \gtrsim 0.1$;][]{Kratter2016}, we cannot rule it out as a possibility. In either case, the resulting planetesimals would have sizes which do not contribute significantly at mm-cm wavelengths, or contribute mostly at frequencies dominated by ionised gas, therefore remaining undetected in our SED analysis.

Ultimately, the precise origin of the reduced masses in Class~II is unclear, potentially stemming from accretion, dust growth or a combination of the two. Nevertheless, even in Class~II we find 10s--100s \Mearth\, of solid material still available, sufficient to create diverse planetary systems. While the specific timescales and planet formation mechanisms remain as open questions, our results point to a consistently more optimistic outlook for planet formation, regardless of evolutionary stage.

\begin{figure*}
    \centering
    \includegraphics[width=\textwidth,keepaspectratio]{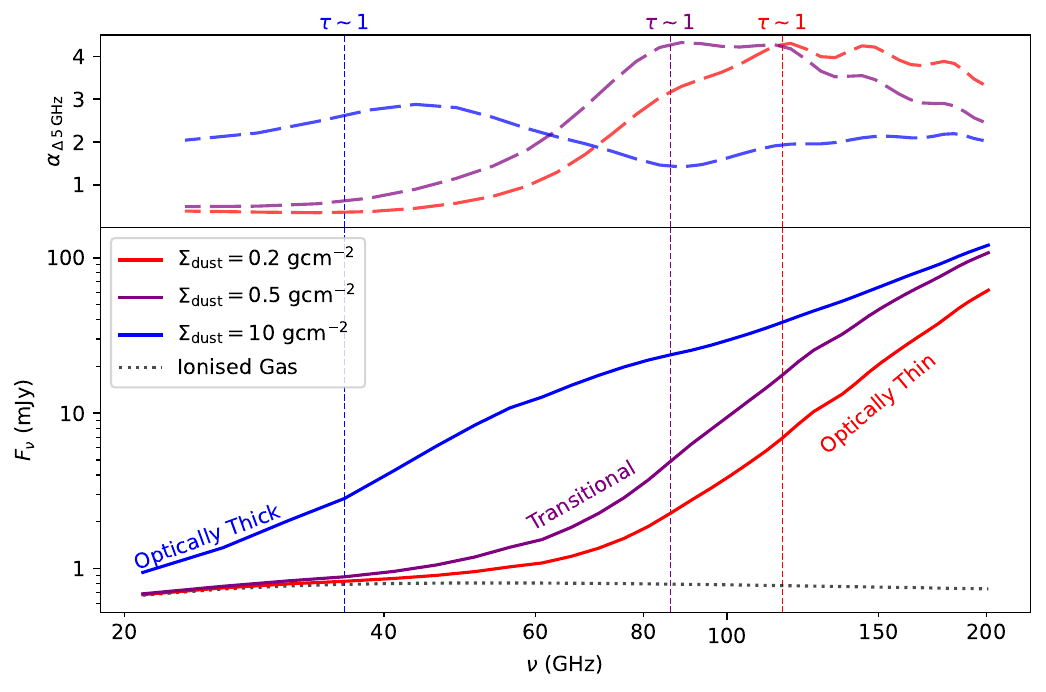}
    \caption{Theoretical radio SEDs (bottom) and their local spectral indices (top) which emulate different optical depth scenarios by assuming a range of $\Sigma_{\rm dust}$ values. The spectral index, $\alpha_{\Delta\rm5\,GHz}$, is calculated at 5\,GHz intervals along the radio SED and plotted against the midpoint of the frequency range. We annotate $\tau\sim1$ in each model with a dashed line, showing the optically thick to optically thin transition. Each model has constant $T_{\rm dust}=$\,21\,K, $a_{\rm max}=$\,0.8\,mm and $\Omega_{\rm dust}=1\times10^{11}$\,Sr for the dust component and, a constant EM$=9\times10^8$cm$^{-6}$pc and $\Omega_{\rm ionised}=13\times10^{-1}$\,Sr for the ionised gas (black dotted line).}
    \label{fig:VarySigmaSED_localSpIx}
\end{figure*}

\subsection{What is the origin of the radio SED discontinuity?}\label{sec:SEDDiscont}
As discussed in Section~\ref{sec:SpIx}, several objects have radio SEDs that cannot be accurately described by a single or double power law model. These objects exhibit a sharp discontinuity in the radio SED between 40--100\,GHz with the observed flux density at 44\,GHz falling significantly below the value extrapolated from the millimetre spectral slope. Using $\alpha_{\rm dust}$ as derived in Section~\ref{sec:SpIx} to estimate the expected dust flux density at 44\,GHz, we find that such extrapolation overpredicts the measured flux density by factors of 300--750. Previous studies of Herbig and T Tauri stars have reported an excess of emission at $\sim40$\,GHz which has been explained by contributions from optically thick ionised gas emission or more recently, the effects of dust scattering \citep[e.g.][]{Ubach12,Ubach2017,Macias2017,Sierra20}. In contrast, the objects in our sample exhibit a reduction, rather than an excess of flux density, suggesting a different physical origin.

One potential explanation for this discontinuity is the transition from optically thick emission at ALMA wavelengths to more optically thin emission observed by the VLA. This transition produces a steepening of the spectral index with $\alpha_{\rm dust}$ increasing from $\sim2$ in the optically thick limit towards values of 3--4 in the optically thin regime \citep[][]{Draine06}. Previous studies have found evidence of more optically thin emission at frequencies below $\sim100$\,GHz through spectral index maps and physical models of the SED \citep[e.g.][]{Xin2023,Radley2025,Zamudio2025}. Similarly, survey studies have revealed a systematic increase in the spectral index at lower frequencies, owing to a transition to optically thin emission \citep[e.g.][]{Chung2025,Painter2025,Garufi2025A&A...694A.290G}. By employing our radio SED model of dust and ionised gas emission, we can investigate whether this optical depth transition is sufficient to reproduce the observed discontinuity.

In Figure~\ref{fig:VarySigmaSED_localSpIx} we vary the dust surface density, $\Sigma_{\rm dust}$, as a proxy for the optical depth. Using the modelling procedure described in Section~\ref{sec:DustPopSED}, we compute integrated radio SEDs including contributions of both dust and ionised gas for a range of $\Sigma_{\rm dust}$ values (Figure~\ref{fig:VarySigmaSED_localSpIx}, bottom panel). We calculate the local spectral index at 5\,GHz intervals for each model, $\alpha_{\Delta\rm5\,GHz}$, and plot the value against the midpoint of the frequency 
range in the top panel. Finally, we derive optical depths using Equation~12 in \citet{Birnstiel2018ApJ...869L..45B} based on the underlying dust model and $\Sigma_{\rm dust}$. We show the optically thick/optically thin transition ($\tau\sim1$) as a dashed line in both plots to highlight its influence on the derived spectral index.

We consider three models representing distinct optical depth regimes across the 40--100\,GHz range: entirely optically thin (0.2\,g\,cm$^{-2}$), a transition from optically thick to thin (0.5\,g\,cm$^{-2}$) and, entirely optically thick (10\,g\,cm$^{-2}$). For the 0.2 and 0.5\,g\,cm$^{-2}$ models, the peak in $\alpha_{\Delta\rm5\,GHz}$ coincides with the $\tau\sim1$ transition, reaching values just above 4, consistent with the optically thin limit for grains $\lesssim3$\,mm \citep[][]{Draine06}. This peak is followed by a rapid decrease to $\alpha_{\Delta\rm5\,GHz}<1$ as ionised gas emission dominates the radio SED at lower frequencies. However, this is not the case for the $\Sigma_{\rm dust}=$10\,g\,cm$^{-2}$ model, where the emission remains optically thick between 40--100\,GHz and the $\tau\sim1$ transition occurs at frequencies which are contaminated by ionised gas emission. As a result, this model shows a shallower decline and broader peak in $\alpha_{\Delta\rm5\,GHz}$. Furthermore, this model does not reproduce the observed sharp discontinuity seen in the broken power law objects.

We can apply these diagnostics to our sample, using the model parameters derived in Section~\ref{sec:DustPopSED}. Broken power law objects have maximum local spectral indices of 3.5--4.4 consistent with those seen in the 0.2 and 0.5\,g\,cm$^{-2}$ models and indicative of a transition between optically thick and thin at these frequencies. The local spectral indices of the double power law objects measured in the 40--100\,GHz range lie between 1.8--3, consistent with the rise or fall seen either side of the $\tau\sim1$ transition. However, considering the turnover frequency directly, we find these objects become optically thin at frequencies $<35$\,GHz, consistent with being still optically thick in the 40-100\,GHz range. Conversely, the broken power law objects transition to optically thin dust emission between 40--60\,GHz. These results are consistent with an optical depth origin for the observed 40--100\,GHz discontinuity, and demonstrates that objects which are best described by a double power law, are likely to remain optically thick up to centimetre wavelengths, with the dust contribution more smoothly transitioning into the ionised gas dominated regime.

Further supporting this idea is the work of \citet[][]{Painter2025} who used very high spectral sampling to characterise the SED morphology in this regime. They found spectral indices of 2.8--4 at 43\,GHz which are substantially steeper than those measured above 200\,GHz and consistent with the steepening expected from the optically thick to thin transition we discuss here. 
From a survey perspective, with less fine spectral sampling, \citet[][]{Garufi2025A&A...694A.290G} finds similar steepening of the spectral index towards lower frequencies. Therefore, the agreement of our model with the observed spectral indices in our survey, the well characterised radio SEDs in \citet[][]{Painter2025} and the survey of \citet[][]{Garufi2025A&A...694A.290G} provides further confidence that this feature is physical in nature, rather than the effect of poorly sampled radio SEDs.

Nevertheless, we require more observations between 40--100\,GHz in order to accurately characterise radio SED morphologies. This in turn will allow us to improve our understanding of dust growth properties and in particular, allow us to understand and account for optical depths effects in the mm-cm regime. Such observations will be achievable using ALMA Band 1 and Band 2, however, the VLA is currently able to achieve better resolution at $\sim40$\,GHz than Band 1. Future improvements such as the upcoming wideband sensitivity upgrade \citep[WSU][]{Carpenter2023} will enable greater access to survey scale observations, enabling a more global understanding of radio SED morphologies and the transition from optically thick to optically thin emission.

\begin{figure}
    \centering
    \includegraphics[width=\linewidth,keepaspectratio]{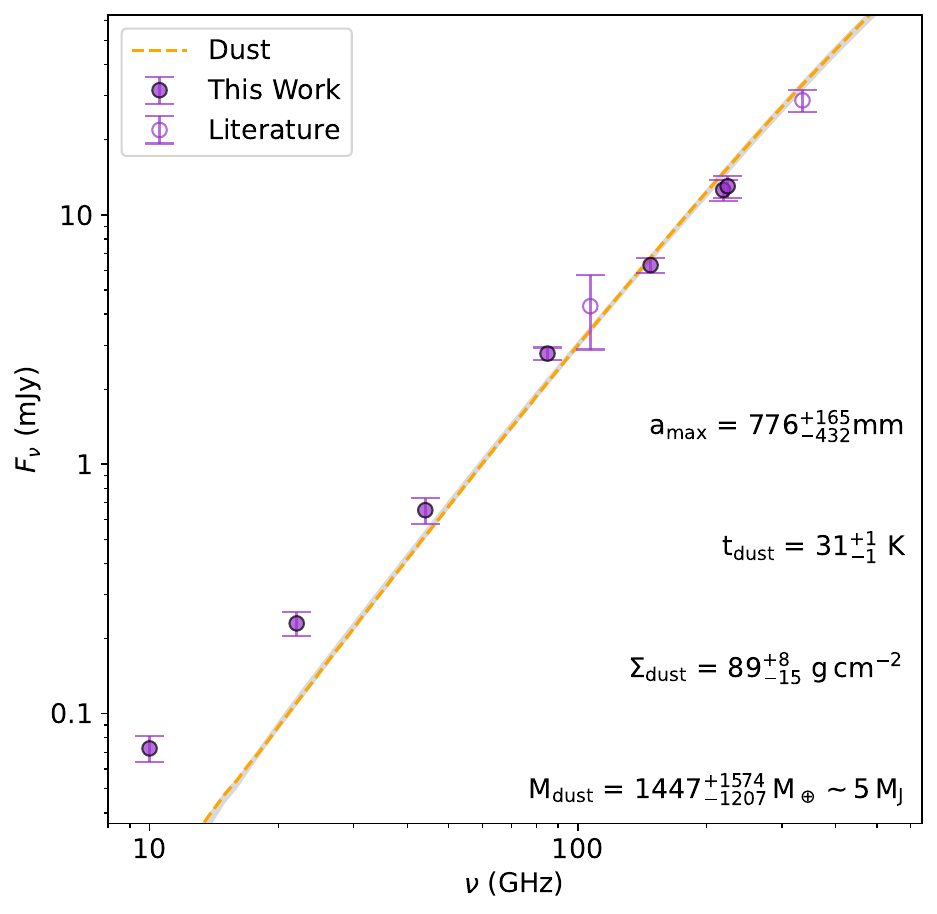}
    \caption{SED model of GSS30-IRS1 considering dust only emission (black solid line), compared to observed flux densities from this work (closed circles) and the literature (open circles). Derived parameters are annotated in the bottom right alongside the 16th and 84th percentile uncertainties in addition to 1000 draws from the posterior as the orange shaded region. Note that even with extreme dust disk properties, the model cannot replicate the radio SED at frequencies $\lesssim40$\,GHz.}
    \label{fig:GSS30-IRS1_DustOnlySED}
\end{figure}

\subsection{Can single power law radio SEDs be explained by dust alone?}
In Section~\ref{sec:SpIx} we find two objects, GSS30-IRS1 and DoAr24Eb, with radio SEDs that are consistent with a single component spectral index of 1.7 and 1.8, respectively. These values lie slightly below the optically thick limit, potentially owing to ionised gas contributions at low frequencies, dust scattering at higher frequencies or a combination of both \citep[e.g.][]{Anglada18,Sierra20}. While we are able to replicate the radio SED using a combination of dust-thermal emission and an ionised gas component in Section~\ref{sec:DustPopSED}, the single power law morphology raises the questions of whether we can reproduce the radio SED using a single dust component. We choose to model GSS30-IRS1 due to its well sampled millimetre radio SED and resolved millimetre continuum, mitigating the effects of degeneracy in the model fitting. Below, we investigate the level of dust growth and mass required for such a dust model and discuss the physical implications.

We follow a similar procedure as in Section~\ref{sec:DustPopSED}, excluding the contributions of the ionised gas component. As larger dust grains contribute more significantly at longer wavelengths, we extend the upper limit of a$_{\rm max}$ to 1\,m to allow a greater exploration of the parameter space. Furthermore, to compensate for the reduced flux density contributions in the low frequency regime, we increase the upper limit of $\Sigma_{\rm dust}$ to 100\,g\,cm$^{-2}$ corresponding to greater quantities of emitting dust. We allow the fitting procedure to run for 5$\times10^4$ iterations with 50 walkers and remove a burn in of 2000. In Figure~\ref{fig:GSS30-IRS1_DustOnlySED} we present the posterior median model, along with derived parameters and uncertainties.

Our results show that despite the approximately linear radio SED, our dust-only model cannot accurately reproduce the emission at frequencies below $\sim$40\,GHz, systematically underpredicting the observed flux densities by a factor of 2 and 4 at 20\,GHz and 10\,GHz, respectively. This can be explained in part due to the contribution of ionised gas emission in the inner disk, which appears to increase with increasing accretion rate \citep[][]{Rota2024,Rota2025}. GSS30-IRS1 is a Class~I object which are often associated with higher accretion rates than more evolved Class~II objects \citep[][]{Fiorellino2023}. Combined with the high ionised gas contamination fractions (f$_{\rm ion}$) found for other objects at frequencies $<40$\,GHz, this provides a potential explanation for the systematic underprediction at centimetre wavelengths.

In order to compensate for the absence of this ionised gas emission, the dust-model converges to values which are potentially unphysical. We find that the model parameters are consistent with dust growth to sizes of nearly 80\,cm, roughly 1700 times bigger than we infer with our dust and ionised gas model (0.46\,mm). Similarly, we find a mass of $\sim$1450\,\Mearth\ corresponding to an increase of almost 4$\times$ relative to our previous estimate (370\,\Mearth). Such a high mass is comparable to the upper mass limit for HL Tau \citep[$\sim2000$\Mearth,][]{GuerraAlvarado2024_HLtau} and well in excess of other previously determined dust masses \citep[250--340,][]{Macias2021,Guidi2022}. In addition, while high surface densities ($\Sigma_{\rm dust}>100$\,g\,cm$^{-2}$) have been observed for Class~I objects, they are exclusively in the inner disk, unlike the average surface density ($\Sigma_{\rm dust}\sim90$\,g\,cm$^{-2}$) presented here \citep[e.g.][]{Guerra-Alvarado2024}. Consequently, we can reject the dust-only model due to its convergence on physically unlikely dust properties. 

Therefore, even with potentially unphysical quantities of large (10s\,cm) dust grains, the dust-only model fails to reproduce the observed radio SED of GSS30-IRS1. In order to account for excess emission at low frequencies, we must invoke an ionised gas mechanism, even when the radio SED can be described by a single power law component. The radio SED of GSS30-IRS1 therefore represents an example of coincidental alignment between the dust and ionised gas spectral indices as opposed to a dust-only origin. Furthermore, this analysis highlights the necessity of low-frequency observations ($\lesssim40$) in order to probe ionised gas emission and constrain the impact of arbitrarily large dust grains on the resulting radio SED.

\section{Conclusions}\label{sec:Conclusions}

We present VLA Q- (44\,GHz), K- (22\,GHz) and X-band (10\,GHz) observations for 20 YSOs from Class 0--III in the Ophiuchus star forming region. Where available, we make comparisons to JWST NIRCam and archival ALMA observations from both a morphological and disk-integrated perspective. We present our key results below:

\begin{itemize}
   
    \item We find multiple examples of alignment between 10\,GHz continuum emission and outflow cavities seen with NIRCam. The aligned nature of these two frequency regimes allows us to morphologically connect outflows to their ionised origins across scales of 10s--1000s of au. 

    \item We find evidence of dust growth to roughly millimetre sizes or beyond for the majority of objects in addition to tens to thousands of Earth masses in solid material available for planet formation.

    \item Discontinuities in the radio SED between 40--100\,GHz can be explained due to the transition to between optically thick ALMA observations and more optically thin VLA observations. These optical depth transitions occur for roughly half of our sample, while the rest remain optically thick.

    \item Radio SEDs covering (sub-)mm and centimetre frequencies cannot be explained solely by dust emission, even when invoking potentially unphysical dust properties. Contributions from ionised gas mechanisms are essential to reconcile the observed flux density with flux densities from dust-only emission at frequencies $<$\,40\,GHz.
    
\end{itemize}

This survey represents one of the highest resolution multiwavelength studies of Ophiuchus spanning the NIR and radio. Nevertheless, our analysis remains limited by the absence of contemporaneous, high resolution observations at frequencies $<40$\,GHz, restricting our ability to characterise the contributions of ionised gas mechanisms such as jets and winds. Future facilities such as the SKA-MID will provide resolutions which are nearly 10$\times$ better than those presented here at frequencies $<20$\,GHz \citep[e.g. 0\farcs04 at 12.5\,GHz][]{Braun2019}, while the ngVLA will provide several orders of magnitude in improvement across the 2--40\,GHz range \citep[][]{Selina2018}. In addition, theses facilities will provide a minimum improvement of 4 times the sensitivity we achieve here, enabling the recovery of faint centimetre emission ($<4\,\mu$Jy\,beam$^{-1}$) and allowing further confirmation of the nature and origin of the 10\,GHz extensions seen in Figure~\ref{fig:10GHzExtensions}. This drastic increase in resolution and sensitivity, will allow for a more accurate morphological and spectral characterisation of the low frequency emission \citep[e.g.][]{Wilner2004,Ilee2020}. In addition, the upcoming ALMA wideband sensitivity upgrade will provide greater access to observations between 60--110\,GHz allowing for survey scale characterisations of YSO dust disk properties \citep[][]{Carpenter2023}. Such observations will yield improved constraints on the spatial extent of dust emission alongside probing the optical depth transition in protoplanetary disks. Together, these facilities will provide the resolution, sensitivity and spectral coverage required to fully characterise the millimetre-centimetre radio SED and, enable a greater understanding of dust and ionised gas emission in YSOs.

\begin{acknowledgments}
We thank the anonymous referee for their constructive and insightful feedback.
I.C.R. is supported by a studentship from the Science and Technology Facilities Council (STFC) of the United Kingdom.
J.D.I. acknowledges support from an STFC Ernest Rutherford Fellowship (ST/W004119/1).
G.B. acknowledges support from the PID2023-146675NB-I00 (MCI-AEI-FEDER, UE) program and CEX2024-001451-M funded by MICIU/AEI/10.13039/501100011033.
G.B. and J.M.G. acknowledge support by Spanish grant PID2023-146675NB-I00 (MCI-AEI-FEDER, UE). This work was also partly supported by the Spanish program Unidad de Excelencia María de Maeztu CEX2020-001058-M, financed by MCIN/AEI/10.13039/501100011033, and by the MaX-CSIC Excellence Award MaX4-SOMMA-ICE.
H.B.L. is supported by the National Science and Technology Council (NSTC) of Taiwan (Grant Nos. 111-2112-M-110-022-MY3, 113-2112-M-110-022-MY3).
A.R. has received funding from the Royal Society through a University Research Fellowship grant number URF\textbackslash R1\textbackslash 241791.
M.A. acknowledges support by the Swiss National Science Foundation, grant 2000-1-240076.
E.B. acknowledges support from the Italian Ministry for Universities and Research under the Italian Science Fund (FIS 2 Call - Ministerial Decree No. 1236 of 1 August 2023) grant FIS-2023-00170.
C.C. and L.P.  acknowledge the project PRIN MUR 2022 FOSSILS (Prot. 2022JC2Y93), the project ASI-Astrobiologia 2023 MIGLIORA (F83C23000800005), the INAF-GO 2023 fundings PROTO-SKA (C13C23000770005).
A.C. received financial support from the European Research Council (ERC) under the European Union’s Horizon 2020 research and innovation programme (ERC Starting Grant “Chemtrip”, grant agreement No 949278.
I.J-.S acknowledges funding from grant PID2022-136814NB-I00 funded by the Spanish Ministry of Science, Innovation and Universities/State Agency of Research MICIU/AEI/ 10.13039/501100011033 and by “ERDF/EU”.
L.L. acknowledges the support of DGAPA-PAPIIT grant IN108324 and SECIHTI grant CBF-2025-I- 109.
We acknowledge the Spanish Prototype of an SRC (espSRC) service and support funded by the Spanish Ministry of Science, Innovation and Universities, by the Regional Government of Andalusia, by the European Regional Development Funds and by the European Union NextGenerationEU/PRTR. The SPSRC acknowledges financial support from the State Agency for Research of the Spanish MCIU through the "Center of Excellence Severo Ochoa" award to the Instituto de Astrof\'isica de Andaluc\'ia (SEV-2017-0709) and from the grant CEX2021-001131-S funded by MCIN/AEI/ 10.13039/501100011033.
This work has been enabled by access to facilities and the scientific and technical support provided by the UK SKA Regional Centre (ukSRC). The ukSRC is a collaboration between the University of Cambridge, University of Edinburgh, Durham University, University of Hertfordshire, University of Manchester, University College London, the UKRI STFC Scientific Computing (STFC) at RAL, and RAL Space. The ukSRC is supported by funding from the UKRI STFC.
A portion of this research was carried out at the Jet Propulsion Laboratory, California Institute of Technology, under a contract with the National Aeronautics and Space Administration (80NM0018D0004). This work is based in part on observations made with the NASA/ESA/CSA James Webb Space Telescope. The data were obtained from the Mikulski Archive for Space Telescopes at the Space Telescope Science Institute, which is operated by the Association of Universities for Research in Astronomy, Inc., under NASA contract NAS 5-03127 for JWST. These observations are associated with program \#2739.
This paper makes use of the following ALMA data: ADS/JAO.ALMA\#2017.1.00107 and ADS/JAO.ALMA\#2019.A.00034. ALMA is a partnership of ESO (representing its member states), NSF (USA) and NINS (Japan), together with NRC (Canada), NSTC and ASIAA (Taiwan), and KASI (Republic of Korea), in cooperation with the Republic of Chile. The Joint ALMA Observatory is operated by ESO, AUI/NRAO and NAOJ.
Some of the data presented in this article were obtained from the Mikulski Archive for Space Telescopes (MAST) at the Space Telescope Science Institute. The specific observations analyzed can be accessed via \dataset[doi:10.17909/hve0-q493]{https://doi.org/10.17909/hve0-q493}.
The National Radio Astronomy Observatory and Green Bank Observatory are facilities of the U.S. National Science Foundation operated under cooperative agreement by Associated Universities, Inc.

\end{acknowledgments}

\facilities{JWST (NIRCam), ALMA, VLA} 

\software{CASA \citep{CASA}, analysisUtils \citep{analysisutils_2023}, JWST Calibration Pipeline \citep{Bushouse23}, Astropy \citep{astropy2013,astropy2018,astropy2022}, PyBDSF \citep{PyBDSF}, emcee \citep{Emcee}, Optool \citep[][]{Optool}}

\bibliography{Bibliography}{}
\bibliographystyle{aasjournalv7}

\appendix
\newpage
\section{VLA Observational Properties}\label{apx:VLA_ObsParams}
\startlongtable
\setlength{\extrarowheight}{0pt}
\begin{deluxetable*}{lcccccc}
\tablecaption{Observational parameters for fiducial Q-, K-, and X-band images and their respective objects \label{tab:obsparams}}
\tabletypesize{\footnotesize}
\tablehead{
\colhead{Object(s)} &
\colhead{Phase Centre} &
\colhead{Beam} &
\colhead{RMS\tablenotemark{a}} &
\colhead{t$_{\rm int}$} &
\colhead{Briggs} &
\colhead{No. of} \\
\colhead{} &
\colhead{(J2000)} &
\colhead{(maj $\times$ min [PA])} &
\colhead{(mJy\,beam$^{-1}$)} &
\colhead{(min)} &
\colhead{Robust} &
\colhead{Epochs}
}
\startdata
\sidehead{\textbf{Q-band:}}
\multicolumn{1}{@{}l@{}}{
    \parbox[t]{4cm}{\raggedright GSS30-IRS1, IRS2, IRS3}} & 16:26:21.56 $-$24.22.55.00 & \phantom{1}\phantom{1}\,\,0\farcs09$\times$0\farcs05 [$-$11$\degr$] &    0.02 &    83 &    0.5 &     2 \\
\multicolumn{1}{@{}l@{}}{
    \parbox[t]{4cm}{\raggedright CRBR36}} & 16:26:25.49 $-$24.23.01.60 & \, 0\farcs09$\times$0\farcs04 [12$\degr$] &    0.02 &         53 &    0.5 &     2 \\
\multicolumn{1}{@{}l@{}}{
    \parbox[t]{4cm}{\raggedright GSS26}} & 16:26:10.32 $-$24.20.54.90 & \phantom{1}\phantom{1}  0\farcs09$\times$0\farcs04 [$-$10$\degr$] &    0.02 &         68 &    0.5 &     2 \\
\multicolumn{1}{@{}l@{}}{
    \parbox[t]{4cm}{\raggedright CRBR12, CRBR15}} & 16:26:18.00 $-$24.24.00.00 &   0\farcs08$\times$0\farcs04 [3$\degr$]  &    0.02 &    83 &    0.5 &     2 \\
\multicolumn{1}{@{}l@{}}{
    \parbox[t]{4cm}{\raggedright DoAr24}} & 16:26:17.06 $-$24.20.21.60 & \phantom{1} 0\farcs09$\times$0\farcs04 [$-$9$\degr$] &    0.02 &    83 &   0.5 &     2 \\
\multicolumn{1}{@{}l@{}}{
    \parbox[t]{4cm}{\raggedright DoAr24Ea, DoAr24Eb}} & 16:26:23.39 $-$24.21.00.00 &   0\farcs09$\times$0\farcs05 [8$\degr$] &    0.03 &         68 &    0.5 &     2 \\
\multicolumn{1}{@{}l@{}}{
    \parbox[t]{4cm}{\raggedright VSSG27}} & 16:26:30.47 $-$24.22.57.10 &   0\farcs08$\times$0\farcs04 [3$\degr$] &    0.02 &    68 &    0.5 &     2 \\
\multicolumn{1}{@{}l@{}}{
    \parbox[t]{4cm}{\raggedright LFAM3, S2}} & 16:26:24.00 $-$24.24.44.00 & \phantom{1}  0\farcs08$\times$0\farcs04 [$-$8$\degr$] &     0.02 &         61 &    0.5 &     2 \\
\multicolumn{1}{@{}l@{}}{
    \parbox[t]{4cm}{\raggedright SM1, VLA~1623~AaAb, B}} & 16:26:27.80 $-$24.24.40.30 & \phantom{1}0\farcs09$\times$0\farcs06 [17$\degr$] &   0.03 &    61 &    0.5 &     2 \\
\multicolumn{1}{@{}l@{}}{
    \parbox[t]{4cm}{\raggedright GSS29}} & 16:26:16.85 $-$24.22.23.50 & \phantom{1} \phantom{1}\, 0\farcs1$\times$0\farcs05 [$-$11$\degr$] &  0.02 &         27 &    2 &     1 \\
\multicolumn{1}{@{}l@{}}{
    \parbox[t]{4cm}{\raggedright VLA~1623~W}} & 16:26:27.80 $-$24.24.40.30 & 0\farcs26$\times$0\farcs14 [8$\degr$] & 0.03 & 61 & 2\tablenotemark{b} & 2\\
\sidehead{\textbf{K-band:}}
\multicolumn{1}{@{}l@{}}{
    \parbox[t]{4cm}{\raggedright
CRBR36, GSS30-IRS1, IRS2, IRS3 }} & 16:26:21.56 $-$24.22.55.00 & 0\farcs18$\times$0\farcs1 [$-$22$\degr$] &   0.01 &    30 &    0.5 &     2 \\
\multicolumn{1}{@{}l@{}}{
    \parbox[t]{4cm}{\raggedright CRBR12, CRBR15}} & 16:26:18.00 $-$24.24.00.00 &   0\farcs15$\times$0\farcs08 [$-$7$\degr$] &     0.01 &    30 &    0.5 &     2 \\
\multicolumn{1}{@{}l@{}}{
    \parbox[t]{4cm}{\raggedright GSS26}} & 16:26:10.32 $-$24.20.54.90 & \phantom{1}0\farcs18$\times$0\farcs09 [$-$19$\degr$] &   0.01 &         30 &    0.5 &     2 \\
\multicolumn{1}{@{}l@{}}{
    \parbox[t]{4cm}{\raggedright DoAr24Ea, DoAr24Eb}} & 16:26:23.39 $-$24.21.00.00 &  0\farcs16$\times$0\farcs08 [$-$1$\degr$] &   0.01 &    30 &    0.5 &     2 \\
\multicolumn{1}{@{}l@{}}{
    \parbox[t]{4cm}{\raggedright GSS35, VSSG27}} & 16:26:30.47 $-$24.22.57.10 &  0\farcs15$\times$0\farcs08 [$-$7$\degr$] &   0.01 &    30 &    0.5 &     2 \\
\multicolumn{1}{@{}l@{}}{
    \parbox[t]{4cm}{\raggedright
LFAM3, S2, SM1,  VLA~1623~AaAb, B }} & 16:26:25.63 $-$24.24.29.40 &  0\farcs18$\times$0\farcs12 [19$\degr$]\, &   0.01 &    29 &    0.5 &     2 \\
\multicolumn{1}{@{}l@{}}{
    \parbox[t]{4cm}{\raggedright GSS29}} & 16:26:16.85 $-$24.22.23.50 & \phantom{1}\phantom{1}0\farcs20$\times$0\farcs09 [$-$16$\degr$]\phantom{1} &   0.01 &    15 &    2 &     1 \\
\multicolumn{1}{@{}l@{}}{
    \parbox[t]{4cm}{\raggedright DoAr24}} & 16:26:17.06 $-$24.20.21.60 & \phantom{1}\phantom{1}0\farcs20$\times$0\farcs09 [$-$17$\degr$]\phantom{1} &   0.03 &    15 &    2 &     1 \\
\multicolumn{1}{@{}l@{}}{
    \parbox[t]{4cm}{\raggedright VLA~1623~W}} & 16:26:27.80 $-$24.24.40.30 & \,0\farcs22$\times$0\farcs16 [24$\degr$]\phantom{1} & 0.01 & 61 & 2 & 2\\
\sidehead{\textbf{X-band:}} 
\multicolumn{1}{@{}l@{}}{
    \parbox[t]{4cm}{\raggedright CRBR12}} &  16:26:16.50 $-$24.22.03.00 &  0\farcs34$\times$0\farcs18 [$-$8$\degr$] &    0.004 &    55 &    0.5 &    15 \\
\multicolumn{1}{@{}l@{}}{
    \parbox[t]{4cm}{\raggedright DoAr24}} &  16:26:16.50 $-$24.22.03.00 &  0\farcs34$\times$0\farcs18 [$-$8$\degr$] &   0.004 &    66 &    0.5 &    18 \\
\multicolumn{1}{@{}l@{}}{
    \parbox[t]{4cm}{\raggedright DoAr24Ea}} & 16:26:16.50 $-$24.22.03.00 &   0\farcs34$\times$0\farcs18 [$-$7$\degr$] &   0.004 &         52 &    0.5 &    14 \\
\multicolumn{1}{@{}l@{}}{
    \parbox[t]{4cm}{\raggedright DoAr24Eb}} &  16:26:16.50 $-$24.22.03.00 &  0\farcs34$\times$0\farcs18 [$-$7$\degr$] &   0.005 &         34 &    0.5 &     9 \\
\multicolumn{1}{@{}l@{}}{
    \parbox[t]{4cm}{\raggedright GSS26 }} &  16:26:16.50 $-$24.22.03.00 &  0\farcs34$\times$0\farcs18 [$-$6$\degr$] &    0.004 &         58 &    0.5 &    16 \\
\multicolumn{1}{@{}l@{}}{
    \parbox[t]{4cm}{\raggedright GSS29}} &   16:26:16.50 $-$24.22.03.00 &  0\farcs34$\times$0\farcs18 [$-$7$\degr$] &   0.004 &    55 &    0.5 &    15 \\
\multicolumn{1}{@{}l@{}}{
    \parbox[t]{4cm}{\raggedright CRBR15}} & 16:26:24.60 $-$24.23.37.00 &  \phantom{1}0\farcs34$\times$0\farcs17 [$-$10$\degr$] &   0.008 &    17 &    0.5 &     5 \\
\multicolumn{1}{@{}l@{}}{
    \parbox[t]{4cm}{\raggedright CRBR36}} & 16:26:24.60 $-$24.23.37.00 &  0\farcs34$\times$0\farcs18 [$-$9$\degr$] &    0.005 &    39 &    0.5 &    11 \\
\multicolumn{1}{@{}l@{}}{
    \parbox[t]{4cm}{\raggedright GSS30-IRS1}} & 16:26:24.60 $-$24.23.37.00 &  0\farcs34$\times$0\farcs18 [$-$6$\degr$] &   0.004 &         61 &    0.5 &    17 \\
\multicolumn{1}{@{}l@{}}{
    \parbox[t]{4cm}{\raggedright GSS30-IRS2}} & 16:26:24.60 $-$24.23.37.00 &  0\farcs34$\times$0\farcs17 [$-$6$\degr$] &    0.006 &    28 &    0.5 &     8 \\
\multicolumn{1}{@{}l@{}}{
    \parbox[t]{4cm}{\raggedright GSS30-IRS3}} &  16:26:24.60 $-$24.23.37.00 &  0\farcs34$\times$0\farcs18 [$-$6$\degr$] &   0.004 &    61 &    0.5 &    17 \\
\multicolumn{1}{@{}l@{}}{
    \parbox[t]{4cm}{\raggedright GSS35}} & 16:26:24.60 $-$24.23.37.00 & \phantom{1}0\farcs34$\times$0\farcs18 [$-$11$\degr$] &    0.01 &    11 &    0.5 &     3 \\
\multicolumn{1}{@{}l@{}}{
    \parbox[t]{4cm}{\raggedright LFAM3}} & 16:26:24.60 $-$24.23.37.00 &  0\farcs34$\times$0\farcs18 [$-$7$\degr$] &   0.004 &    64 &    0.5 &    18 \\
\multicolumn{1}{@{}l@{}}{
    \parbox[t]{4cm}{\raggedright S2}} & 16:26:24.60 $-$24.23.37.00 &  0\farcs34$\times$0\farcs18 [$-$6$\degr$] &   0.004 &         61 &    0.5 &    17 \\
\multicolumn{1}{@{}l@{}}{
    \parbox[t]{4cm}{\raggedright SM1}} & 16:26:24.60 $-$24.23.37.00 &  0\farcs34$\times$0\farcs18 [$-$6$\degr$] &   0.004 &         61 &    0.5 &    17 \\
\multicolumn{1}{@{}l@{}}{
    \parbox[t]{4cm}{\raggedright VLA~1623~Aa}} & 16:26:24.60 $-$24.23.37.00 &  0\farcs34$\times$0\farcs18 [$-$7$\degr$] &   0.004 &    64 &    0.5 &    18 \\
\multicolumn{1}{@{}l@{}}{
    \parbox[t]{4cm}{\raggedright VLA~1623~Ab}} & 16:26:24.60 $-$24.23.37.00 &  0\farcs34$\times$0\farcs18 [$-$7$\degr$] &   0.004 &    64 &    0.5 &    18 \\
\multicolumn{1}{@{}l@{}}{
    \parbox[t]{4cm}{\raggedright VLA~1623~B}} &  16:26:24.60 $-$24.23.37.00 &  0\farcs34$\times$0\farcs18 [$-$7$\degr$] &   0.004 &         57 &    0.5 &    16 \\
\multicolumn{1}{@{}l@{}}{
    \parbox[t]{4cm}{\raggedright VLA~1623~W}} & 16:26:24.60 $-$24.23.37.00 &  0\farcs34$\times$0\farcs18 [$-$7$\degr$] &   0.004 &    64 &    0.5 &    18 \\
\multicolumn{1}{@{}l@{}}{
    \parbox[t]{4cm}{\raggedright VSSG27}} & 16:26:24.60 $-$24.23.37.00 &  0\farcs34$\times$0\farcs18 [$-$7$\degr$] &   0.004 &    64 &    0.5 &    18\\
\enddata
\tablenotetext{a}{Measured at the centre of the primary beam.}
\tablenotetext{b}{UVTaper = 0\farcs2$\times$0\farcs08 [15$\degr$], see \citet{Radley2025}}
\end{deluxetable*}

\section{Archival ALMA Image Properties}\label{apx:ALMA_ObsParams}
\startlongtable
\setlength{\extrarowheight}{0pt}
\begin{deluxetable*}{lcccc}
\tablecaption{Image properties for observations within archival ALMA Projects \label{tab:ALMAProjects}}
\tabletypesize{\footnotesize}
\tablehead{
\colhead{Object(s)} &
\colhead{Frequency} &
\colhead{Phase Centre} &
\colhead{Beam} &
\colhead{RMS\tablenotemark{a}} \\
\colhead{} &
\colhead{(GHz)} &
\colhead{(J2000)} &
\colhead{(maj $\times$ min [PA])} &
\colhead{(mJy\,beam$^{-1}$)} 
}
\startdata
\sidehead{\textbf{2017.1.00107.S}} 
\multicolumn{1}{@{}l@{}}{
    \parbox[t]{4cm}{\raggedright CRBR12}} & 224 & 16:26:17.23 $-$24.23.45.79 & 1\farcs21$\times$0\farcs82 [$-$80$\degr$] & 0.1  \\
\multicolumn{1}{@{}l@{}}{
    \parbox[t]{4cm}{\raggedright GSS30-IRS1, IRS3}} & 224 & 16:26:21.34 $-$24.23.03.30 & 1\farcs21$\times$0\farcs82 [$-$80$\degr$] & 0.2  \\
\multicolumn{1}{@{}l@{}}{
    \parbox[t]{4cm}{\raggedright LFAM3, S2}} & 224 & 16:26:23.53 $-$24.24.38.55 & 1\farcs21$\times$0\farcs82 [$-$80$\degr$] & 0.2  \\
\multicolumn{1}{@{}l@{}}{
    \parbox[t]{4cm}{\raggedright CRBR36}} & 224 & 16:26:25.39 $-$24.23.01.13 & 1\farcs21$\times$0\farcs82 [$-$80$\degr$] & 0.2  \\
\multicolumn{1}{@{}l@{}}{
    \parbox[t]{4cm}{\raggedright VLA~1623~W}} & 224 & 16:26:25.59 $-$24.24.27.80 & 1\farcs20$\times$0\farcs82 [$-$80$\degr$] & 0.3  \\
\multicolumn{1}{@{}l@{}}{
    \parbox[t]{4cm}{\raggedright CRBR15}} & 224 & 16:26:18.93 $-$24.24.14.04 & 1\farcs21$\times$0\farcs82 [$-$80$\degr$] & 0.1  \\
\multicolumn{1}{@{}l@{}}{
    \parbox[t]{4cm}{\raggedright CRBR12}} & 148 & 16:26:17.21 $-$24.23.45.68 & 1\farcs16$\times$0\farcs95 [$-$88$\degr$] & 0.1  \\
\multicolumn{1}{@{}l@{}}{
    \parbox[t]{4cm}{\raggedright GSS30-IRS1, IRS2, IRS3}} & 148 & 16:26:21.33 $-$24.23.04.42 & 1\farcs16$\times$0\farcs95 [$-$88$\degr$] & 0.2  \\
\multicolumn{1}{@{}l@{}}{
    \parbox[t]{4cm}{\raggedright LFAM3, S2}} & 148 & 16:26:23.52 $-$24.24.39.81 & 1\farcs16$\times$0\farcs95 [$-$88$\degr$] & 0.2  \\
\multicolumn{1}{@{}l@{}}{
    \parbox[t]{4cm}{\raggedright CRBR36}} & 148 & 16:26:25.43 $-$24.23.01.29 & 1\farcs16$\times$0\farcs95 [$-$88$\degr$] & 0.1  \\
\multicolumn{1}{@{}l@{}}{
    \parbox[t]{4cm}{\raggedright VLA~1623~W}} & 148 & 16:26:25.60 $-$24.24.27.64 & 1\farcs16$\times$0\farcs95 [$-$87$\degr$] & 0.2  \\
\multicolumn{1}{@{}l@{}}{
    \parbox[t]{4cm}{\raggedright CRBR15}} & 148 & 16:26:18.96 $-$24.24.14.55 & 1\farcs16$\times$0\farcs95 [$-$88$\degr$] & 0.1  \\
\multicolumn{1}{@{}l@{}}{
    \parbox[t]{4cm}{\raggedright SM1}} & 85 & 16:26:25.96 $-$24.23.39.12 & 1\farcs22$\times$0\farcs99 [$-$82$\degr$] & 0.1  \\
\multicolumn{1}{@{}l@{}}{
    \parbox[t]{4cm}{\raggedright CRBR12}} & 85 & 16:26:17.25 $-$24.23.45.64 & 1\farcs23$\times$0\farcs99 [$-$83$\degr$] & 0.05  \\
\multicolumn{1}{@{}l@{}}{
    \parbox[t]{4cm}{\raggedright GSS30-IRS1, IRS2, IRS3}} & 85 & 16:26:21.36 $-$24.23.03.55 & 1\farcs23$\times$1\farcs00 [$-$82$\degr$] & 0.06  \\
\multicolumn{1}{@{}l@{}}{
    \parbox[t]{4cm}{\raggedright LFAM3, S2}} & 85 & 16:26:23.55 $-$24.24.38.12 & 1\farcs23$\times$0\farcs99 [$-$82$\degr$] & 0.05  \\
\multicolumn{1}{@{}l@{}}{
    \parbox[t]{4cm}{\raggedright CRBR36}} & 85 & 16:26:25.45 $-$24.23.01.42 & 1\farcs23$\times$0\farcs99 [$-$82$\degr$] & 0.05  \\
\multicolumn{1}{@{}l@{}}{
    \parbox[t]{4cm}{\raggedright VLA~1623~W}} & 85 & 16:26:25.65 $-$24.24.27.77 & 1\farcs22$\times$0\farcs99 [$-$82$\degr$] & 0.06  \\
\multicolumn{1}{@{}l@{}}{
    \parbox[t]{4cm}{\raggedright CRBR15}} & 85 & 16:26:18.96 $-$24.24.13.99 & 1\farcs23$\times$0\farcs99 [$-$82$\degr$] & 0.04  \\ 
    \sidehead{\textbf{2019.A.00034.S}}
\multicolumn{1}{@{}l@{}}{
    \parbox[t]{4cm}{\raggedright GSS30-IRS1, IRS3}} & 219 & 16:26:21.72 $-$24.22.50.86 & 0\farcs38$\times$0\farcs28 [82$\degr$] & 0.03  \\
\enddata
\tablenotetext{a}{Measured at the centre of the primary beam.}
\end{deluxetable*}

\section{VLA Images}\label{apx:VLA_Obs}
\begin{figure*}[ht]
    \centering
    \includegraphics[height=0.88\textheight,width=\textwidth,keepaspectratio]{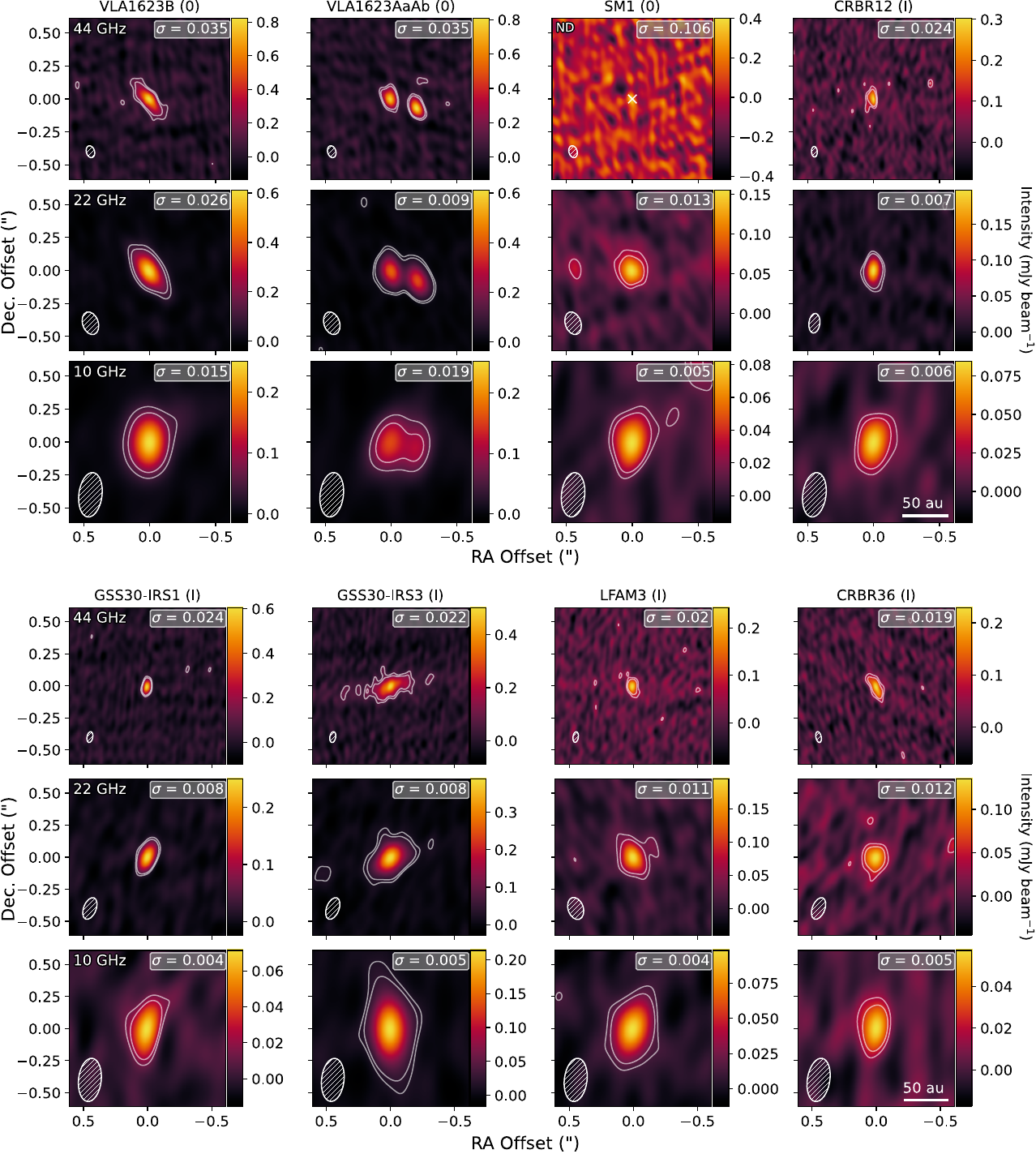}
    \caption{VLA Continuum images for Q (44\,GHz), K (22\,GHz) and X band (10\,GHz) and respective 3- and 5-$\sigma$ contours broken into two blocks. RMS values in mJy\,beam$^{-1}$ are shown in the top right of each panel. For the non-detection (ND) in SM1 we indicate the expected position with a white cross. Image beam sizes and position angles are shown as the hatched ellipse in the bottom left of each panel. We include a 50\,au scale bar in the bottom right panel of each block shared across all images. Each row corresponds to a single frequency and each column, per block, to a single object. Object names and classes are found in the column title.}
    \label{fig:VLA_Cont_0}
\end{figure*}

\begin{figure*}[ht]
    \centering
    \includegraphics[height=\textheight,width=\textwidth,keepaspectratio]{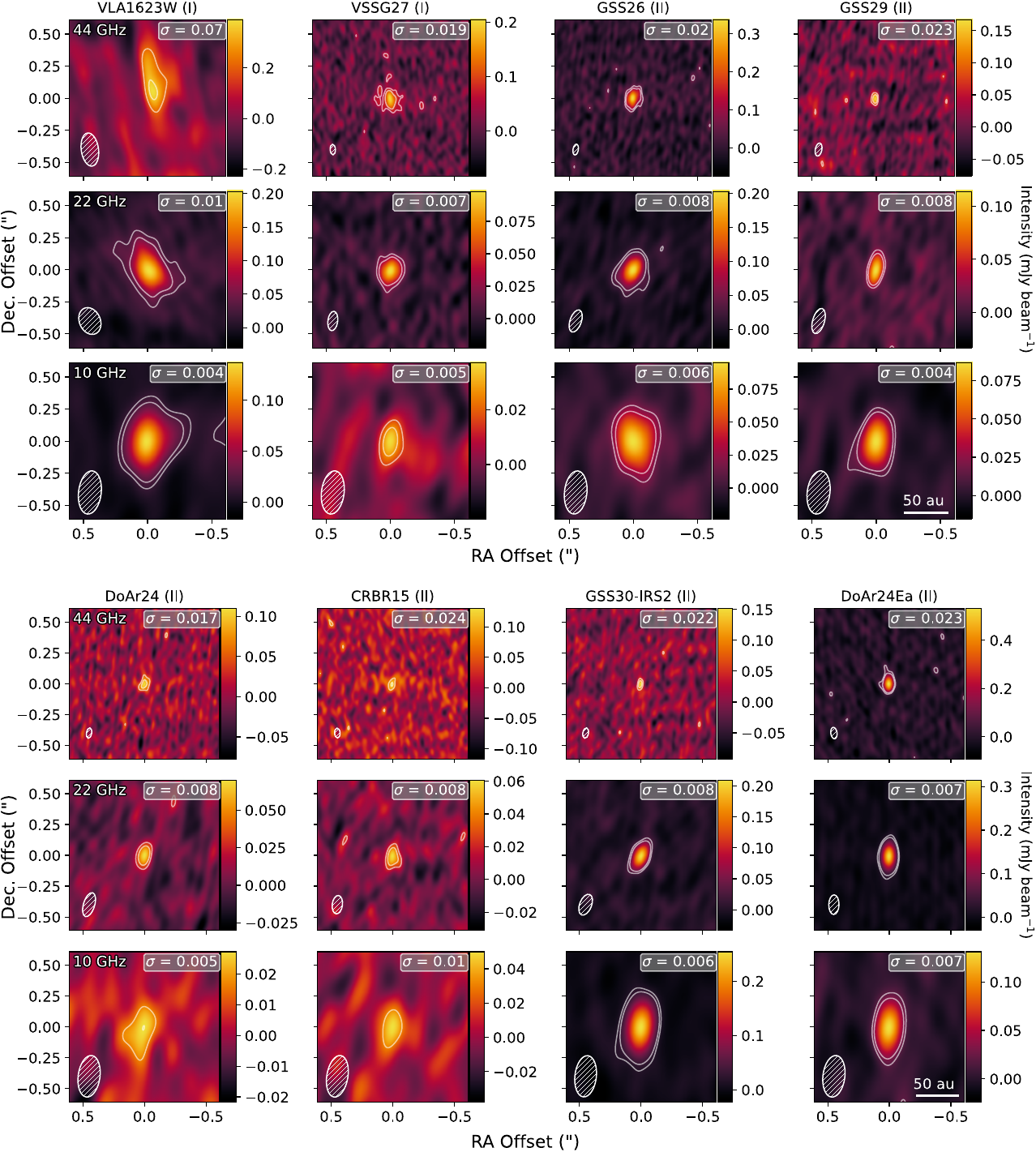}
    \caption{Continued from Figure~\ref{fig:VLA_Cont_0}}
    \label{fig:VLA_Cont_1}
\end{figure*}

\begin{figure*}[ht]
    \centering
    \includegraphics[width=\textwidth,keepaspectratio]{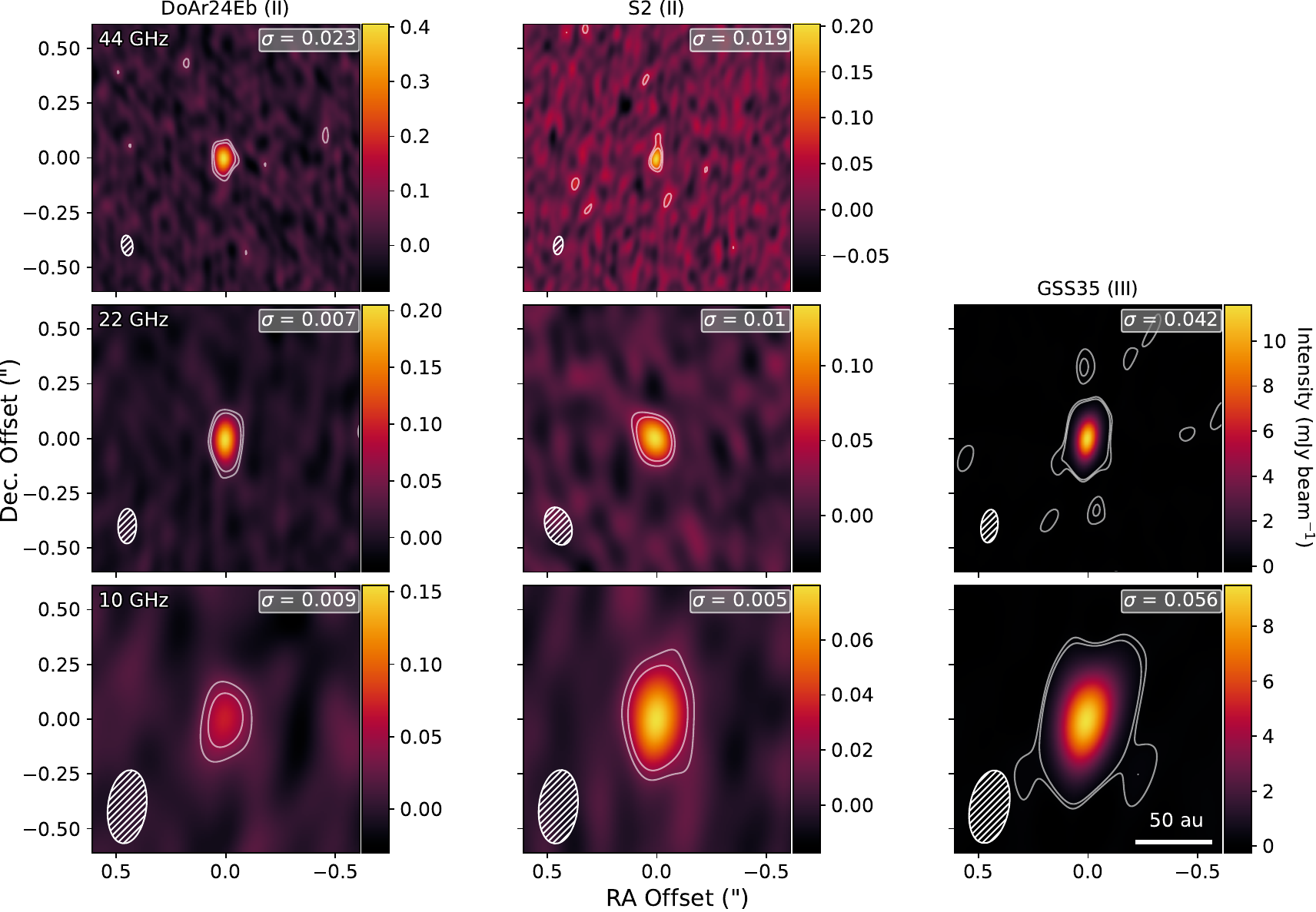}
    \caption{Continued from Figure~\ref{fig:VLA_Cont_0}. Note that GSS35 is beyond the FOV of our 44\,GHz observations.}
    \label{fig:VLA_Cont_2}
\end{figure*}

\clearpage

\section{GSS30-IRS1}\label{apx:IRS1_Bowshock}

\begin{figure*}[htbp]
    \centering
    \includegraphics[width=\textwidth,keepaspectratio]{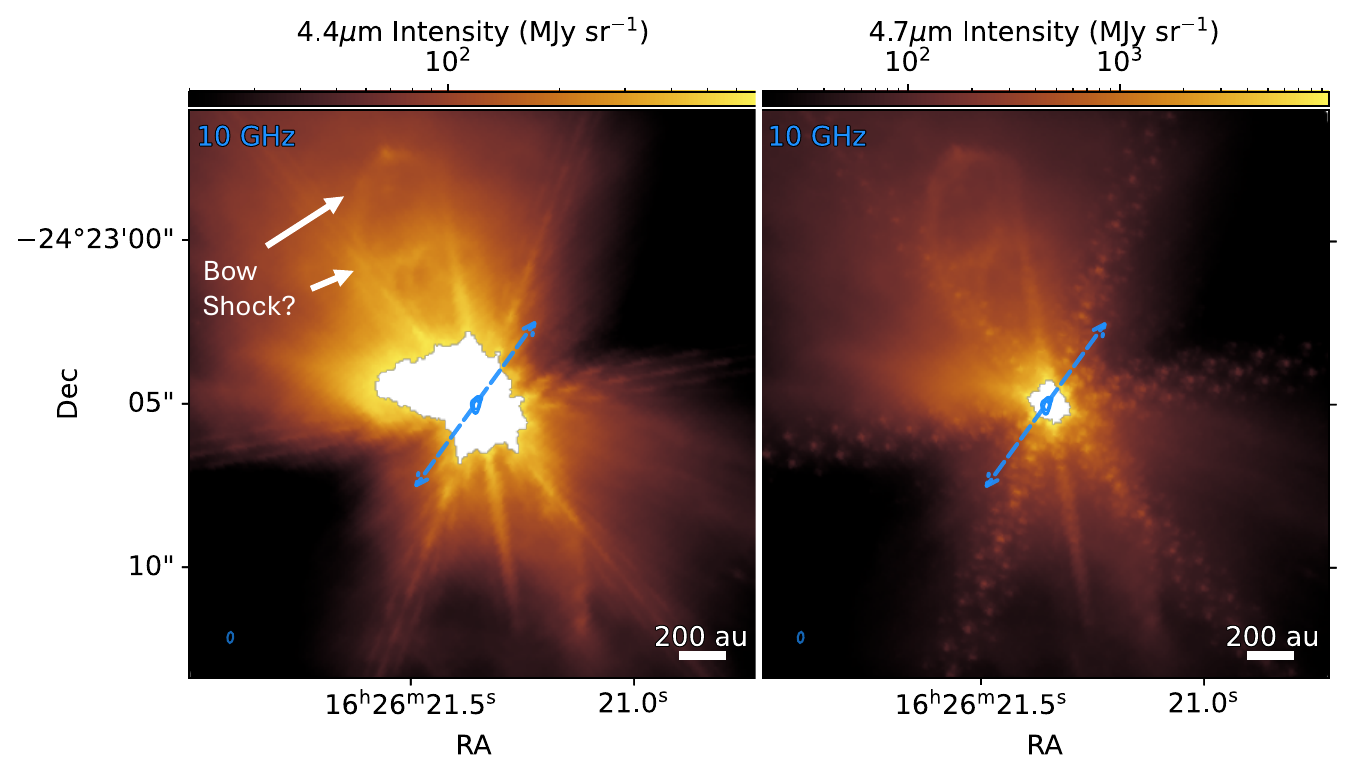}
    \caption{4.4\,\micron\, (left) and 4.7\,\micron\, (right) colour maps focused on GSS30-IRS1. We have applied a lognormal stretch with a minimum of 20\,MJy\,sr$^{-1}$ in order to accentuate potential structures in the image. We show VLA 10\,GHz contours in blue where contours are 3-, 5-, 10-, 20- and 100-$\sigma$ with $\sigma\sim\,4\,\mu$Jy. Beam sizes are shown as open ellipses in the bottom left and a 200\,au scale bar is shown in the bottom right of each panel. Finally, we indicate the position angle and its inverse of the 10\,GHz continuum for GSS30-IRS1 with blue dashed arrows.}
    \label{fig:GSS30-IRS1_BowShocks}
\end{figure*}

\clearpage

\section{Flux Densities Used in MCMC Fitting Procedures}\label{apx:MCMC_Fluxes}
\startlongtable
\begin{deluxetable*}{cccccc}
\tablecaption{Flux measurements and deconvolved morphologies measured for each object and compiled across this work, literature and archival observations. \label{tab:AllUsedFluxes}}
\tabletypesize{\footnotesize}
\tablehead{
\colhead{Frequency} &
\colhead{F$_{\text{int}}$} &
\colhead{RMS} &
\colhead{Deconvolved Diameter} &
\colhead{Deconvolved PA} &
\colhead{Reference/Project Code} \\
\colhead{(GHz)} &
\colhead{(mJy)} &
\colhead{(mJy\,beam$^{-1}$)} &
\colhead{(au)} &
\colhead{($\degr$)} &
\colhead{} 
}
\startdata
\sidehead{\textbf{VLA1623B}}
345 & 321$\pm$32 & $\cdots$ & $\cdots$ & $\cdots$ & \citealt{Harris18} \\
  217 & 107$\pm$11 & $\cdots$ & 38.8$\times$10.89 [$\pm\,$0.1$\times$0.03] & 41.9$\pm$0.2 & \citealt{Radley2025} \\
  93 & 21.6$\pm$1 & $\cdots$ & 36.7$\times$9.08 [$\pm\,$0.1$\times$0.03] & 41.6$\pm$0.2 & \citealt{Radley2025} \\
  44 & 1.8$\pm$0.2 & 0.035 & 19$\times$5.5 [$\pm\,$1$\times$0.4] & 45$\pm$4 & This Work \\
  22 & 0.84$\pm$0.09 & 0.026 & 18.6$\times$6.3 [$\pm\,$0.6$\times$0.3] & 47$\pm$2 & This Work \\
  10 & 0.35$\pm$0.02 & 0.015 & 25$\times$14.4 [$\pm\,$1$\times$0.6] & 79$\pm$3 & This Work \\ 
\sidehead{\textbf{VLA1623Aa}}
345 & 193$\pm$22 & $\cdots$ & $\cdots$ & $\cdots$ & \citealt{Harris18} \\
  217 & 45.4$\pm$5 & $\cdots$ & 14.73$\times$8.21 [$\pm\,$0.08$\times$0.04] & 30.9$\pm$0.5 & \citealt{Radley2025} \\
  93 & 9$\pm$0.5 & $\cdots$ & 13.02$\times$6.58 [$\pm\,$0.08$\times$0.04] & 30.8$\pm$0.5 & \citealt{Radley2025} \\
  44 & 1.2$\pm$0.1 & 0.035 & 10.2$\times$6.55 [$\pm\,$0.9$\times$0.4] & 17$\pm$6 & This Work \\
  22 & 0.47$\pm$0.05 & 0.009 & 8.8$\times$5.3 [$\pm\,$0.8$\times$0.4] & 163$\pm$3 & This Work \\
  10 & 0.21$\pm$0.02 & 0.019 & 29$\times$5.0 [$\pm\,$2$\times$0.8] & 124$\pm$3 & This Work \\
\sidehead{\textbf{VLA1623Ab}}
345 & 152$\pm$16 & $\cdots$ & $\cdots$ & $\cdots$ & \citealt{Harris18} \\
  217 & 39.7$\pm$4 & $\cdots$ & 13.61$\times$6.96 [$\pm\,$0.08$\times$0.04] & 28.2$\pm$0.5 & \citealt{Radley2025} \\
  93 & 8.8$\pm$0.4 & $\cdots$ & 12.12$\times$6.26 [$\pm\,$0.07$\times$0.04] & 30.0$\pm$0.5 & \citealt{Radley2025} \\
  44 & 1.2$\pm$0.1 & 0.035 & 10.5$\times$4.9 [$\pm\,$0.9$\times$0.4] & 33$\pm$5 & This Work \\
  22 & 0.45$\pm$0.05 & 0.009 & 11.8$\times$4.4 [$\pm\,$0.9$\times$0.4] & 42$\pm$3 & This Work \\
  10 & 0.11$\pm$0.01 & 0.019 & 23$\times$0.0 [$\pm\,$3$\times$0.8] & 142$\pm$3 & This Work \\ 
\sidehead{\textbf{SM1}}
370 & 366$\pm$5 & $\cdots$ & $\cdots$ & $\cdots$ & \citealt{Kawabe2018} \\
  359 & 327$\pm$1 & $\cdots$ & $\cdots$ & $\cdots$ & \citealt{Kawabe2018} \\
  345 & 350$\pm$10 & $\cdots$ & $\cdots$ & $\cdots$ & \citealt{Kawabe2018} \\
  230 & 118$\pm$2 & $\cdots$ & $\cdots$ & $\cdots$ & \citealt{Kawabe2018} \\
  107 & 28.8$\pm$2 & $\cdots$ & $\cdots$ & $\cdots$ & \citealt{Kirk2017} \\
  85 & 16.1$\pm$0.9 & 0.23 & 56$\times$36 [$\pm\,$3$\times$2] & 116$\pm$3 & 2017.1.00107.S \\
  44 & $<$0.32$\pm$0.1 & 0.11 & $\cdots$ & $\cdots$ & This Work \\
  22 & 0.25$\pm$0.04 & 0.013 & 18$\times$14 [$\pm\,$3$\times$2] & 105$\pm$33 & This Work \\
  10 & 0.098$\pm$0.01 & 0.0047 & $\cdots$ & $\cdots$ & This Work \\ 
\sidehead{\textbf{CRBR12}}
230 & 67.8$\pm$7 & $\cdots$ & $\cdots$ & $\cdots$ & \citealt{Cieza_2019} \\
  224 & 71.3$\pm$7 & 0.085 & 42.2$\times$32.3 [$\pm\,$0.3$\times$0.1] & 171.3$\pm$0.2 & 2017.1.00107.S \\
  148 & 30.6$\pm$2 & 0.11 & 43.8$\times$24.2 [$\pm\,$0.7$\times$0.5] & 163$\pm$1 & 2017.1.00107.S \\
  107 & 15.2$\pm$0.9 & $\cdots$ & $\cdots$ & $\cdots$ & \citealt{Kirk2017} \\
  85 & 10.7$\pm$0.5 & 0.04 & 35.4$\times$16.5 [$\pm\,$0.7$\times$0.5] & 166$\pm$1 & 2017.1.00107.S \\
  44 & 0.56$\pm$0.09 & 0.024 & 12$\times$6.4 [$\pm\,$2$\times$0.7] & 2$\pm$9 & This Work \\
  22 & 0.24$\pm$0.03 & 0.0074 & 10$\times$5.1 [$\pm\,$1$\times$0.4] & 13$\pm$3 & This Work \\
  10 & 0.092$\pm$0.01 & 0.0059 & $\cdots$ & $\cdots$ & This Work \\ 
\sidehead{\textbf{GSS30-IRS1}}
335 & 28.8$\pm$3 & $\cdots$ & $\cdots$ & $\cdots$ & \citealt{Encalada2021} \\
  224 & 13$\pm$1 & 0.29 & $\cdots$ & $\cdots$ & 2017.1.00107.S \\
  219 & 12.6$\pm$1 & 0.078 & 15.7$\times$8.7 [$\pm\,$0.4$\times$0.2] & 103$\pm$1 & 2019.A.00034.S \\
  148 & 6.3$\pm$0.4 & 0.17 & 52$\times$39 [$\pm\,$5$\times$4] & 168$\pm$8 & 2017.1.00107.S \\
  107 & 4.3$\pm$1 & $\cdots$ & $\cdots$ & $\cdots$ & \citealt{Kirk2017} \\
  85 & 2.8$\pm$0.2 & 0.049 & 50$\times$10 [$\pm\,$3$\times$2] & 54$\pm$4 & 2017.1.00107.S \\
  44 & 0.65$\pm$0.08 & 0.024 & 3.6$\times$1.4 [$\pm\,$0.6$\times$0.2] & 8$\pm$3 & This Work \\
  22 & 0.23$\pm$0.03 & 0.0078 & $\cdots$ & $\cdots$ & This Work \\
  10 & 0.073$\pm$0.008 & 0.0045 & 19$\times$0.0 [$\pm\,$4$\times$0.9] & 144$\pm$4 & This Work \\
\sidehead{\textbf{GSS30-IRS3}} 
225 & 124$\pm$12 & $\cdots$ & $\cdots$ & $\cdots$ & \citealt{SantamariaMiranda_2024} \\
  224 & 166$\pm$17 & 0.3 & 86.0$\times$32.4 [$\pm\,$0.4$\times$0.2] & 108.0$\pm$0.2 & 2017.1.00107.S \\
  219 & 161$\pm$16 & 0.026 & 76.1$\times$23.4 [$\pm\,$0.8$\times$0.2] & 109.4$\pm$0.3 & 2019.A.00034.S \\
  148 & 68.6$\pm$3 & 0.17 & 81.8$\times$31.0 [$\pm\,$0.6$\times$0.3] & 111.1$\pm$0.5 & 2017.1.00107.S \\
  107 & 31$\pm$2 & $\cdots$ & $\cdots$ & $\cdots$ & \citealt{Kirk2017} \\
  85 & 23$\pm$1 & 0.042 & 72.4$\times$25.7 [$\pm\,$0.5$\times$0.3] & 107.0$\pm$0.4 & 2017.1.00107.S \\
  44 & 2.1$\pm$0.2 & 0.022 & 25$\times$8.8 [$\pm\,$1$\times$0.6] & 116$\pm$5 & This Work \\
  22 & 0.82$\pm$0.09 & 0.0078 & 19.1$\times$4.7 [$\pm\,$0.7$\times$0.4] & 99$\pm$3 & This Work \\
  10 & 0.3$\pm$0.02 & 0.0045 & 31$\times$14.1 [$\pm\,$2$\times$0.5] & 22$\pm$2 & This Work \\
\sidehead{\textbf{LFAM3}}
230 & 139$\pm$14 & $\cdots$ & $\cdots$ & $\cdots$ & \citealt{Cieza_2019} \\
  224 & 132$\pm$13 & 0.11 & 106.1$\times$33.2 [$\pm\,$0.2$\times$0.1] & 49.0$\pm$0.2 & 2017.1.00107.S \\
  148 & 50.8$\pm$3 & 0.12 & 98.1$\times$29.3 [$\pm\,$0.6$\times$0.4] & 47.5$\pm$0.6 & 2017.1.00107.S \\
  91 & 18$\pm$0.9 & 0.03 & 85.7$\times$24.5 [$\pm\,$0.7$\times$0.2] & 48.3$\pm$0.6 & Ribas et al. (in prep.) \\
  85 & 15.1$\pm$0.8 & 0.045 & 88.0$\times$24.0 [$\pm\,$0.7$\times$0.5] & 49.4$\pm$0.7 & 2017.1.00107.S \\
  44 & 0.49$\pm$0.08 & 0.02 & 11$\times$5.9 [$\pm\,$2$\times$0.7] & 29$\pm$10 & This Work \\
  22 & 0.28$\pm$0.04 & 0.011 & 16$\times$12 [$\pm\,$2$\times$1] & 38$\pm$9 & This Work \\
  10 & 0.14$\pm$0.01 & 0.0041 & 25$\times$15 [$\pm\,$3$\times$1] & 121$\pm$5 & This Work \\
\sidehead{\textbf{CRBR36}}
335 & 99.3$\pm$10 & $\cdots$ & $\cdots$ & $\cdots$ & \citealt{Encalada2021} \\
  230 & 47.4$\pm$5 & $\cdots$ & $\cdots$ & $\cdots$ & \citealt{Cieza_2019} \\
  224 & 43.9$\pm$4 & 0.13 & 34.0$\times$19.1 [$\pm\,$0.6$\times$0.3] & 28.2$\pm$0.4 & 2017.1.00107.S \\
  148 & 18.3$\pm$0.9 & 0.12 & $\cdots$ & $\cdots$ & 2017.1.00107.S \\
  107 & 8.9$\pm$0.9 & $\cdots$ & $\cdots$ & $\cdots$ & \citealt{Kirk2017} \\
  91 & 7.4$\pm$0.4 & 0.029 & 27.5$\times$11.4 [$\pm\,$0.3$\times$0.2] & 27$\pm$1 & Ribas et al. (in prep.) \\
  85 & 6.5$\pm$0.3 & 0.039 & 35$\times$0.0 [$\pm\,$1$\times$0.8] & 36$\pm$2 & 2017.1.00107.S \\
  44 & 0.72$\pm$0.1 & 0.019 & 22$\times$7.1 [$\pm\,$3$\times$0.7] & 30$\pm$7 & This Work \\
  22 & 0.22$\pm$0.04 & 0.012 & 16$\times$8 [$\pm\,$3$\times$2] & 60$\pm$20 & This Work \\
  10 & 0.058$\pm$0.009 & 0.0051 & $\cdots$ & $\cdots$ & This Work \\
\sidehead{\textbf{VLA1623W}}
345 & 159$\pm$17 & $\cdots$ & $\cdots$ & $\cdots$ & \citealt{Harris18} \\
  224 & 62.7$\pm$6 & 0.28 & 91.7$\times$24.1 [$\pm\,$0.8$\times$0.6] & 10$\pm$1 & 2017.1.00107.S \\
  217 & 51.8$\pm$5 & $\cdots$ & 80.5$\times$12.2 [$\pm\,$0.8$\times$0.1] & 10$\pm$1 & \citealt{Radley2025} \\
  148 & 25.1$\pm$1 & 0.15 & 91$\times$32 [$\pm\,$1$\times$1] & 6$\pm$10 & 2017.1.00107.S \\
  107 & 11.2$\pm$1 & $\cdots$ & $\cdots$ & $\cdots$ & \citealt{Kirk2017} \\
  93 & 8.2$\pm$0.4 & $\cdots$ & 83$\times$11.0 [$\pm\,$1$\times$0.2] & 10$\pm$1 & \citealt{Radley2025} \\
  85 & 7.8$\pm$0.4 & 0.043 & 76$\times$15.8 [$\pm\,$1$\times$0.9] & 10$\pm$3 & 2017.1.00107.S \\
  44 & 0.9$\pm$0.2 & 0.07 & 77$\times$5 [$\pm\,$19$\times$2] & 9$\pm$7 & This Work \\
  22 & 0.37$\pm$0.06 & 0.01 & 35$\times$18 [$\pm\,$3$\times$1] & 29$\pm$6 & This Work \\
  10 & 0.2$\pm$0.02 & 0.0044 & 26$\times$19 [$\pm\,$2$\times$1] & 102$\pm$5 & This Work \\ 
\sidehead{\textbf{VSSG27}}
230 & 44.6$\pm$4 & $\cdots$ & 21.84$\times$21.84 [$\pm\,$0.3$\times$6] & 38$\pm$2 & \citealt{Cieza_2019} \\
  107 & 11.9$\pm$1 & $\cdots$ & $\cdots$ & $\cdots$ & \citealt{Kirk2017} \\
  44 & 0.78$\pm$0.2 & 0.019 & 17$\times$12 [$\pm\,$4$\times$3] & 10$\pm$30 & This Work \\
 22 & 0.21$\pm$0.03 & 0.0066 & 17$\times$13 [$\pm\,$2$\times$1] & 90$\pm$20 & This Work \\
  10 & 0.033$\pm$0.009 & 0.0053 & $\cdots$ & $\cdots$ & This Work \\ 
\sidehead{\textbf{GSS26}}
230 & 173$\pm$17 & $\cdots$ & 62.5$\times$62.5 [$\pm\,$0.2$\times$0.6] & 0.0$\pm$1.4 &  \citealt{Cieza_2019} \\
  107 & 27.7$\pm$2 & $\cdots$ & $\cdots$ & $\cdots$ & \citealt{Kirk2017} \\
  44 & 0.9$\pm$0.1 & 0.02 & 13$\times$9.7 [$\pm\,$1$\times$0.7] & 155$\pm$9 & This Work \\
  22 & 0.32$\pm$0.04 & 0.0075 & 15$\times$10.4 [$\pm\,$1$\times$0.7] & 93$\pm$5 & This Work \\
  10 & 0.15$\pm$0.02 & 0.0062 & 30$\times$16 [$\pm\,$4$\times$2] & 50$\pm$9 & This Work \\ 
\sidehead{\textbf{GSS29}}
230 & 3.4$\pm$0.4 & $\cdots$ & $<$27.68$\times$27.68\tablenotemark{a}  & $\cdots$ & \citealt{Cieza_2019} \\
  44 & 0.14$\pm$0.04 & 0.023 & $\cdots$ & $\cdots$ & This Work \\
  22 & 0.11$\pm$0.02 & 0.0082 & $\cdots$ & $\cdots$ & This Work \\
  10 & 0.087$\pm$0.008 & 0.0041 & $\cdots$ & $\cdots$ & This Work \\ 
\sidehead{\textbf{DoAr24}}
230 & 6.1$\pm$0.7 & $\cdots$ & $<$27.68$\times$27.68\tablenotemark{a}  & $\cdots$ & \citealt{Cieza_2019} \\
  44 & 0.24$\pm$0.06 & 0.017 & 11$\times$7 [$\pm\,$3$\times$2] & 126$\pm$23 & This Work \\
  22 & 0.055$\pm$0.01 & 0.0082 & $\cdots$ & $\cdots$ & This Work \\
  10 & 0.043$\pm$0.01 & 0.0054 & 35$\times$14 [$\pm\,$15$\times$5] & 140$\pm$20 & This Work \\
\sidehead{\textbf{CRBR15}}
230 & 21.2$\pm$2 & $\cdots$ & $\cdots$ & $\cdots$ & \citealt{Cieza_2019} \\
  224 & 29.1$\pm$3 & 0.086 & 29.1$\times$18.8 [$\pm\,$0.7$\times$0.3] & 150.0$\pm$0.4 & 2017.1.00107.S \\
  148 & 12.9$\pm$0.7 & 0.1 & 31$\times$16 [$\pm\,$2$\times$1] & 144$\pm$2 & 2017.1.00107.S \\
  107 & 5.4$\pm$0.6 & $\cdots$ & $\cdots$ & $\cdots$ & \citealt{Kirk2017} \\
  85 & 4.3$\pm$0.2 & 0.039 & $\cdots$ & $\cdots$ & 2017.1.00107.S \\
  44 & 0.18$\pm$0.06 & 0.024 & 10$\times$0.0 [$\pm\,$4$\times$0.8] & 150$\pm$10 & This Work \\
  22 & 0.086$\pm$0.02 & 0.0075 & 11$\times$9 [$\pm\,$4$\times$2] & 140$\pm$20 & This Work \\
  10 & 0.05$\pm$0.02 & 0.01 & $\cdots$ & $\cdots$ & This Work \\
\sidehead{\textbf{GSS30-IRS2}}
148 & 3.1$\pm$0.5 & 0.22 & 82$\times$43 [$\pm\,$19$\times$11] & 80$\pm$20 &  2017.1.00107.S \\
  85 & 0.31$\pm$0.1 & 0.059 & $\cdots$ & $\cdots$ & 2017.1.00107.S \\
  44 & 0.16$\pm$0.05 & 0.022 & $\cdots$ & $\cdots$ & This Work \\
  22 & 0.2$\pm$0.02 & 0.0084 & $\cdots$ & $\cdots$ & This Work \\
  10 & 0.26$\pm$0.02 & 0.006 & $\cdots$ & $\cdots$ & This Work \\ 
\sidehead{\textbf{DoAr24Ea}}
230 & 20.3$\pm$2 & $\cdots$ & 15$\times$8 [$\pm\,$2$\times$6] & 160$\pm$20 &  \citealt{Cieza_2019} \\
  44 & 0.74$\pm$0.09 & 0.023 & 7.3$\times$3.6 [$\pm\,$0.8$\times$0.3] & 142$\pm$5 & This Work \\
  22 & 0.34$\pm$0.04 & 0.0068 & 7.3$\times$3.3 [$\pm\,$0.7$\times$0.2] & 172$\pm$2 & This Work \\
  10 & 0.12$\pm$0.01 & 0.0065 & $\cdots$ & $\cdots$ & This Work \\ 
\sidehead{\textbf{DoAr24Eb}}
230 & 17.2$\pm$2 & $\cdots$ & $<$27.68$\times$27.68\tablenotemark{a}  & $\cdots$ & \citealt{Cieza_2019} \\
  44 & 0.66$\pm$0.09 & 0.023 & 9$\times$6.0 [$\pm\,$1$\times$0.5] & 155$\pm$7 & This Work \\
  22 & 0.23$\pm$0.03 & 0.0075 & 8$\times$3.6 [$\pm\,$1$\times$0.3] & 4$\pm$3 & This Work \\
  10 & 0.073$\pm$0.02 & 0.0091 & $\cdots$ & $\cdots$ & This Work \\ 
\sidehead{\textbf{S2}}
230 & 3.6$\pm$0.4 & $\cdots$ & $<$27.68$\times$27.68\tablenotemark{a} & $\cdots$ & \citealt{Cieza_2019} \\
  224 & 4$\pm$0.6 & 0.22 & $\cdots$ & $\cdots$ & 2017.1.00107.S \\
  148 & 2$\pm$0.3 & 0.16 & $\cdots$ & $\cdots$ & 2017.1.00107.S \\
  91 & 0.84$\pm$0.07 & 0.034 & $\cdots$ & $\cdots$ & Ribas et al. (in prep.) \\
  85 & 0.81$\pm$0.08 & 0.039 & $\cdots$ & $\cdots$ & 2017.1.00107.S \\
  44 & 0.24$\pm$0.04 & 0.019 & 6$\times$2.1 [$\pm\,$2$\times$0.4] & 4$\pm$7 & This Work \\
  22 & 0.14$\pm$0.02 & 0.01 & $\cdots$ & $\cdots$ & This Work \\
  10 & 0.1$\pm$0.01 & 0.0047 & 23$\times$14 [$\pm\,$4$\times$1] & 22$\pm$6 & This Work \\ 
\sidehead{\textbf{GSS35}}
22 & 12.6$\pm$1 & 0.042 & 5.92$\times$2.99 [$\pm\,$0.08$\times$0.02] & 154.7$\pm$0.2 & This Work \\
  10 & 10.5$\pm$0.5 & 0.056 & $\cdots$ & $\cdots$ & This Work \\
\enddata
\tablenotetext{a}{Unresolved in ODISEA observations so sizes are calculated from a beam of 0\farcs2}
\tablecomments{We derive the integrated flux, $F_{\rm int}$, from 2D Gaussian fits in image plane as described in Sections~\ref{sec:VLA_DataReduction} and \ref{sec:ALMA_DataReduction}. $\sigma_{\rm RMS}$ is calculated using annuli centred on each object with dimensions as described in Sections~\ref{sec:ALMA_DataReduction} and \ref{sec:VLA_obs}. Deconvolved object diameters and position angles (PA) are reported based on the FWHM of 2D Gaussian fits from PyBDSF with diameter uncertainties shown in square brackets.}
\end{deluxetable*}

\end{document}